\documentclass[useAMS,usegraphicx,usenatbib]{mn2e}

\usepackage{graphicx}
\usepackage{times}
\usepackage{amssymb}
\usepackage{amsmath}
\usepackage{euscript}
\usepackage{longtable,lscape}
\usepackage{color}
\def\210keV{{\rm\thinspace 2--10 keV}}
\usepackage{wasysym}

\title[The power output of local obscured and unobscured AGN]{The power output of local obscured and unobscured AGN: crossing the absorption barrier with Swift/BAT and IRAS}
\author[R. V. Vasudevan, A. C. Fabian, P. Gandhi, Winter, L. M. \& Mushotzky, R. F.]
{\parbox[]{6.in} {   R. V. Vasudevan$^1$, A. C. Fabian$^1$, P. Gandhi$^2$, Winter, L. M.$^3$ \& Mushotzky, R. F.$^4$ 
    \\
    \footnotesize
    $^{1}$Institute of Astronomy, Madingley Road, Cambridge CB3 0HA\\
    $^{2}$RIKEN Cosmic Radiation Lab, 2-1 Hirosawa, Wakoshi, Saitama 351-0198, Japan\\
    $^{3}$Center for Astrophysics and Space Astronomy, University of Colorado at Boulder, 440 UCB, Boulder, CO 80309-0440, USA\\
    $^{4}$Laboratory for High Energy Astrophysics, NASA/GSFC, Greenbelt, MD 20771, USA\\
  }}

\begin{document}

\maketitle

\begin{abstract}
The Swift/BAT 9-month catalogue of active galactic nuclei (AGN) provides an unbiased census of local supermassive black hole accretion, and probes to all but the highest levels of absorption in AGN.  We explore a method for characterising the bolometric output of both obscured and unobscured AGN by combining the hard X-ray data from the \emph{Swift}/BAT instrument (14--195keV) with the reprocessed IR emission as seen with the Infrared Astronomical Satellite (\emph{IRAS}) all-sky surveys.  This approach bypasses the complex modifications to the SED introduced by absorption in the optical, UV and 0.1--10 keV regimes and provides a long-term, average picture of the bolometric output of these sources.  We broadly follow the approach of Pozzi et al. for calculating the bolometric luminosities by adding nuclear IR and hard X-ray luminosities, and consider different approaches for removing non-nuclear contamination in the large-aperture \emph{IRAS} fluxes.  Using mass estimates from the black hole mass-host galaxy bulge luminosity relation, we present the Eddington ratios $\lambda_{\rm Edd}$ and 2--10 keV bolometric corrections for a subsample of 63 AGN (35 obscured and 28 unobscured) from the Swift/BAT catalogue, and confirm previous indications of a low Eddington ratio distribution for both samples.  Importantly, we find a tendency for low bolometric corrections (typically 10--30) for the obscured AGN in the sample (with a possible rise from $\sim15$ for $\lambda_{\rm Edd}<0.03$ to $\sim32$ above this), providing a hitherto unseen window onto accretion processes in this class of AGN. This finding is of key importance in calculating the expected local black hole mass density from the X-ray background since it is composed of emission from a significant population of such obscured AGN.  Analogous studies with high resolution IR data and a range of alternative models for the torus emission will form useful future extensions to this work.
\end{abstract}


\section{Introduction}
\label{Intro}

Accretion onto supermassive black holes (with masses of $\sim10^{6}-10^{9}$ solar masses) is responsible for the power output of active galactic nuclei (AGN).  Thermal emission from an accretion disc emerges in the optical--UV regime and manifests as the `Big Blue Bump' \citep{1978Natur.272..706S}, and inverse Compton scattering of UV disc photons by a corona above the disc is responsible for the X-ray emission \citep{1990ApJ...363L...1Z}.  This optical--to--X-ray emission constitutes the bulk of the intrinsic accretion emission in AGN.  In the standard paradigm for AGN, a dusty torus surrounding the accretion disc partially absorbs this emission and re-radiates it in the infrared \citep{1992ApJ...401...99P}, manifesting as an IR bump \citep{1994ApJS...95....1E}.  If the AGN is viewed through the torus, the optical--to--X-ray emission may be significantly absorbed along the line-of-sight, presenting difficulties in determining the true, bolometric accretion luminosity $L_{\rm bol}$ from the central engine.  \cite{2007MNRAS.381.1235V} and \cite{2009MNRAS.392.1124V} present $L_{\rm bol}$ (along with bolometric conversion ratios and accretion rates) determined from optical--to--X-ray spectral energy distributions (SEDs) for AGN, but both studies highlight the difficulties associated with determining $L_{\rm bol}$ when spectral complexity associated with absorption is present.  In optical surveys of AGN, it is possible for objects to be missed altogether because such absorption can reduce the flux severely enough (reddening) to be below the detection limit.  Although X-ray surveys are more effective at providing a census of AGN activity \citep{2004ASSL..308...53M}, significant columns of dusty gas (more than $\sim10^{22} \rm cm^{-2}$) will heavily reduce the X-ray (0.1--10 keV) flux and cause similar problems.  Heavily Compton-thick objects (those with column densities $N_{\rm H}\gg 10^{24} \rm cm^{-2}$) in particular can have X-ray fluxes reduced by many orders of magnitude from their intrinsic flux in this energy band \citep{2009ApJ...692..608I}.

The Burst Alert Telescope (BAT) on board the \emph{Swift} satellite is extremely useful at addressing these considerations to provide a more complete census of AGN activity.  The very hard X-ray bandpass of this instrument (14--195 keV) misses absorption signatures for moderate-to-high levels of absorption.  The 9-month catalogue of BAT-detected AGN (hereafter the \emph{Swift}/BAT catalogue, \citealt{2008ApJ...681..113T}) therefore provides an unprecedented level of completeness when surveying the AGN population, since it contains all but the most heavily absorbed objects.  The X-ray properties of the 153 AGN in the catalogue are presented in \cite{2009ApJ...690.1322W}, using data from a variety of X-ray missions to determine their intrinsic luminosities, spectral shapes (including measures of the spectral complexity) and absorbing column densities ($N_{\rm H}$).  \cite{2009arXiv0907.2272V} present simultaneous optical--to--X-ray SEDs from \emph{Swift}'s X-ray telescope (XRT) and UV--optical telescope (UVOT) and calculate the total accretion output $L_{\rm bol}$ and associated quantities for a subsample of 26 local, low-absorption, low-spectral-complexity AGN from the \emph{Swift}/BAT catalogue.  For their subsample of objects, the aforementioned problems associated with absorption are minimised or straightforwardly corrected for.  Their work provides hints that the local AGN population is dominated by AGN in which X-ray emission processes contribute significantly to the bolometric emission, manifesting as low bolometric corrections $\kappa_{\rm 2-10 keV}$ ($\kappa_{\rm 2-10 keV}=L_{\rm bol}/L_{\rm 2-10 keV}\sim10-30$ for 2--10 keV luminosity $L_{\rm 2-10 keV}$) and also suggests that accretion rates are low (Eddington ratios $\lambda_{\rm Edd}\lesssim 0.1$, where $\lambda_{\rm Edd}=L_{\rm bol}/L_{\rm Edd}$ for Eddington luminosity $L_{\rm Edd}= 1.3\times10^{38}(M_{\rm BH}/M_{\odot})$ for a black hole of mass $M_{\rm BH}$).  While such an approach is of particular utility for determining the total power output of unobscured AGN, an alternative approach is needed to address this question more generally for AGN of all absorption levels.  This study explores a method for crossing the `absorption barrier', and aims to characterise the bolometric output of local AGN in the Swift/BAT 9-month catalogue sources across the range of absorption properties probed by the catalogue. 

The hard X-ray data from the BAT instrument gathered over many months are available for the \emph{Swift}/BAT catalogue sources and are presented online.  These data provide a useful starting point for our absorption-unbiased study on the bolometric AGN power output.  However, instead of using optical--UV emission to constrain the thermal emission from the accretion disc, we use the reprocessed infrared (IR) emission to estimate the non-X-ray component of the bolometric luminosity $L_{\rm bol}$.  We take the re-processed infrared (IR) emission to be a severely averaged proxy for the intrinsic disc emission, since the timescale for transferral of energy from the high energy flux at the inner edge of the torus to emission as IR at its outer edge is of the order of several years in standard models \citep{1987ApJ...320..537B}; whereas optical, UV and X-ray variability in AGN is known to occur on far shorter timescales (a few minutes to hours).  The averaged accretion output represented by the IR can be used in conjunction with the long-term hard X-ray observations to estimate $L_{\rm bol}$.  We use the archival mid-to-far infrared observations from the Infrared Astronomical Satellite (\emph{IRAS}) all-sky catalogues to measure this reprocessed output; mid-infrared (MIR) observations in particular are well suited to our purposes since they capture the emission in the range where reprocessed AGN emission is expected to peak ($\sim12 \rm \mu m$) and the all-sky surveys readily available in the archives are ideal for identifying IR counterparts to X-ray sources in the all-sky \emph{Swift}/BAT catalogue.   Our method is a low-redshift, absorption-`neutral' extension of the work of \cite{2007A&A...468..603P} on high-redshift luminous obscured quasars, with the key difference being the use of hard X-ray BAT data instead of 2--10 keV X-ray data.

When considering the effects of absorption in AGN, it is important to note that the absorbing columns measured in different wavebands do not always match.  Under the standard unified scheme for AGN discussed above, one would expect an AGN with significant X-ray gas absorption to exhibit significant dust absorption in the optical--UV (if one assumes a relatively constant dust-to-gas ratio across the AGN population).  However, there are many classes of object which do not obey this simple picture, as discussed by \cite{2007ASPC..373..447M}.  However, statistically speaking, X-ray and optical absorptions are broadly correlated, and significant mismatch occurs in only about 10--20 per cent of the AGN populations studied in various surveys (e.g. \citealt{2007A&A...474..473G}, \citealt{2005A&A...444...79M}, \citealt{2005ApJ...618..123S}).  In any case, these considerations do not not significantly affect the approach used here, but we return to this issue when calculating the reprocessed IR emission.  In this paper, whenever the labels 'absorbed/obscured' or 'unabsorbed/unobscured' are used, they consistently refer to X-ray absorption.

The use of the IR emission as an indication of AGN activity is supported by numerous works. \cite{2008arXiv0807.4695M} find a strong correlation between the hard X-ray BAT luminosity and the Two-Micron All-Sky Survey (2MASS) J-band luminosity. \cite{2009MNRAS.394..491M} use analyses of X-ray observations from the literature and \emph{IRAS} data to calculate X-ray--to--IR luminosity ratios, and identify from these that the parameters specifying the physical AGN environment such as the torus geometry must span a narrow range.  The study of \cite{2009A&A...502..457G} discusses the correlation between X-ray (2--10 keV) emission and IR emission in detail for a sample of Seyfert nuclei, again using X-ray analyses from the literature and new IR data from the Very Large Telescope's Imager and Spectrometer for mid-Infrared (VISIR/VLT), taken specifically for addressing the issue of nuclear emission in local Seyferts.  Their data are the best estimates of the nuclear (non-stellar) IR flux in AGN to date, and show a strong correlation between intrinsic, uncontaminated nuclear IR monochromatic luminosity (at 12.3 $\rm \mu m$) and 2--10 keV X-ray luminosity.  Their work reinforces the idea that the uncontaminated MIR continuum is an accurate proxy for the intrinsic power of AGN, and reinforces and improves constraints on previous determinations of the correlation (\citealt{2008A&A...479..389H}, \citealt{2006A&A...457L..17H}).  They also highlight the usefulness of extending this approach to a very hard X-ray selected sample such as the one used here.  Their correlation also provides one way to estimate non-nuclear contamination in IR observations taken with larger apertures, as is the case with the \emph{IRAS} data used here.  The other method employed for correcting for non-nuclear contamination in this study is fitting of host galaxy and nuclear SED templates to the mid-to-far IR SEDs.


In this paper, we use the \emph{IRAS} and BAT data to generate SEDs from which the reprocessed IR and intrinsic X-ray emission can be determined, for calculation of $L_{\rm bol}$.  We calibrate this approach against integration of the optical--to--X-ray SEDs as presented in \cite{2009arXiv0907.2272V}, using the sources overlapping between the two studies for this purpose.  We note that although our \emph{IRAS} and BAT data are not contemporaneous, they both represent an averaged, long-term picture of accretion whereas the simultaneous optical--to--X-ray data presented in \cite{2009arXiv0907.2272V} are a `snapshot' of the accretion emission at a particular time.  That approach has numerous advantages (as discussed in \citealt{2009MNRAS.392.1124V}), but the average, long-term picture should be found to agree with `snapshots' when considering a reasonably large sample of objects, since the effects of variability should be averaged out over many sources.  Certainly, `quasi-simultaneous' data between hard X-ray and IR are preferable since ideally we want the IR to capture a reprocessed version of the same long-term state seen at higher energies, but the IR and hard X-ray bands can afford to be separated by much longer time periods to qualify as `quasi-simultaneous' due to the large timescales involved in reprocessing. We present the bolometric luminosities, Eddington ratios and bolometric corrections determined for all the local ($z<0.1$) objects in the \emph{Swift}/BAT catalogue with good quality \emph{IRAS} and BAT data, using black hole mass estimates calculated from K-band host bulge luminosity estimates via the method of \cite{2009arXiv0907.2272V} to determine accretion rates (apart from for Cyg A, where a known dynamical mass estimate is used).

This study can offer perspectives on unification schemes for AGN \citep{1993ARA&A..31..473A}, as it provides an idea of which type of accretion processes dominate in AGN (as parameterised by the X-ray bolometric correction) over a range of absorbing column densities and hence orientations.  A flat Universe ($\Omega_{M}+\Omega_{\Lambda}=1$) with a cosmology $H_{\rm 0}= 71 \rm km$ $\rm s^{-1} Mpc^{-1}$  and $\Omega_{\rm M}=0.27$ has been assumed throughout this work.

\section{Sample selection}
\label{sampleselection}

We firstly apply a redshift cut ($z<0.1$) on the 145 sources with spectral properties presented in \cite{2009ApJ...690.1322W} and exclude sources which would exhibit prominent jet contributions to their luminosity such as BL Lacs or blazars (see \citealt{2009ApJ...690.1322W}), yielding 116 potential objects.  We then exclude the two merging galaxies NGC 6921 and MCG +04-48-002 since AGN at such close proximity are very likely to be confused in the low-resolution IRAS images. We also exclude the galaxy NGC 6814 which is known to have a prominent cataclysmic variable in the foreground, yielding 113 potential objects.  

In assembling \emph{IRAS} data for this sample, we take care to ensure confident positional matches between the \emph{IRAS} catalogues and the NED positions.  We preferentially use the \emph{IRAS} Point Source Catalogue (PSC) as our source for IR photometry; if data was not available in this catalogue we then turned to the \emph{IRAS} Faint Source Catalogue (FSC), which also catalogues point sources.  There was some overlap between the PSC and FSC; in many cases the FSC provided greater wavelength coverage and data quality than the PSC.  The 12, 25, 60 and 100 $\rm \mu m$ photometry for these catalogues are available online \footnote{http://irsa.ipac.caltech.edu/applications/Gator/} along with useful diagnostic and data quality information.  The positional accuracy of the \emph{IRAS} photometry is dependent on a number of factors. We first of all restrict our searches to within 180 arcsec of the NED positions of our AGN (bearing in mind the PSF FWHM of $\sim1-5$ arcmin for IRAS), yielding 70 matches in the PSC and 73 matches in the FSC (with significant overlap between the two).  We then employ the information on the positional error ellipses for the \emph{IRAS} photometry provided for each object, to gauge the goodness of the positional match.  The semimajor and semiminor axes ($\sigma_{\rm maj}$, $\sigma_{\rm min}$) for each error ellipse are provided (in arcsec), and we determine an error circle of radius $\sigma_{\rm pos}=(\sigma_{\rm maj}\sigma_{\rm min})^{1/2}$ with the same area as the ellipse defined by these axes.  We require the \emph{IRAS} photometry to have a position within $3\sigma_{\rm pos}$ of the NED position, taking care to ensure that only one \emph{IRAS} source matched for each object to avoid confusion with other sources in the field.  For the purposes of probing the nuclear emission, we require that there are \emph{IRAS} data available at least one of 12 or 25 $\rm \mu m$ (both are preferred, to get an impression of the shape of the IR AGN SED).

We then enforce a quality criterion on the \emph{IRAS} photometry, requiring at least one of the 12--100 $\rm \mu m$ points to have a signal-to-noise ratio (SNR) greater than 4.0, and not using any data with a SNR $<$ 3.0.  The photometry satisfying the above stringent criteria were used in further analyses.  This yielded 71 out of 113 objects in the low-redshift sample. We identify the sources for which processed 9-month BAT data are available online \footnote{http://swift.gsfc.nasa.gov/docs/swift/results/bs9mon/} (see \citealt{2008ApJ...681..113T}).  We also require that an estimate of the black hole mass from the 2MASS Point Source Catalogue (PSC) magnitudes is available, using the method presented in \cite{2009arXiv0907.2272V}.  The 2MASS magnitude for the cD galaxy Cyg A yields a very low estimate of the black hole mass (log $M_{\rm BH}/M_{\odot}$=7.9), since the bulge is unusually large in this particular galaxy.  We therefore use the more reliable dynamical mass estimate from \cite{2003MNRAS.342..861T} for this object (log $M_{\rm BH}/M_{\odot}$=9.4). This finally yields 64 objects for further analysis, listed in Table~\ref{sampletable} (29 low absorption and 35 high absorption objects), constituting 57 per cent of the available low redshift sample.  Luminosity distances are computed from redshifts for most of the sample.  Due to the large number of NGC galaxies in the sample, we use the redshift-independent distance estimates from the Nearby Galaxies Catalogue where available \citep{1994yCat.7145....0T} and scale luminosities appropriately.  This only produces a pronounced change for two objects, NGC 4388 (where the change in luminosity from the redshift-estimated distance is a factor of $\sim 4$) and NGC 4051 (where the luminosity changes by a factor of $\sim 3$).  For all other NGC galaxies, the changes are negligible and less than the random errors from other sources.  To allow easy comparison with previous work in this area and on the 9-month catalogue, we finally bifurcate the sample on $N_{\rm H}$, producing two subsamples of low (log $N_{\rm H}<$22, 51 AGN) and high (log $N_{\rm H}>$22, 61 AGN) absorption (using the dividing line between `obscured' and `unobscured' sources suggested by \citealt{2009ApJ...690.1322W}).


\begin{table*}
\begin{tabular}{llll}

\hline
AGN&$N_{\rm H}/(10^{22}\rm cm^{-2})$&redshift&comments\\\hline
3C 382&--&$0.05787$&radio galaxy\\
ESO 548-G081&--&$0.01448$\\
Mrk 590&--&$0.026385$\\
Mrk 766&--&$0.012929$\\
Mrk 841&--&$0.036422$\\
NGC 4051&--&$0.002336$\\
NGC 7469&--&$0.016317$\\
NGC 985&--&$0.043143$\\
IRAS 05589+2828&$0.00_{}^{+0.04}$&$0.033$\\
Mrk 79&$0.0063$ (upper limit)&$0.022189$\\
Mrk 279&$0.013$ (upper limit)&$0.030451$\\
Mrk 509&$0.015_{-0.008}^{+0.008}$&$0.034397$\\
NGC 7213&$0.025_{-0.012}^{+0.011}$&$0.005839$\\
NGC 4593&$0.031_{-0.012}^{+0.011}$&$0.009$\\
2MASX J21140128+8204483&$0.047_{-0.021}^{+0.023}$&$0.084$&radio galaxy\\
NGC 5548&$0.07_{-0.05}^{+0.04}$&$0.017175$\\
ESO 511-G030&$0.098_{-0.021}^{+0.021}$&$0.022389$\\
3C 390.3&$0.12_{-0.03}^{+0.03}$&$0.0561$&radio galaxy\\
Mrk 290&$0.15_{-0.05}^{+0.03}$&$0.029577$\\
3C 120&$0.16_{-0.01}^{+0.01}$&$0.03301$&radio galaxy\\
MCG -06-30-015&$0.19_{-0.01}^{+0.03}$&$0.007749$\\
MCG +08-11-011&$0.25_{-0.015}^{+0.016}$&$0.020484$\\
ESO 490-G026&$0.33_{-0.02}^{+0.04}$&$0.02485$\\
NGC 3516&$0.353_{-0.12}^{+0.32}$&$0.008836$\\
NGC 931&$0.36_{-0.08}^{+0.08}$&$0.016652$\\
NGC 3783&$0.57_{-0.14}^{+0.21}$&$0.00973$\\
IC 4329A&$0.61_{-0.03}^{+0.03}$&$0.016054$\\
NGC 4945&$0.793_{-0.495}^{+0.957}$*&$0.001878$\\
IRAS 09149-6206&$0.85_{-0.17}^{+0.26}$&$0.0573$\\
NGC 7314&$1.16_{-0.14}^{+0.01}$&$0.004763$\\
NGC 2992&$1.19_{-0.96}^{+2.21}$&$0.00771$\\
Mrk 3&$1.24_{-1.24}^{+5.67}$&$0.013509$\\
4U 1344-60&$1.45_{-0.19}^{+0.20}$&$0.012879$\\
NGC 526A&$1.50_{-0.14}^{+0.14}$&$0.019097$\\
NGC 3227&$1.74_{-0.09}^{+0.12}$&$0.003859$\\
EXO 055620-3820.2&$2.57_{-0.14}^{+0.14}$&$0.03387$\\
NGC 5506&$2.78_{-0.05}^{+0.05}$&$0.006181$\\
NGC 2110&$2.84_{-0.16}^{+0.19}$&$0.007789$\\
Mrk 6&$3.26_{-1.19}^{+1.33}$&$0.018813$\\
2MASX J04440903+2813003&$3.39_{-0.25}^{+0.31}$&$0.011268$\\
NGC 6860&$4.53_{-1.30}^{+1.33}$&$0.014884$\\
ESO 005-G004&$5.58_{-0.16}^{+0.16}$&$0.006228$\\
NGC 7582&$7.39_{-1.00}^{+1.46}$&$0.005254$\\
NGC 7172&$8.19_{-3.30}^{+3.42}$&$0.008683$\\
Cyg A&$11.0_{-6.0}^{+21.0}$&$0.056075$&cD galaxy, radio galaxy\\
PGC 13946&$14.4_{-5.9}^{+7.4}$&$0.036492$\\
Mrk 348&$16_{-3}^{+4}$&$0.015034$\\
Mrk 1498&$17.84_{-1.82}^{+2.37}$&$0.0547$\\
Mrk 18&$18.25_{-2.71}^{+3.64}$&$0.011088$\\
NGC 6300&$21.5_{-0.9}^{+0.8}$&$0.003699$\\
ESO 103-035&$21.6_{-2.5}^{+2.6}$&$0.013286$\\
IC 5063&$21.78_{-2.06}^{+2.24}$&$0.011348$\\
NGC 4507&$34.28_{-4.57}^{+4.50}$&$0.011801$\\
NGC 4388&$36.17_{-3.82}^{+3.81}$&$0.008419$\\
MCG -03-34-064&$40.73_{-4.30}^{+4.79}$&$0.016541$\\
ESO 297-018&$41.71_{-2.90}^{+4.70}$&$0.0252$\\
\hline
\emph{continued on next page...}\\
\end{tabular}
\caption{Objects used in the sample.  Values of absorbing column density $N_{\rm H}$ are taken from \protect\cite{2009ApJ...690.1322W}\label{sampletable}.  * For NGC 4945, observations spanning 2--20 keV from the \emph{Ginga} satellite \protect\citep{1993ApJ...409..155I} yield a much higher value for $N_{\rm H}$ of $10^{24.7}\rm cm^{-2}$, placing it in the heavily `obscured' category; the optical images reveal a heavily dust-extincted edge-on galaxy.}
\end{table*}

\begin{table*}
\begin{tabular}{llll}

\hline
AGN&$N_{\rm H}/(10^{22}\rm cm^{-2})$&redshift&comments\\\hline

3C 403&$45.0_{-6.0}^{+7.0}$&$0.059$&radio galaxy\\
NGC 788&$46.89_{-4.47}^{+4.68}$&$0.013603$\\
ESO 506-G027&$76.82_{-6.79}^{+7.37}$&$0.025024$\\
NGC 1142&$79.75_{-3.05}^{+5.81}$&$0.028847$\\
NGC 5728&$82.0_{-5.0}^{+5.3}$&$0.009353$\\
NGC 3281&$86.3_{-16.12}^{+16.32}$&$0.010674$\\
NGC 1365&$104.8_{}^{}$&$0.005457$\\
NGC 612&$129.70_{-8.30}^{+12.90}$&$0.029771$&radio galaxy\\

\hline

\end{tabular}
\begin{center}
Table~\ref{sampletable} (continued)\\
\end{center}

\end{table*}

\section{Generating SEDs from BAT and IRAS data}

\subsection{BAT data}
\label{batdata}

The 4-channel BAT data are available readily processed online (see \S\ref{sampleselection}) for analysis with the X-ray analysis software \textsc{xspec}.  The key goal of using the BAT data is to obtain an accurate estimate of the intrinsic X-ray continuum.  The BAT data become particularly useful at determining the intrinsic spectral shape and normalisation in the more obscured sources (provided they are not too Compton thick, $N_{\rm H}\gg 10^{24} \rm cm^{-2}$).

For our purposes, we wish to extrapolate the 2--10 keV luminosities of these sources from their BAT data in order to calculate bolometric corrections.  These luminosities also allow comparison with the $L_{\rm 12.3 \mu m}-L_{\rm 2-10keV}$ correlation presented in \cite{2009A&A...502..457G}.  The key assumption needed when extrapolating to calculate $L_{\rm 2-10keV}$ is the spectral shape across the X-ray continuum, parameterised by the photon index $\Gamma$, assuming an intrinsic spectrum consisting of a power-law of the form $N(E)\propto E^{-\Gamma}$ across the whole 0.1--200 keV range.  The analysis of \cite{2008ApJ...681..113T} suggests an average photon index of $\sim2$ in the BAT energy band with a root-mean-square spread of $\sim0.3$.  We first consider the approach of calculating the intrinsic 2--10 keV luminosities by employing $\Gamma=2$ for all sources, but this yields systematically higher values of $L_{\rm 2-10keV}$ than those from X-ray analyses presented in \cite{2009ApJ...690.1322W}.  Although the BAT 9-month data are averaged over many months and the X-ray analyses in \cite{2009ApJ...690.1322W} are from individual shorter observations, statistically one expects better agreement, assuming the X-ray analyses have managed to recover the true intrinsic luminosity.

\begin{figure}
\includegraphics[width=8cm]{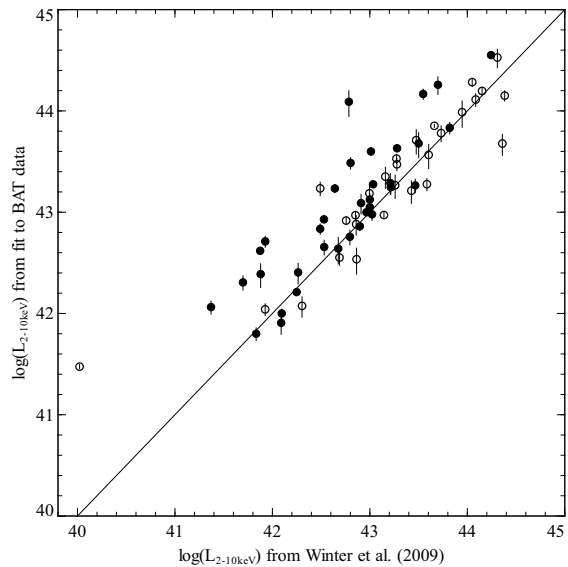}
    \caption{Comparison between 2--10keV X-ray luminosity extrapolated from fitting to the BAT data with the values presented in \protect\cite{2009ApJ...690.1322W}. The photon indices ($\Gamma$) from \protect\cite{2009ApJ...690.1322W} were used in fitting the BAT data, but were restricted to lie within $1.5<\Gamma<2.2$ on physical grounds.  Filled circles and empty circles represent objects with high and low absorption respectively (with objects above log($N_{\rm H}$)=22 classed as having high absorption).}
\label{LX_BAT_Winter09_comparison}
\end{figure}

A second approach involves using the values of the 2--10 keV photon indices presented in \cite{2009ApJ...690.1322W} to constrain the \emph{shape} of the overall X-ray spectrum, but using the BAT data to constrain the \emph{normalisation} and hence the luminosity.  The X-ray variability seen in the well-studied Seyfert 1 galaxy MCG-06-30-15 supports such an approach, as the observations are consistent with variation in normalisation while maintaining a relatively constant spectral shape (\citealt{2004MNRAS.348.1415V}, \citealt{2007PASJ...59S.315M}).  This produces better agreement and less of a systematic shift above the \cite{2009ApJ...690.1322W} values, but still results in a large scatter.  Part of this could be attributable to extreme values of $\Gamma$ found for some of these sources in the X-ray analyses.  On physical grounds, one expects a minimal photon index of $\sim1.5$ on canonical inverse-Comptonization scattering models for the coronal emission, and similarly, photon indices above $\sim2.2$ are likely to indicate complex absorption is preventing the recovery of the intrinsic shape, despite the inclusion of some absorption components.  In our final approach, we therefore constrain $\Gamma$ to lie within the hard limits $1.5 < \Gamma < 2.2$ when the values from \cite{2009ApJ...690.1322W} lie outside this range.  This correction is seen to be necessary for a number of the high absorption, `complex' spectrum sources identified in \cite{2009ApJ...690.1322W}. The scatter between the BAT values and the Winter et al. values is significantly reduced, and the results of this comparison are presented in Fig.~\ref{LX_BAT_Winter09_comparison}.  We note that there are two objects which deviate significantly from the one-to-one correspondence line: these are NGC 4945 at log($L_{\rm 2-10keV}^{\rm (Winter)}$)$\sim40.1$ (classified as unobscured in \citealt{2009ApJ...690.1322W}) and 3C 403 at log($L_{\rm 2-10keV}^{\rm (Winter)}$)$\sim42.8$ (classified as obscured).  Despite the identification of the former as unobscured in \cite{2009ApJ...690.1322W}, these are both known to have very complex spectra in which the 2--10 keV luminosity is prone to being underestimated with simple absorbed power-law fits (see e.g. \citealt{1993ApJ...409..155I}).  The BAT-extrapolated values for $L_{\rm 2-10keV}$ are likely to be more representative in these cases.

Having thus determined a sensible X-ray spectrum, we determine the total X-ray luminosity by integrating between 0.5 and 100 keV.  A cut-off of 500 keV (as used by \citealt{2007A&A...468..603P}) would not produce an appreciable difference in total X-ray luminosities for objects with $\Gamma \sim2$.  The effect would be $\sim20$ per cent in the X-ray luminosity, but since the dominant component of $L_{\rm bol}$ is the accretion disc emission (reprocessed or otherwise), this choice of cut-off does not have a big impact.  Additionally, a 500 keV cut would be outside the BAT bandpass. Recent work by Mushotzky et al. (in prep) suggests that the majority of high-energy cut-offs in BAT spectra for the 22-month catalogue AGN are located at $\sim 100$ keV, which would support the use of the upper integration limit energy used here.

We note that two objects in our high absorption sample could be borderline Compton-thick ($N_{\rm H}>10^{24}\rm cm^{-2}$) in \cite{2009ApJ...690.1322W}, namely NGC 612 and NGC 1365.  In the case of NGC 1365, this source has been identified to have variable absorption, with the source switching between Compton-thick and Compton-thin states on timescales of a few tens of kiloseconds \citep{2009ApJ...696..160R}.  There are additionally some sources which are identified as Compton-thick in the literature due to different approaches of modelling the spectrum (e.g. Mrk 3, \citealt{2008PASJ...60S.293A}, NGC 4945, \citealt{1993ApJ...409..155I}).  The galaxy NGC 4945 in particular may have $N_{\rm H}$ as high as $10^{24.7} \rm cm^{-2}$ \citep{1993ApJ...409..155I}, making it unsuitable for the type of $L_{\rm bol}$ calculation used here, and we therefore exclude it from our final bolometric correction results.  The other very high-absorption objects represent the limits of the approach outlined here and their results must be treated with caution.

\subsection{IRAS data}

The next step in the process is to determine the total nuclear IR luminosity, which, under the reprocessing paradigm, constitutes the remaining major component of the bolometric luminosity.  The \emph{IRAS} 12 and 25 $\rm \mu m$ fluxes are thought to be dominated by the nuclear component, but they are typically calculated for apertures significantly larger than the expected torus dimensions - indeed for the vast majority of local AGN, the torus cannot be resolved even with very high resolution imaging.  There is therefore a likelihood of significant non-nuclear flux contaminating the photometry, which we now turn to address.

\subsubsection{Host galaxy/starburst contamination in IRAS fluxes (1): using the $L_{\rm 12 \mu m}-L_{\rm 2-10keV}$ relation}
\label{IRAScorr_L12LX}

\begin{figure}
\includegraphics[width=8cm]{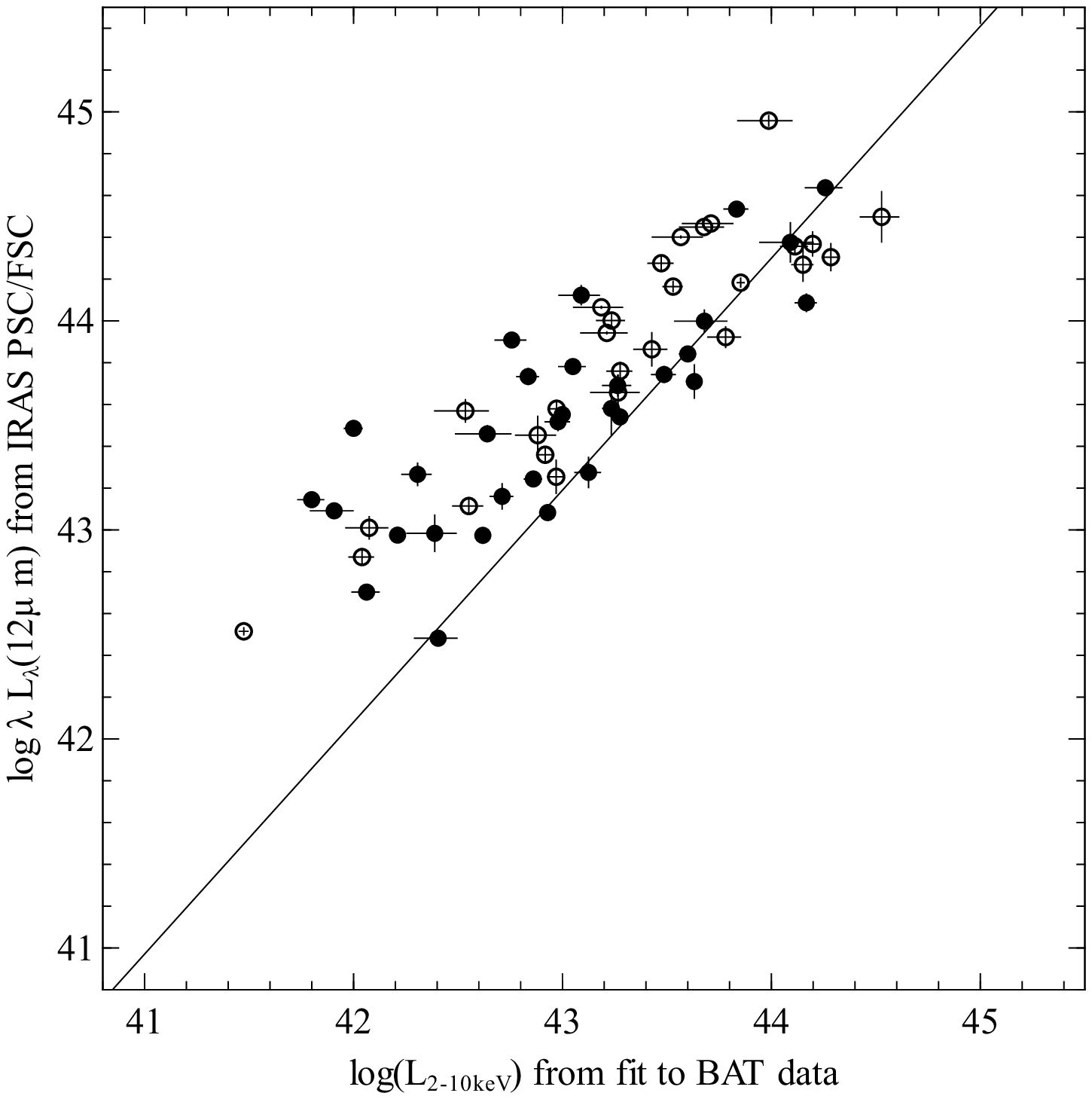}
    \caption{Infrared 12 $\rm \mu m$ luminosity against 2--10keV luminosity from fits to BAT data.  The solid line represents the correlation found for well-resolved sources from \protect\cite{2009A&A...502..457G}.  Filled circles and empty circles represent objects with high and low absorption respectively (with objects above log($N_{\rm H}$)=22 classed as having high absorption).}
\label{L12micron_vs_Xray_nogalaxycorr}
\end{figure}

\cite{2009A&A...502..457G} present a strong correlation between nuclear 12.3 $\rm \mu m$ and 2--10 keV luminosities for AGN.  Assuming the colour correction between 12 $\rm \mu m$ and 12.3 $\rm \mu m$ is negligible, we can compare the relationship between our \emph{IRAS} 12 $\rm \mu m$ luminosity and BAT-derived 2--10 keV luminosity with that from \cite{2009A&A...502..457G}, to get a statistcal estimate of the degree of host galaxy contamination present in the \emph{IRAS} fluxes at different X-ray luminosities.  We present these data in Fig.~\ref{L12micron_vs_Xray_nogalaxycorr}, in addition to the correlation found for well-resolved sources from \cite{2009A&A...502..457G}.  This comparison also assumes that the flux discrepancies introduced by the different filters on \emph{IRAS} and the VLT are negligible (the broad IRAS 12 $\rm \mu m$ filter could include many features contaminating the continuum, whereas the much narrower 12.3 $\rm \mu m$ VLT filter is likely to provide a better sample of the continuum).

\begin{figure}
\includegraphics[width=8cm]{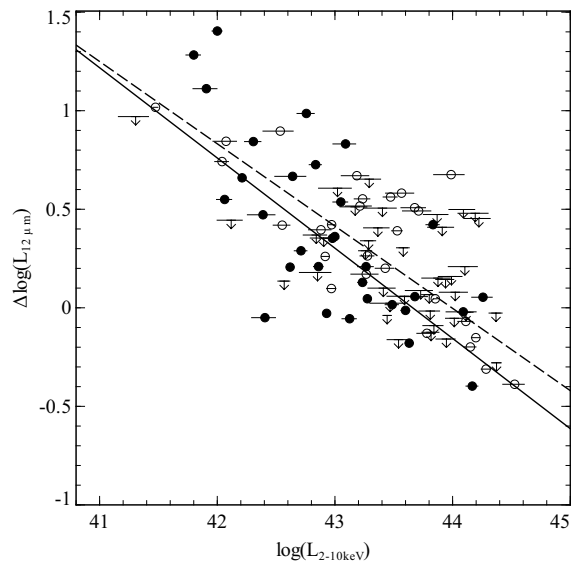}
    \caption{Offset $\Delta \rm log{L_{\rm 12 \mu m}}$ between \emph{IRAS} 12 $\rm \mu m$ fluxes and the relation of \protect\cite{2009A&A...502..457G}, plotted against $L_{\rm 2-10 keV}$. Detected objects are represented using black filled circles (obscured objects) and empty circles (unobscured objects), and upper limits are shown using downward pointing arrows.  The solid line is the best-fit taking upper limits into account (using the EM algorithm in the \textsc{asurv} utility), and the dashed line shows the fit from ignoring non-detections.  Over the luminosity range of the sample, the differences are not significant.}
\label{deltaL12micron_vs_Xray}
\end{figure}

It is clear that the non-nuclear excess in the 12 $\rm \mu m$ luminosity increases with lower X-ray luminosity; this is expected since at lower intrinsic nuclear luminosities (as traced by $L_{\rm 2-10 keV}$), it is more difficult for the nucleus to outshine the host galaxy.  This excess is likely to be due to star-formation.  We determine the excess with respect to the \cite{2009A&A...502..457G} relation for all the AGN with detections in \emph{IRAS} and plot them against $L_{\rm 2-10keV}$ in Fig.~\ref{deltaL12micron_vs_Xray}.  We also include upper limits for 39 AGN without \emph{IRAS} detections, taking care to check whether the objects in this list were in a part of the sky surveyed by \emph{IRAS}.  We take the completeness limit of the \emph{IRAS} FSC to be the upper limiting flux for these objects, $\sim 0.2$ Jy \footnote{http://irsa.ipac.caltech.edu/IRASdocs/surveys/fsc.html}.

We use the \emph{Astronomy Survival Analysis} (\textsc{asurv}) package from the \textsc{StatCodes} suite of utilities developed by Eric Feigelson \footnote{http://astrostatistics.psu.edu/statcodes/sc\_censor.html} to determine the correlation between the excess $\Delta$log$(L_{\rm 12 \mu m})$ and log$L_{\rm 2-10keV}$ including the effects of the upper limits.  The `EM' and `Buckley-James' algorithms available in \textsc{asurv} yield almost identical results.  We obtain the following non-nuclear 12 $\rm \mu m$ excess:

\begin{equation}
\Delta \rm log (L_{\rm 12 \mu m})= 19.988 - 0.4578 \thinspace \rm log(L_{\rm 2-10keV}).
\label{12micexcess}
\end{equation}

\begin{figure}
\includegraphics[width=8cm]{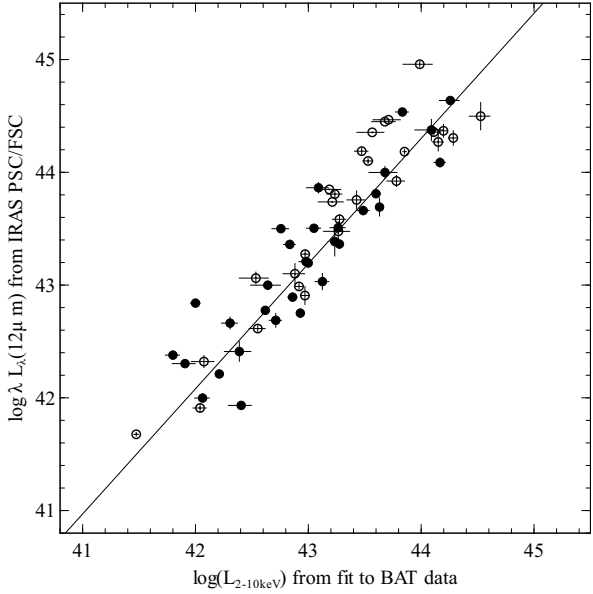}
    \caption{Infrared 12 $\rm \mu m$ luminosity against 2--10keV luminosity from fits to BAT data, corrected for the excess starburst/host galaxy flux seen in the large apertures for IRAS. The excess host contribution was obtained by comparing the correlation from our data with that from \protect\cite{2009A&A...502..457G}.}
\label{L12micron_vs_Xray_WITHgalaxycorr}
\end{figure}

The correction exhibits a significant dependence on luminosity: for AGN with $\rm log(L_{\rm 2-10keV})\sim 41$, the total IR 12 $\rm \mu m$ flux is estimated to be $\sim 20$ times larger than the nuclear flux alone, reducing to zero at X-ray luminosities of $\rm log(L_{\rm 2-10keV})\sim 43.5$.  We then correct the 12 $\rm \mu m$ luminosities according to this correction (up until the threshold at which the correction becomes negative) and re-plot the results in Fig.~\ref{L12micron_vs_Xray_WITHgalaxycorr}.

In this approach, we assume that the 25 $\rm \mu m$ luminosity needs to be corrected by the same factor, despite a possible larger non-nuclear contribution at longer wavelengths.  The larger excess expected at 25 $\rm \mu m$ would further decrease the total IR nuclear luminosities obtained.  However, since we do not have any information on the 25$\rm \mu m$--$L_{\rm 2-10keV}$ correlation, an estimation of the extra correction required is beyond the scope of this paper.

The $L_{\rm 12 \mu m}-L_{\rm 2-10 keV}$ correlation in Seyferts itself presents a possibility for predicting the approximate range of bolometric corrections under the reprocessing paradigm, since the $L_{\rm 12 \mu m}-L_{\rm 2-10 keV}$ relation links the two bands used in our calculation of $L_{\rm bol}$.

We take the bolometric luminosity $L_{\rm bol}$ to be the sum of the total integrated IR luminosity and the total X-ray luminosity, i.e. $L_{\rm bol}=L_{\rm IR,tot} + L_{\rm X, tot}$.  Assuming that the monochromatic 12 $\rm \mu m$ luminosity scales up to the total integrated luminosity by some factor $K$, dependent on the IR SED and including the effects of geometry and anisotropy (see \S\ref{totalIRlumin} for details), we can re-write this as:

\begin{equation}
L_{\rm bol} = K\thinspace L_{\rm 12 \mu m} + L_{\rm X,tot}.
\end{equation}
This can be converted into an expression for the bolometric correction $\kappa_{\rm 2-10keV}$ by dividing through by $L_{\rm 2-10keV}$:

\begin{equation}
\kappa_{\rm 2-10 keV} = K\thinspace \frac{L_{\rm 12 \mu m}}{L_{\rm 2-10keV}} + \frac{L_{\rm X,tot}}{L_{\rm 2-10 keV}}.
\end{equation}
The final part of the expression $L_{\rm X,tot}/L_{\rm 2-10 keV}$ is constrained to lie within 3.1 and 5.3 by the range of photon indices $1.5<\Gamma<2.2$ adopted here as discussed in \S\ref{batdata}.  We can then include the $L_{\rm 12 \mu m}-L_{\rm 2-10 keV}$ relation as $L_{\rm 12 \mu m} = 10^{a} L_{\rm 2-10keV}^{b}$ to finally obtain

\begin{equation}
\kappa_{\rm 2-10 keV} = K\thinspace 10^{a} L_{\rm 2-10keV}^{(b-1)} + \frac{L_{\rm X,tot}}{L_{\rm 2-10 keV}}.
\label{bolcor_predictedbyL12LX}
\end{equation}
The distribution of bolometric corrections seen will therefore depend on the diversity of IR SED shapes and assumptions about torus geometry and emission (via $K$), the degree of spread intrinsic to the measured $L_{\rm 12 \mu m}-L_{\rm 2-10 keV}$ relation (via $a$ and $b$) and the amount of variation in the X-ray spectral shape (via $L_{\rm X,tot}/L_{\rm 2-10 keV}$).  The nuclear SED templates from \cite{2004MNRAS.355..973S} used in this paper, in conjunction with assumptions about torus geometry and anisotropy of emission (\S\ref{totalIRlumin}) yield a range in $K$ of $4-15$.  Combining this with the spread in $a$ and $b$ reported in \cite{2009A&A...502..457G} and the range of X-ray SED shapes discussed above, this predicts bolometric corrections ranging from 6-60 over the luminosity range $41.0<\rm log(L_{\rm 2-10 keV})<45.0$ spanned by the objects considered here, with an increase in $\kappa_{\rm 2-10keV}$ with $L_{\rm 2-10 keV}$ as predicted by equation (\ref{bolcor_predictedbyL12LX}).  Therefore, by requiring that our objects lie on the $L_{\rm 12 \mu m}-L_{\rm 2-10 keV}$ relation, we naturally expect a certain distribution of bolometric corrections, with the degree of spread related to the tightness of the $L_{\rm 12 \mu m}-L_{\rm 2-10 keV}$ relation observed.  The tighter the distribution, the smaller the range of bolometric corrections will be, centering around 10--30.  Since, under the assumptions used in our paper, the requirement of a tight $L_{\rm 12 \mu m}-L_{\rm 2-10 keV}$ relation can limit the range of bolometric corrections allowed, we also consider an approach for removal of the non-nuclear component which is independent of the $L_{\rm 12 \mu m}-L_{\rm 2-10 keV}$ relation.

\subsubsection{Host galaxy/starburst contamination in IRAS fluxes (2): using host galaxy IR SED templates}
\label{IRAScorr_Silva04}

\begin{figure}
\includegraphics[width=8cm]{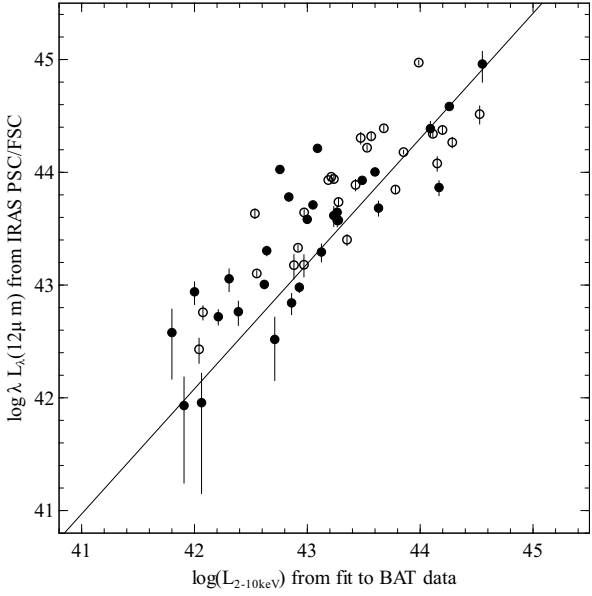}
    \caption{Nuclear 12 $\rm \mu m$ luminosity against 2--10keV luminosity from fits to BAT data, from combined host and nuclear IR SED template fitting to the IRAS 12--100 $\rm \mu m$ data.}
\label{L12micron_vs_Xray_Silva04hostgalcor}
\end{figure}

We also employ the strategy used by \cite{2007A&A...468..603P} to account for the host galaxy, namely fitting the nuclear and host galaxy SED templates from \cite{2004MNRAS.355..973S} to the broad-band (12-100 $\rm \mu m$) IR data.  This does not presuppose any relation between $L_{IR}$ and $L_{\rm 2-10 keV}$, but we can also determine whether this approach to correcting for the host galaxy reproduces the $L_{\rm 12 \mu m}-L_{\rm 2-10 keV}$ relation.  We select from the host galaxy SED templates for different luminosity bins presented in \cite{2004MNRAS.355..973S} using our estimate of $L_{\rm 2-10 keV}$ from the BAT data.  In a few cases where 60 or 100 $\rm \mu m$ data were not available, the normalisation of the host template was tied to the nuclear template normalisation, to provide a sensible estimate of host galaxy contamination based on the observations of \cite{2004MNRAS.355..973S} in their sample of AGN.  Some example SEDs with both host and nuclear SED templates fitted are shown in Fig.~\ref{example_SEDs_silvaetal04hostcorr}.  We also present the distribution of $L_{\rm 12 \mu m}$ against $L_{\rm 2-10 keV}$ in Fig.~\ref{L12micron_vs_Xray_Silva04hostgalcor}, with the nuclear $L_{\rm 12 \mu m}$ determined from the 12 $\rm \mu m$ flux of the nuclear part of the SED fit only.

\begin{figure}
\includegraphics[width=8.5cm]{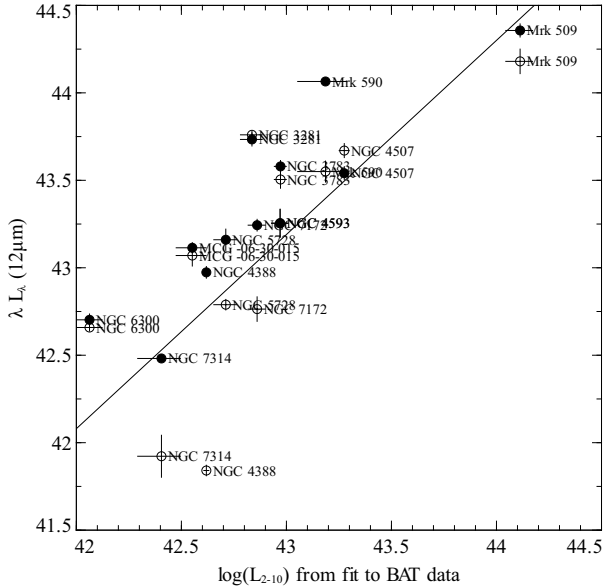}
    \caption{Infrared 12 $\rm \mu m$ luminosity against 2--10keV luminosity from fits to BAT data.  Black points are IRAS data uncorrected for host galaxy contamination, and empty circles show the nuclear 12 $\rm \mu m$ luminosity from the VISIR/VLT data presented in \protect\cite{2009A&A...502..457G}.}
\label{L12micron_vs_Xray_Gandhi09compare_IRASandVISIR}
\end{figure}

In the absence of high-resolution data for a large fraction of our sample, these methods for removing the host galaxy can only go so far in tackling this complex issue.  In Fig.~\ref{L12micron_vs_Xray_Gandhi09compare_IRASandVISIR} we present a comparison of our uncorrected \emph{IRAS} fluxes with the accurate nuclear VLT/VISIR fluxes from \cite{2009A&A...502..457G}.  It can be seen that in some cases, the degree of host contamination in the \emph{IRAS} fluxes can be very large (e.g. NGC 4388); while in others the \emph{IRAS} fluxes are almost identical to the nuclear fluxes from VISIR/VLT (e.g. NGC 6300), indicating that no correction for contamination is necessary.  The two methods presented here go some way in accounting for aperture effects, but as discussed by \cite{2008A&A...479..389H}, higher resolution IR data which isolates the nuclear emission is ultimately much preferred.

\subsubsection{Total nuclear IR luminosity}
\label{totalIRlumin}

We broadly follow the approach of \cite{2007A&A...468..603P} outlined in their study on high-redshift, luminous, obscured quasars and fit the nuclear and host galaxy templates from \cite{2004MNRAS.355..973S} to our data.  The appropriate nuclear template from \cite{2004MNRAS.355..973S} is selected based on the absorbing column $N_{\rm H}$ reported in \cite{2009ApJ...690.1322W}.  The fitting is performed using a simple least-squares algorithm from the \textsc{scipy} suite of functions available for use with the \textsc{python} programming language.
If the host galaxy contamination is removed via method 1 (\S\ref{IRAScorr_L12LX}) above, we only fit the 12-25 $\rm \mu m$ data and correct the integrated nuclear IR luminosity by the fraction given by equation (\ref{12micexcess}).  If method 2 (\S\ref{IRAScorr_Silva04}) is used, only the nuclear component of the fit is integrated.  The observed IR luminosity is then obtained by integration of the nuclear template between 1 and 1000 $\rm \mu m$.

We reiterate briefly here that the absorptions seen in X-rays and the amouts of dust responsible for reddening the optical--UV SED/producing the reprocessed IR do not necessarily match (e.g. \citealt{2007ASPC..373..447M}).  This calls into question the use of the X-ray $N_{\rm H}$ for selecting the template to use from \cite{2004MNRAS.355..973S}; however, the fact that \cite{2004MNRAS.355..973S} find from observations that the IR SED shapes of AGN can be broadly grouped on $N_{\rm H}$ would argue for at least a statistical correlation between the two types of absorption, in addition to the findings from surveys discussed in \S\ref{Intro}.  In this first-order approach, the selection of nuclear SED template based on $N_{\rm H}$ is therefore sensible.

Under method (2) for host galaxy removal, a few objects (Mrk 841, NGC 612, PGC 13946, 2MASX J04440903+2813003, EXO 055620-3820.2, 4U 1344-60,  ESO 297-018, NGC 7314) yield a fit with zero nuclear component and only a host galaxy component.  These probably represent cases where the host galaxy contamination is very large and the nucleus is too buried to get an accurate estimate of its presence from our simple SED fitting alone.  We exclude these objects from the results from method (2) due to these uncertainties.  A detailed exploration of the different host and nuclear SED models may be able to yield more satisfatory fits in these cases, but is beyond the scope of this paper.

To convert this observed IR luminosity into a measure of the reprocessed nuclear accretion disc luminosity, correction factors are required to account for the geometry of the torus and the anisotropy of the torus emission.  We adopt the same approach as \cite{2007A&A...468..603P} in correcting for these effects.  The correction for geometry relates to the covering factor $f$ which obscures the optical--UV emission from the accretion disc from view.  Their geometry correction is based on a statistical argument, employing the ratio of obscured to unobscured quasars as found by recent X-ray background synthesis models \citep{2007A&A...463...79G} to infer a typical torus covering fraction of $f\approx0.67$.  This value is also consistent with the covering fraction obtained from recent detailed clumpy torus models \citep{2008ApJ...685..160N}, for typical torus parameters (e.g. number of line-of-sight clouds $\sim5$, with an opening angle of $\sim30-45^{\circ}$). Inverting this, we obtain a factor $\sim1.5$ by which the observed IR nuclear luminosity needs to be multiplied to obtain the total optical--UV accretion emission. \cite{2007A&A...468..603P} estimate the anisotropy correction factors by computing the ratio of the luminosities from face-on vs. edge-on AGN (with the column density $N_{\rm H}$ parameterising the inclination of the torus).  They normalise the \cite{2004MNRAS.355..973S} templates for all different $N_{\rm H}$ values to have the same 30--100 $\rm \mu m$ luminosities.  The luminosity ratios for face-on to edge-on AGN are then calculated in the 1--30 $\rm \mu m$ range, where the effects of anisotropy are most pronounced.  They obtain values of $\sim 1.2-1.3$ for $10^{22}<N_{\rm H}<10^{24} \rm cm^{-2}$ and $\sim 3-4$ for $N_{\rm H}> 10^{24} \rm cm^{-2}$ sources.  For face on (low absorption $\rm log N_{\rm H}<22$) sources, no correction is necessary.  We adopt a simple approach, using values of $1.3$ for $\rm 22<log N_{\rm H}<24$ sources and $3.5$ for $\rm{log} N_{\rm H} > 24$ sources.

\section{Calculating the total power output}

\begin{figure*}
\includegraphics[width=4.5cm]{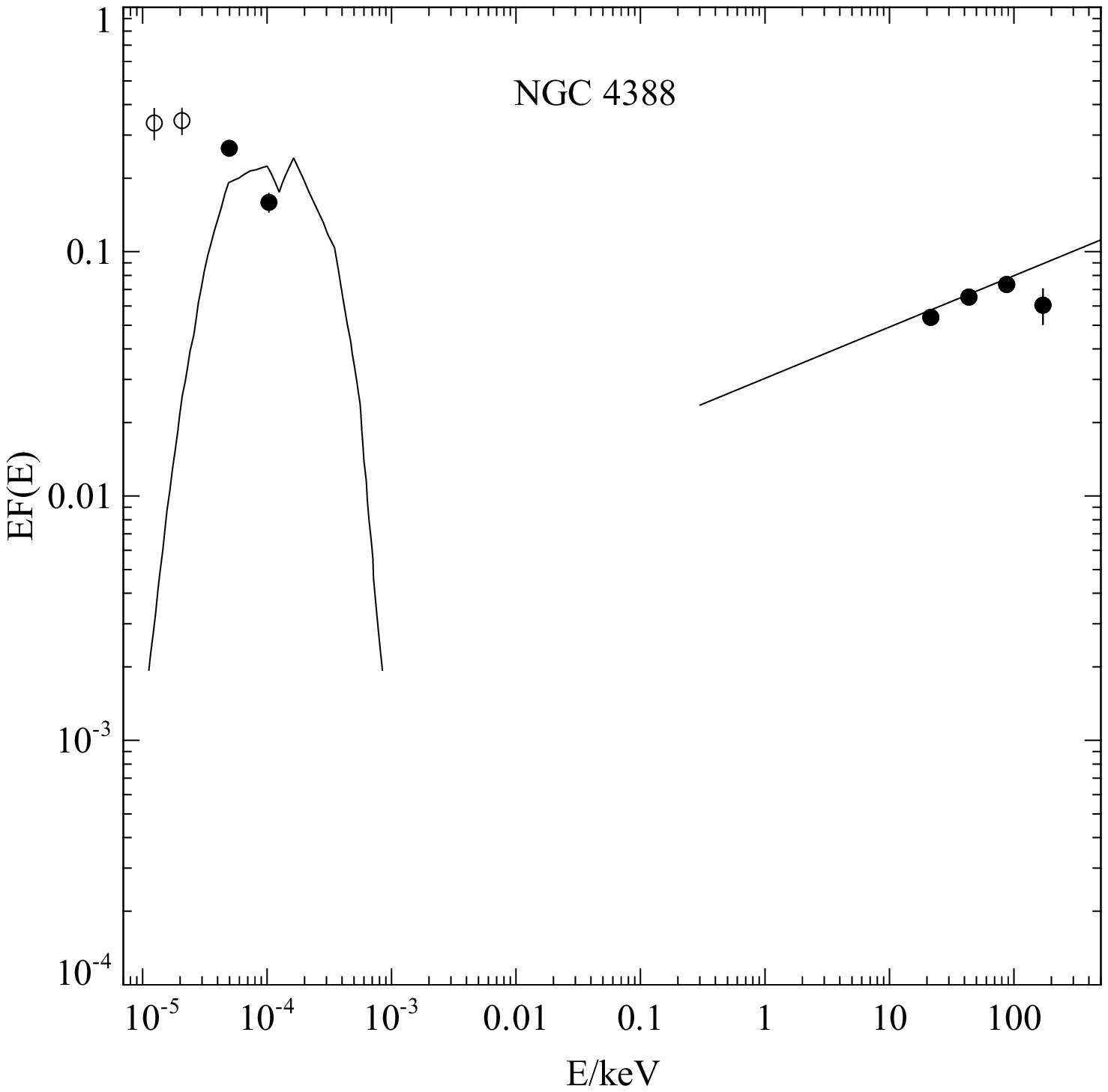}
\includegraphics[width=4.5cm]{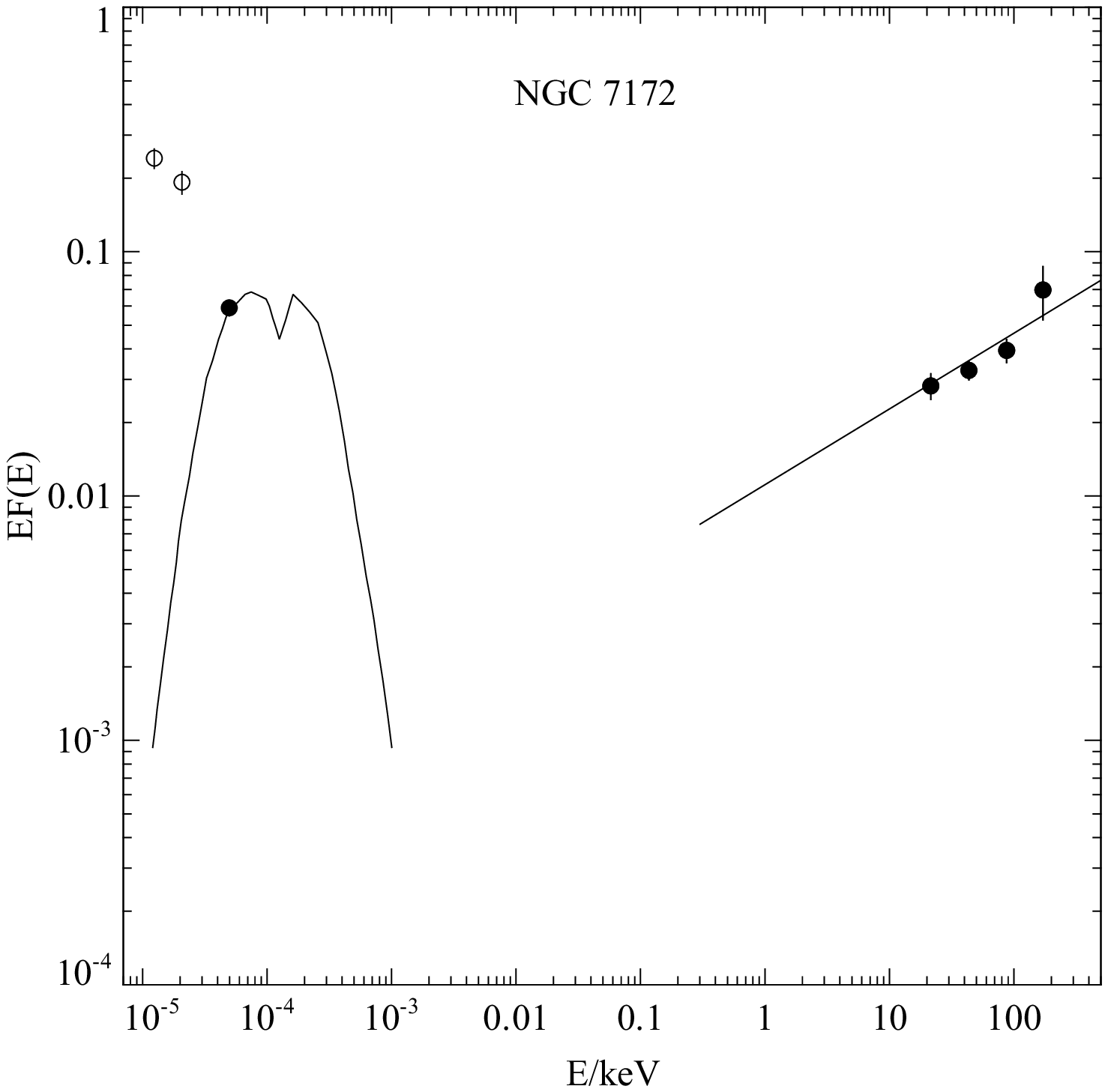}
\includegraphics[width=4.5cm]{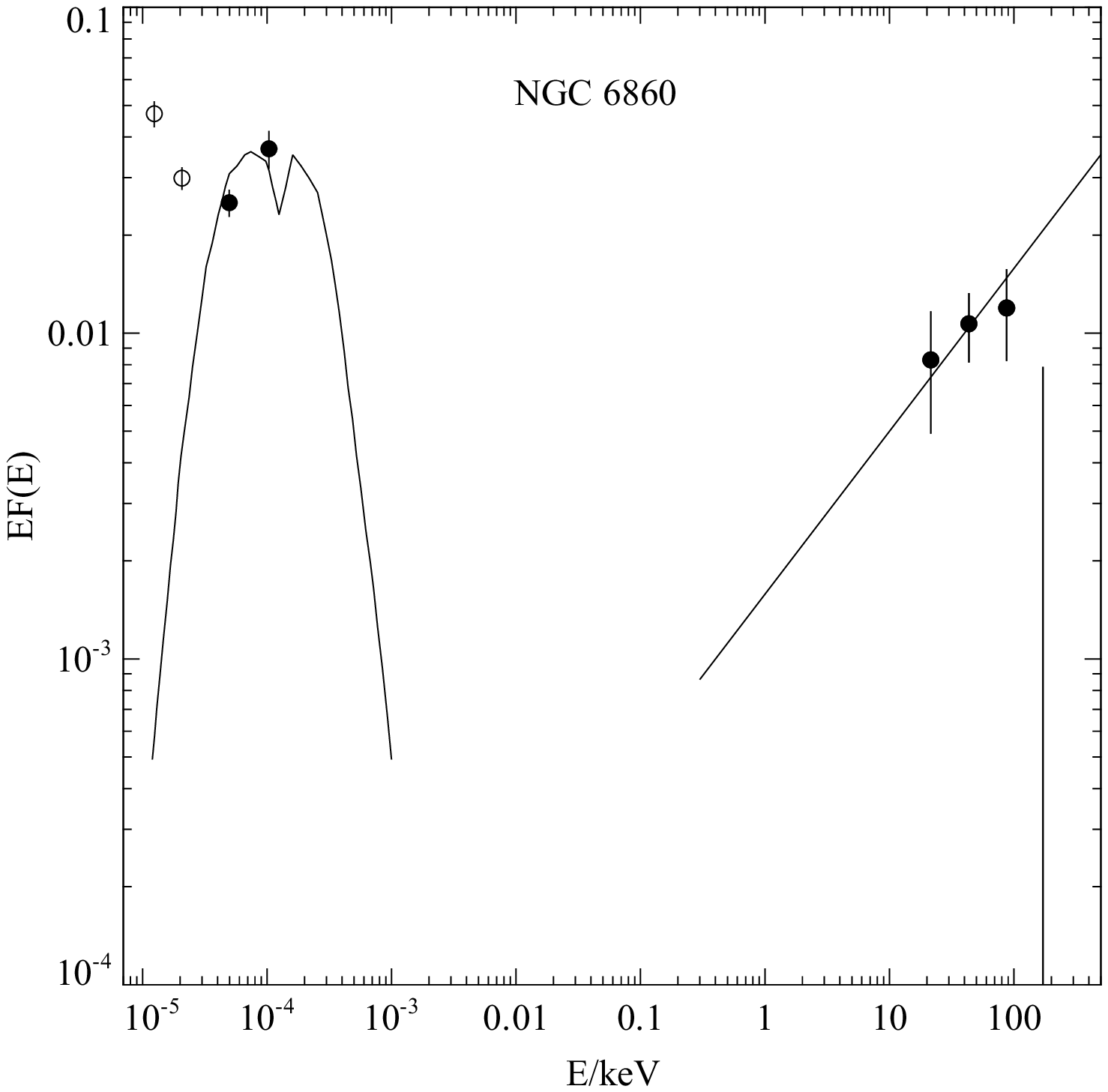}
\includegraphics[width=4.5cm]{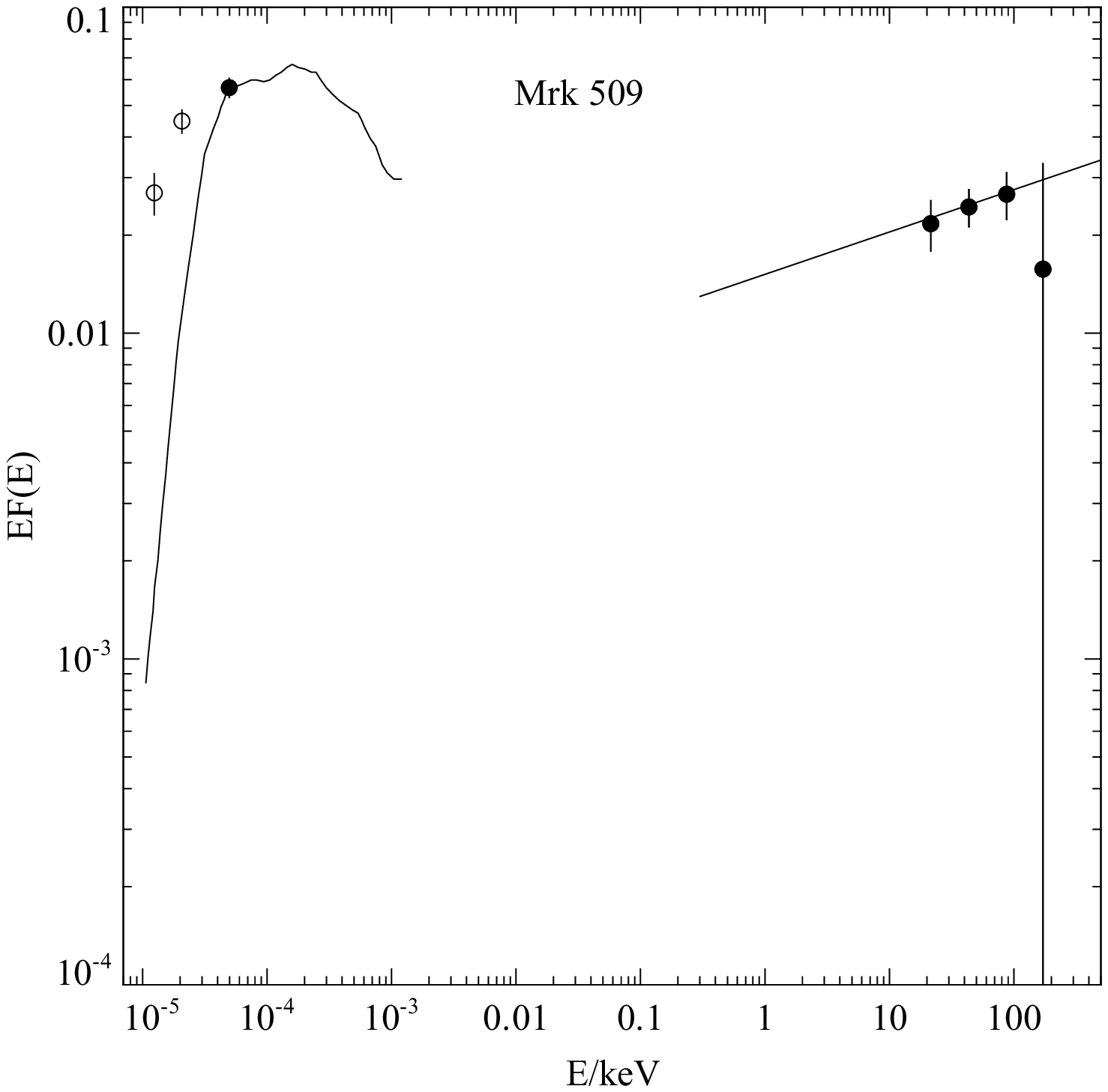}
\includegraphics[width=4.5cm]{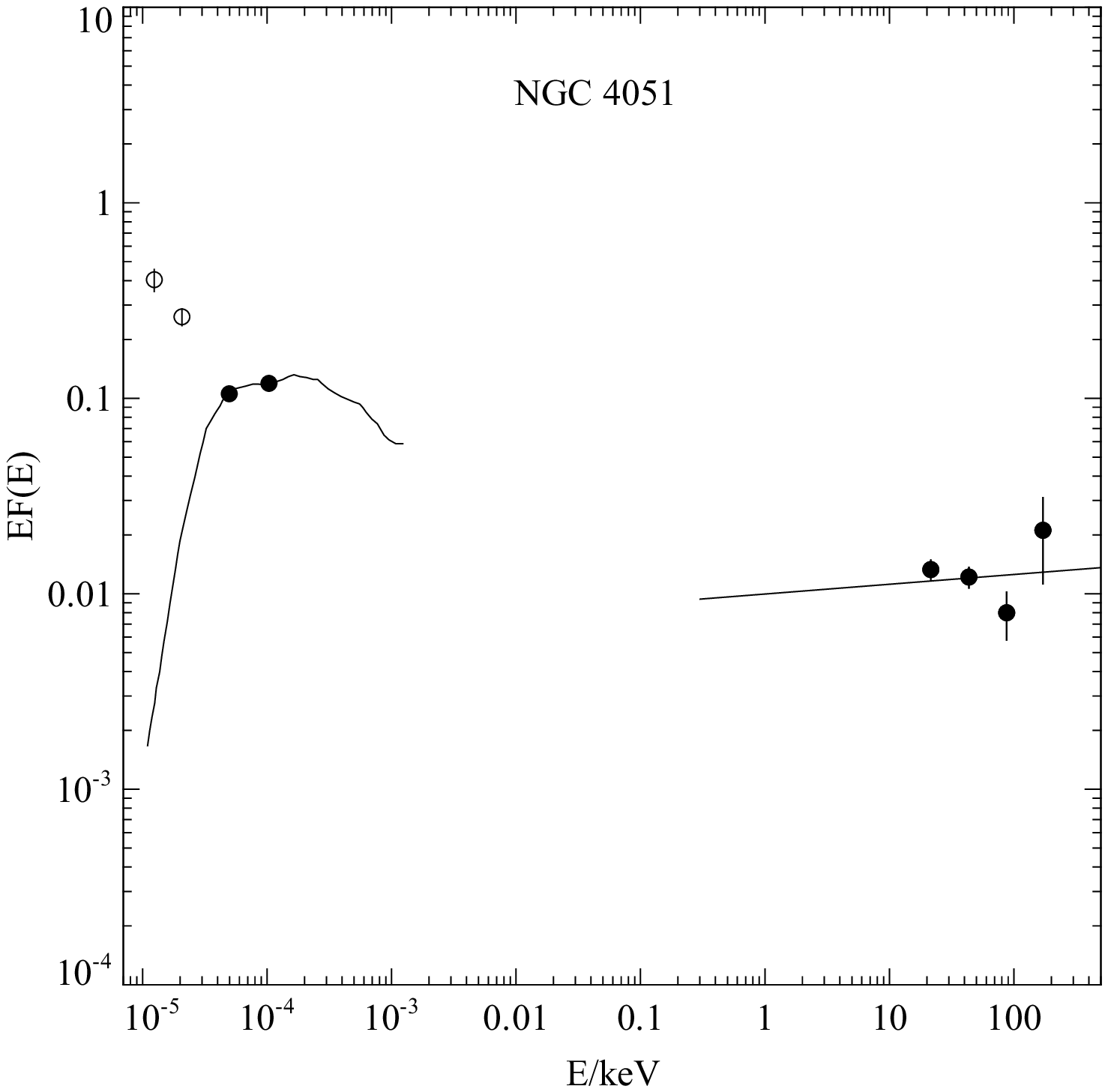}
\includegraphics[width=4.5cm]{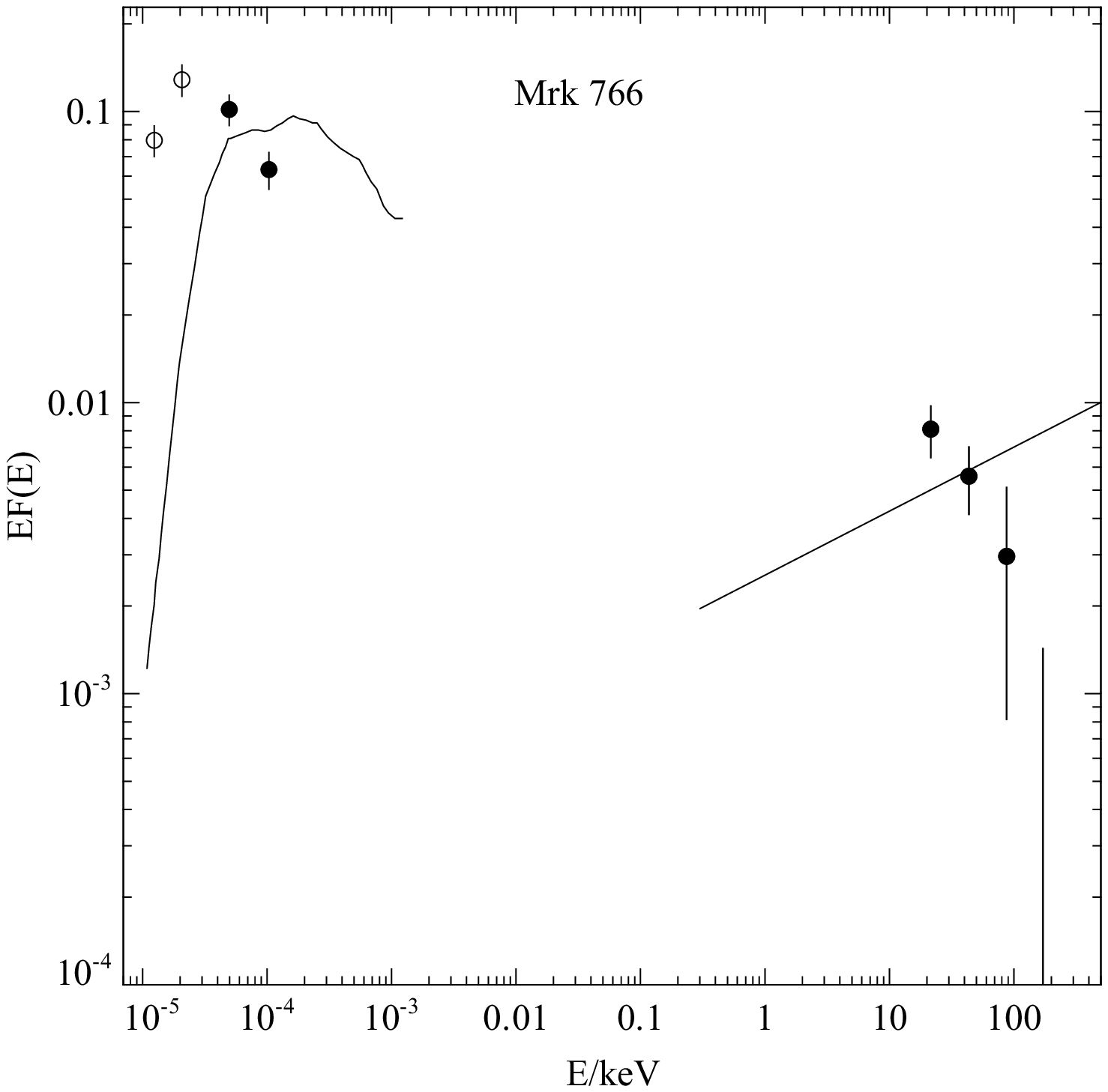}
    \caption{Some example SEDs with \emph{IRAS} and BAT data.  Data points used in the fit (IRAS and BAT) are shown with filled black circles.  Model fits are represented by the solid black lines.  Only 12-25 $\rm \mu m$ \emph{IRAS} data are used to fit the IR nuclear templates; longer wavelength (lower energy) 60 and 100 $\rm \mu m$ data are shown for information (unfilled circles).  The nuclear luminosities from the fit are corrected for host contribution using method 1 (\S\protect\ref{IRAScorr_L12LX}).}
\label{example_SEDs}
\end{figure*}

\begin{figure*}
\includegraphics[width=4.5cm]{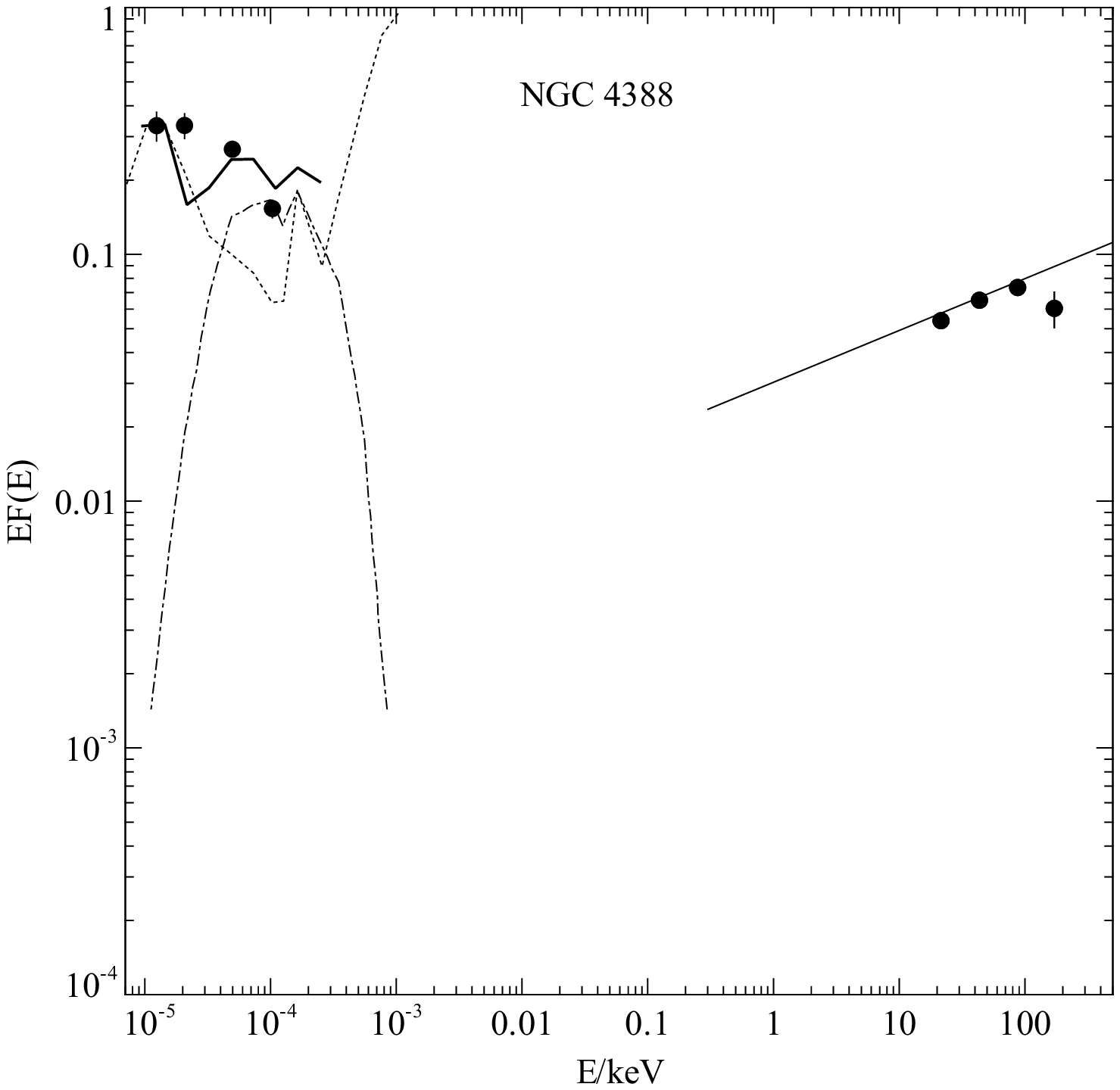}
\includegraphics[width=4.5cm]{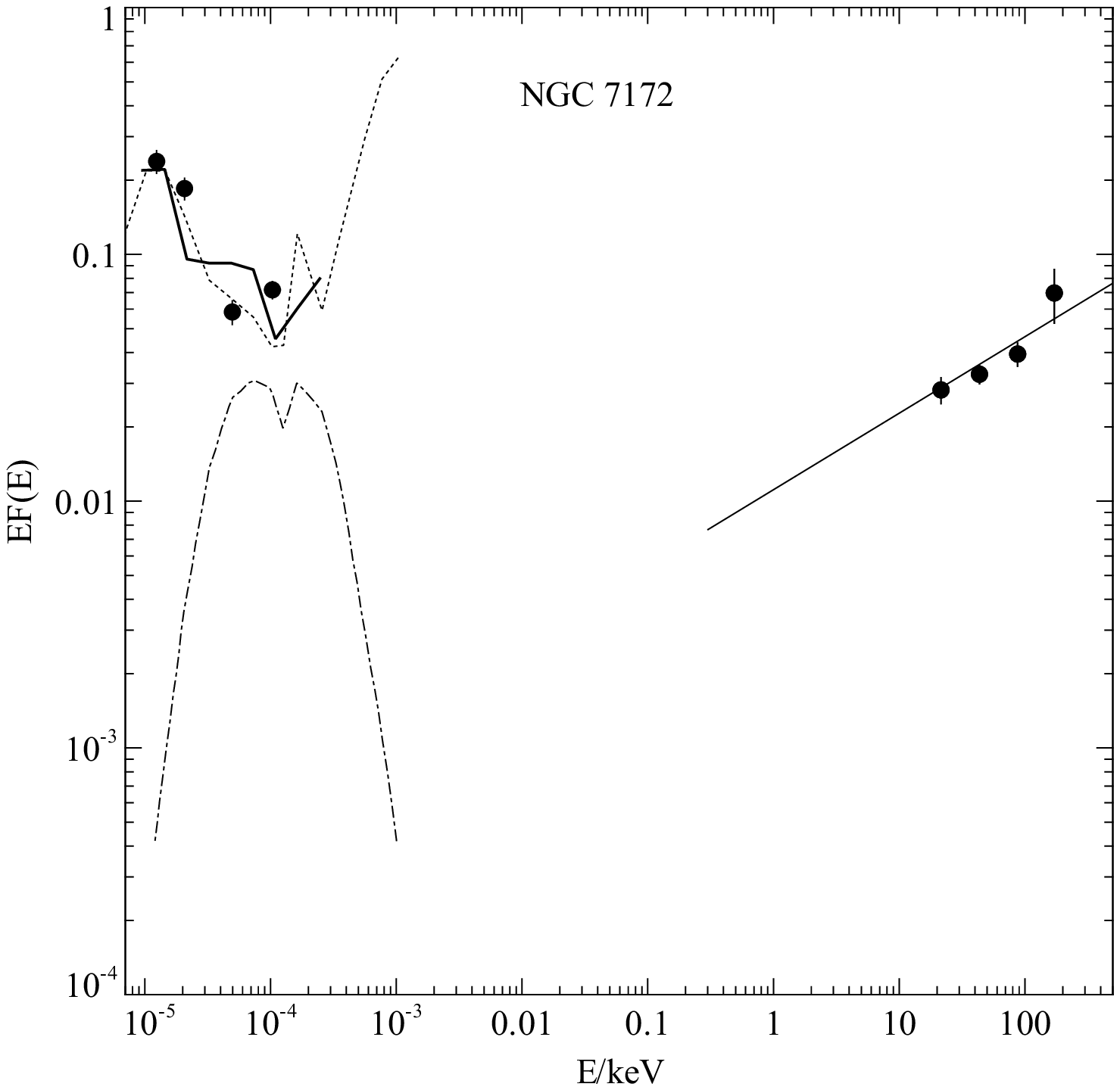}
\includegraphics[width=4.5cm]{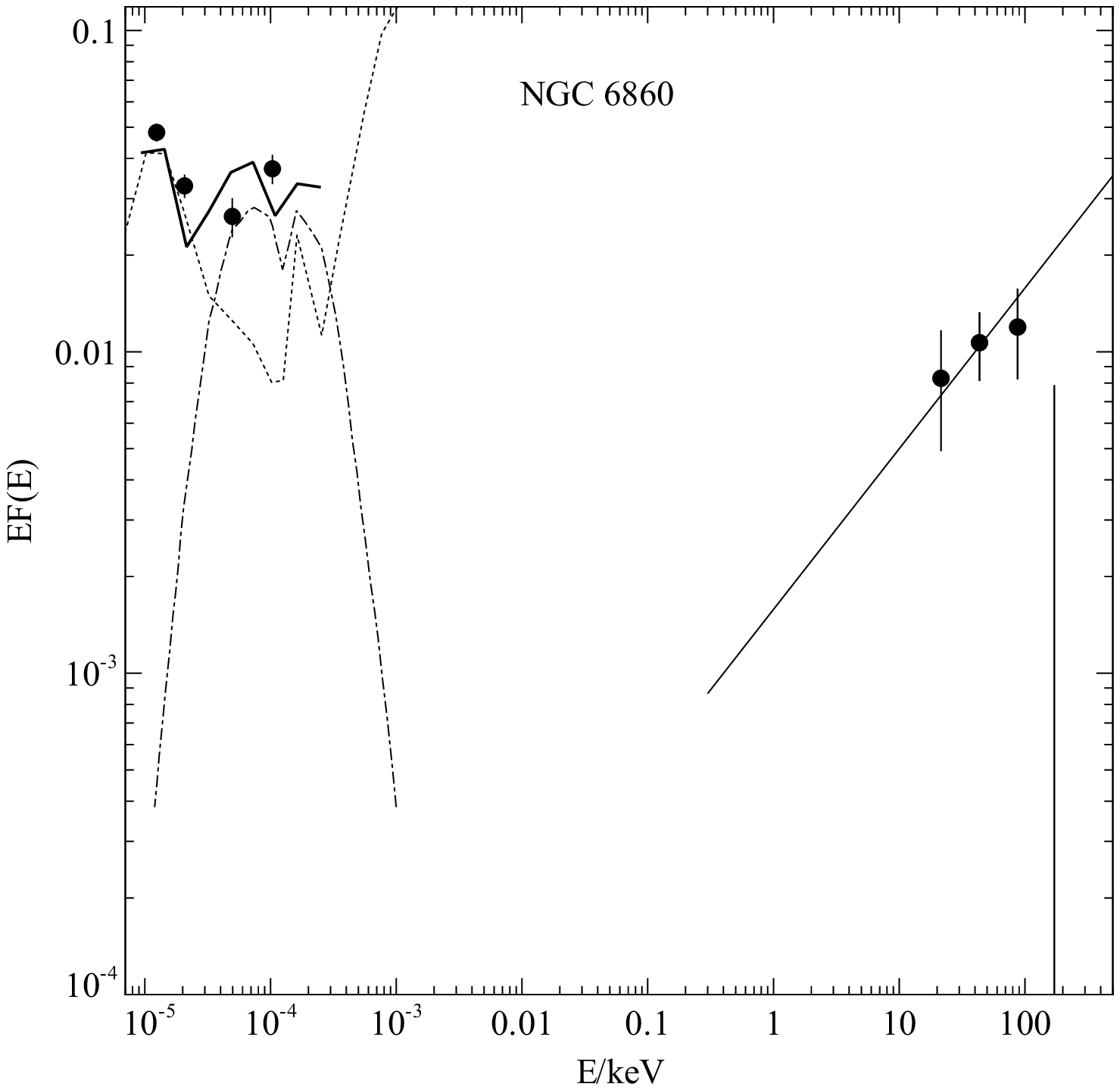}
\includegraphics[width=4.5cm]{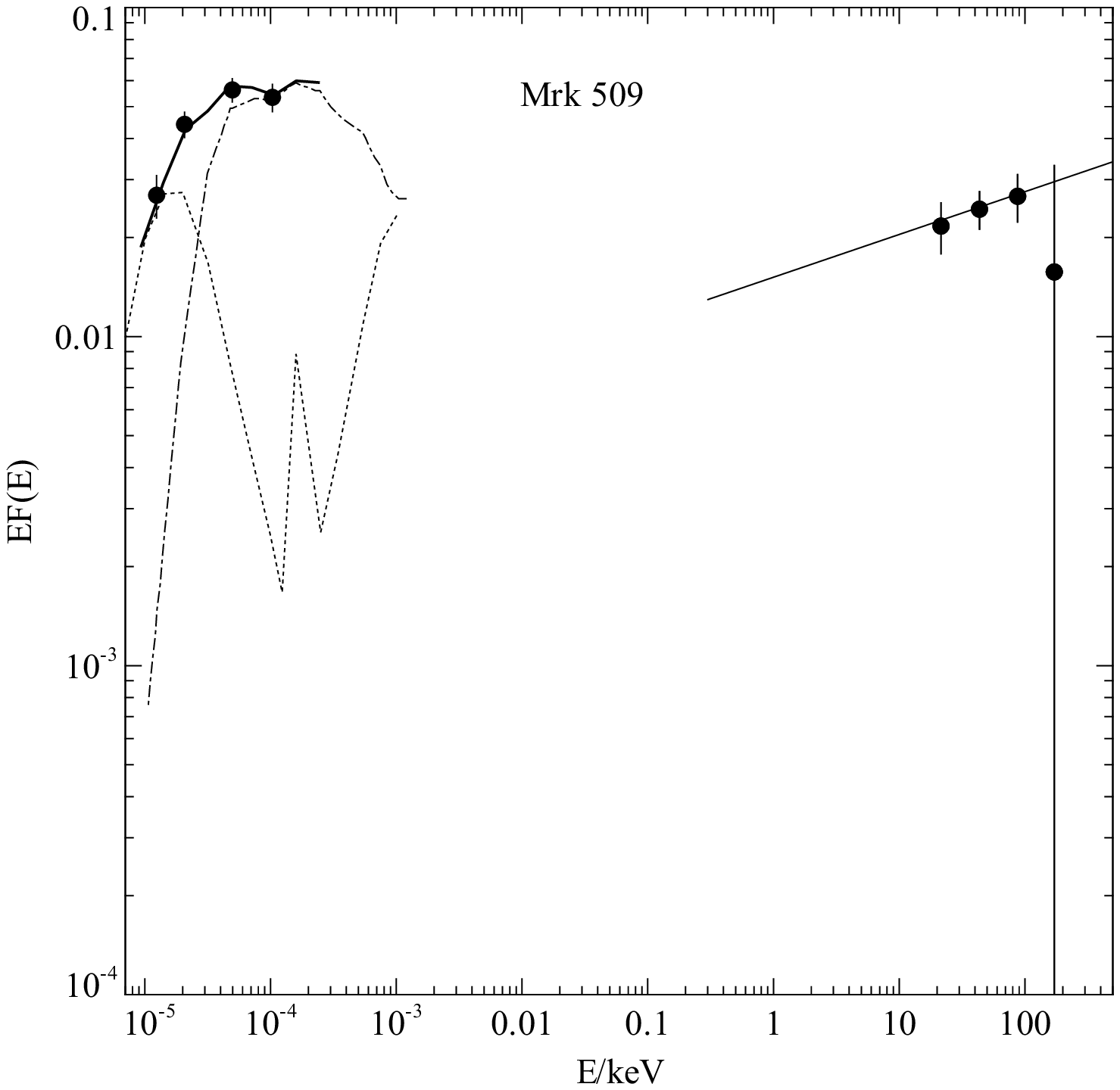}
\includegraphics[width=4.5cm]{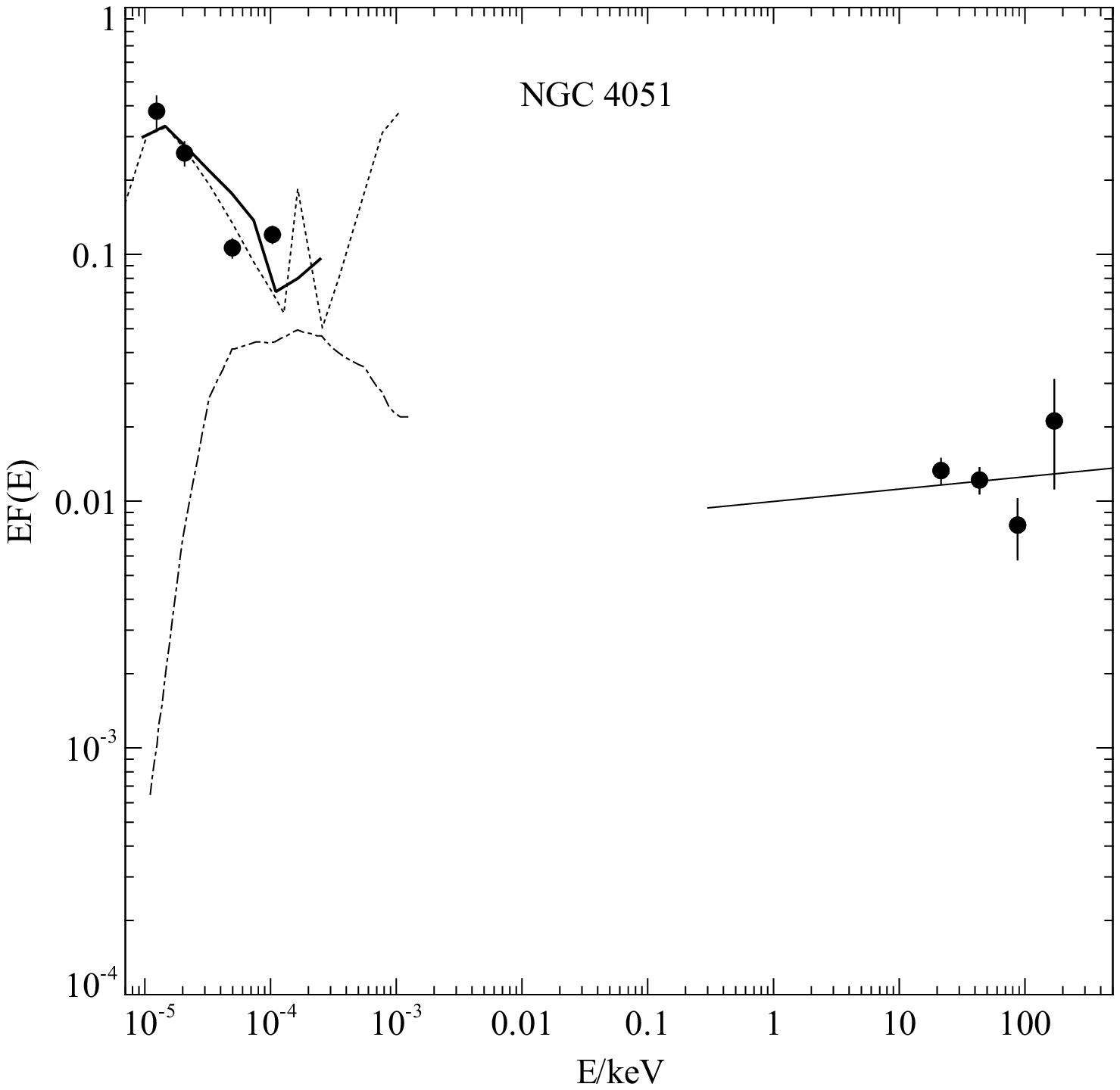}
\includegraphics[width=4.5cm]{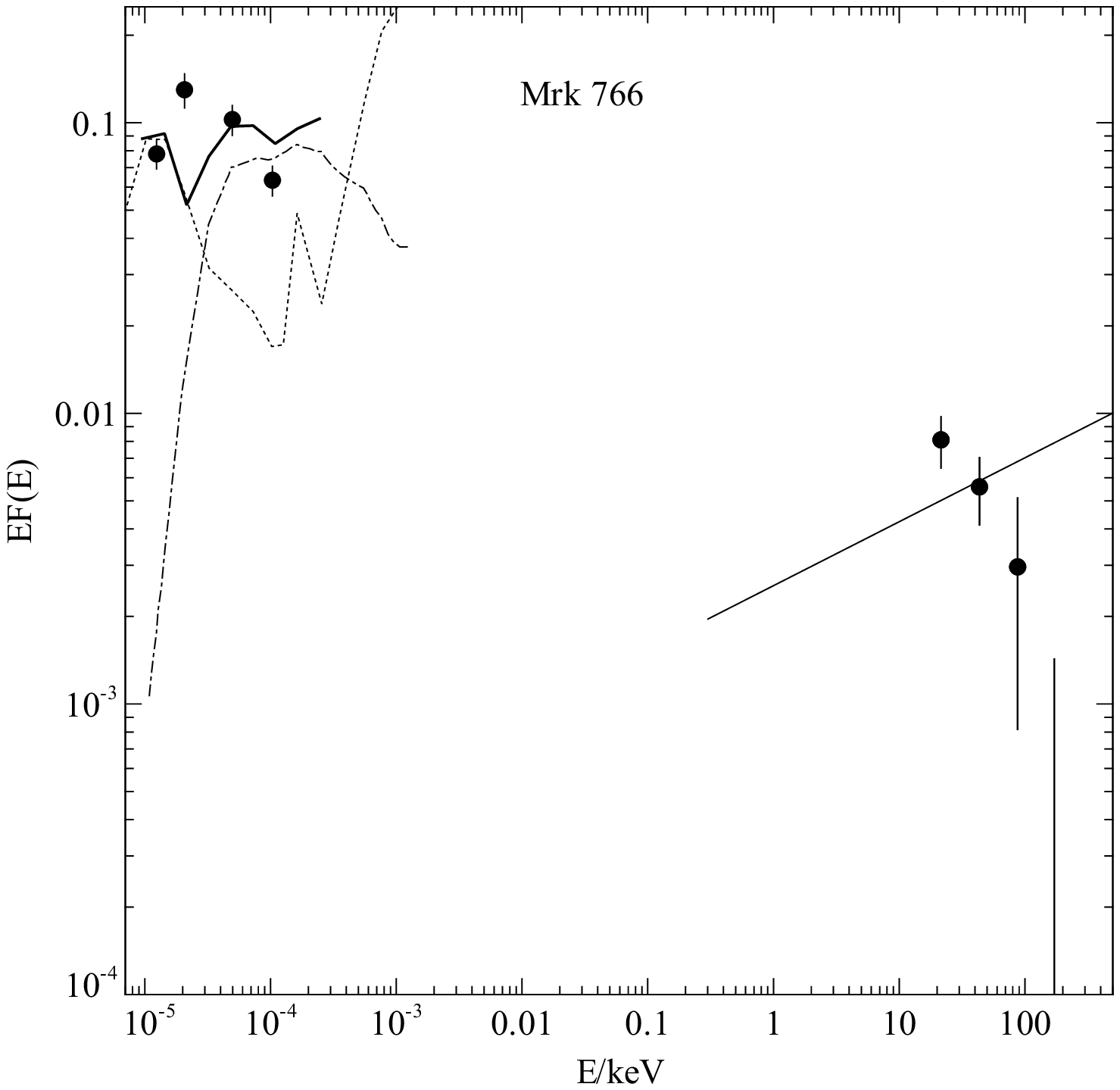}
    \caption{Some example SEDs with \emph{IRAS} and BAT data, with both host and nuclear SED fitting (\S\protect\ref{IRAScorr_Silva04}).  Data points (IRAS and BAT) are shown with filled black circles.  The full model fit is represented by solid black lines; the dotted and dot-dashed lines show the host and nuclear components respectively.}
\label{example_SEDs_silvaetal04hostcorr}
\end{figure*}

We present some of the SEDs constructed from the \emph{IRAS} and BAT data in Figs.~\ref{example_SEDs} and \ref{example_SEDs_silvaetal04hostcorr}.  The bolometric luminosity $L_{\rm bol}$ is calculated as the sum of the nuclear IR luminosity $L_{\rm IR,corr}$ (corrected for torus geometry, anisotropy of emission and non-nuclear contamination as detailed above) and the total X-ray luminosity, $L_{\rm 0.5-100keV}$.  For the high-absorption objects, we multiply the power-law by an intrinsic absorption component (the \textsc{wabs} model in \textsc{xspec}, using the value of $N_{\rm H}$ from \cite{2009ApJ...690.1322W}) before calculating $L_{\rm 0.5-100keV}$ to avoid double-counting part of the X-ray luminosity which is reprocessed to the IR, in line with the approach of \cite{2007A&A...468..603P}.  For low-absorption objects, the difference between the intrinsic and absorption-corrected 0.5--100 keV luminosities are found to be negligible, so we leave out this step for that class. We also extract the absorption-corrected $L_{\rm 2-10keV}$ for calculating bolometric corrections $\kappa_{\rm 2-10keV}=L_{\rm bol}/L_{\rm 2-10keV}$.

We also calculate Eddington ratios for our sample using black hole mass ($M_{\rm BH}$) estimates. For all but one object (Cyg A), these are calculated from the $M_{\rm BH}-L_{\rm K,bulge}$ relation (where $L_{\rm bulge}$ is the host galaxy K-band bulge luminosity), for consistency with the work of \cite{2009arXiv0907.2272V}.   We use the \cite{2003ApJ...589L..21M} formulation for this relation, using an identical method to determine the bulge luminosity as that outlined in \cite{2009arXiv0907.2272V}.  For Cyg A, the dynamical mass estimate from \cite{2003MNRAS.342..861T} is used as discussed in \S\ref{sampleselection}.  Eddington ratios are calculated using $\lambda_{\rm Edd}=L_{\rm bol}/L_{\rm Edd}$ for Eddington luminosities $L_{\rm Edd}=1.3 \times 10^{38} (M_{\rm BH}/M_{\odot}) \rm erg s^{-1}$.

\subsection{Calibrating the method}

\begin{figure*}
    \includegraphics[width=7cm]{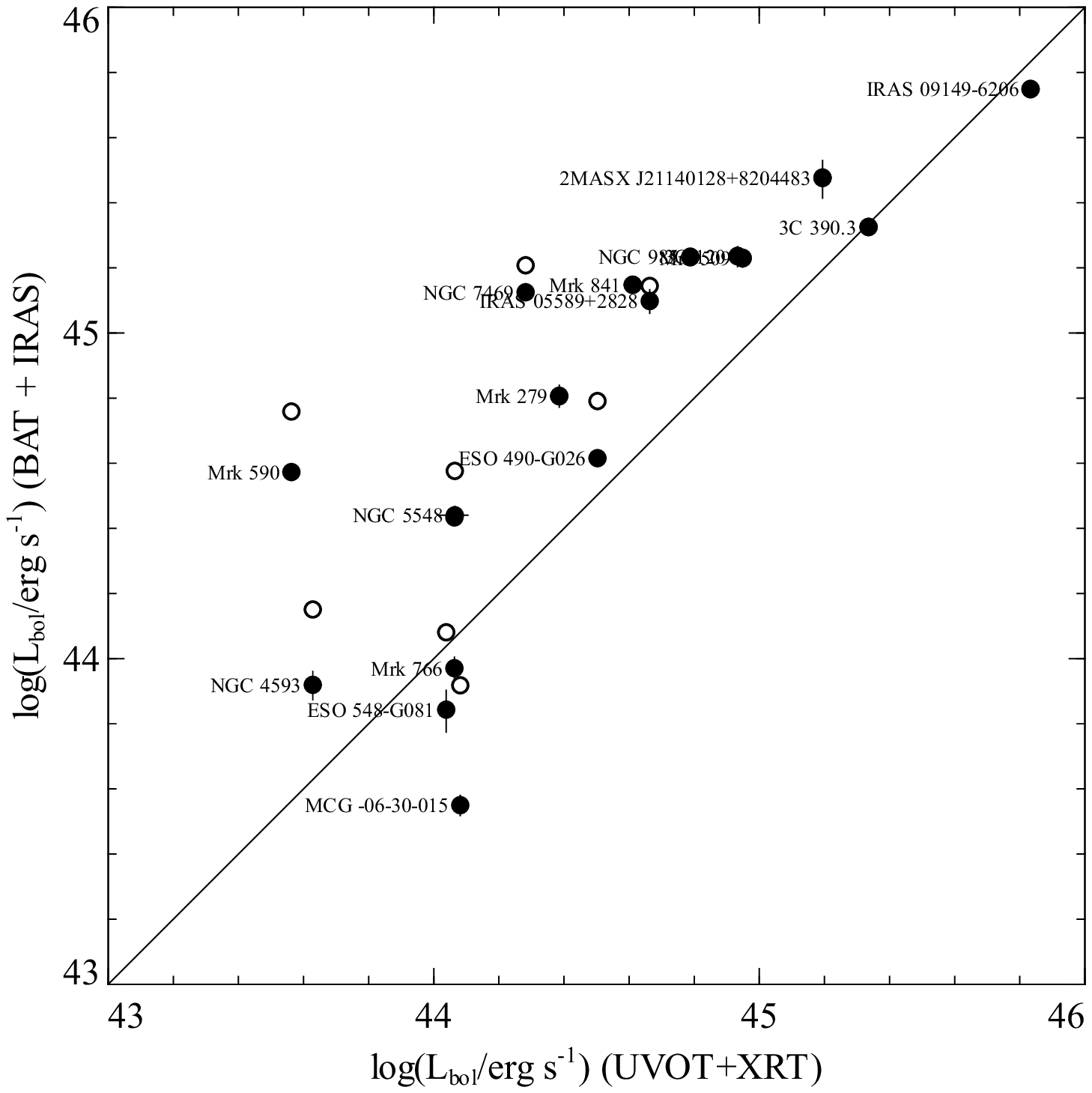}
    \includegraphics[width=7cm]{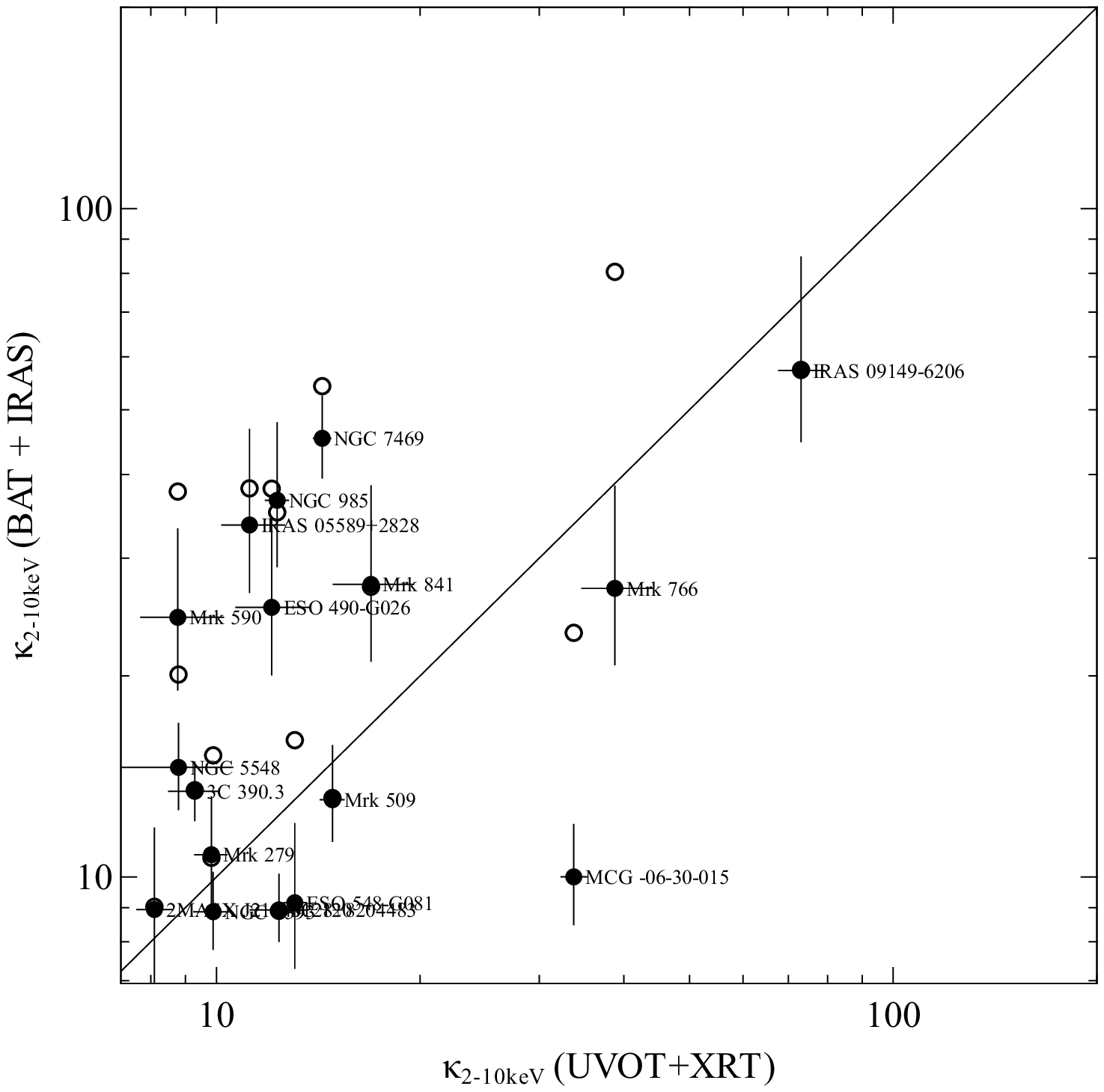}
    \includegraphics[width=7cm]{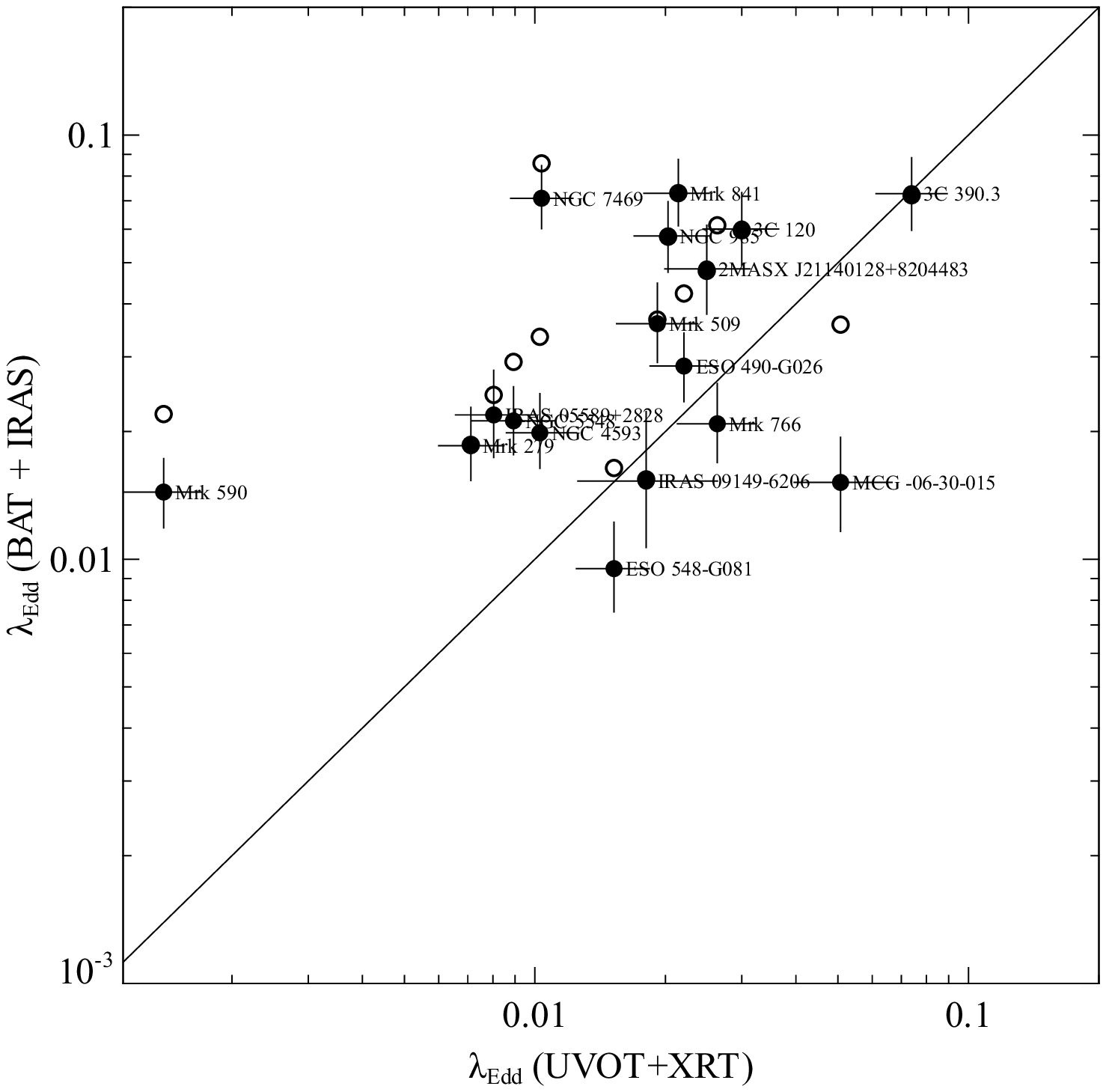}
    \includegraphics[width=7cm]{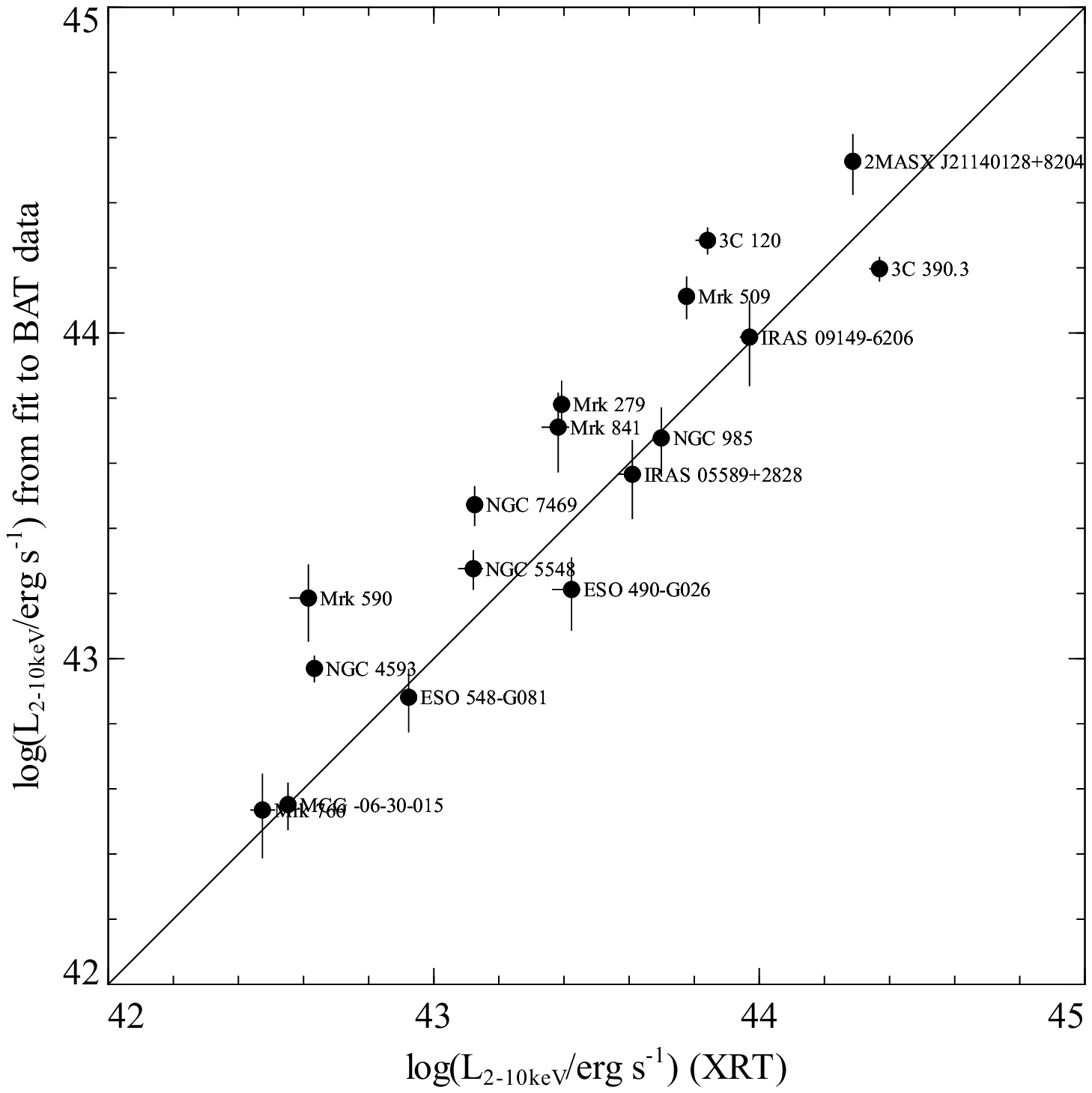}
    \caption{Comparison between values of bolometric luminosity, 2--10keV bolometric corrections, Eddington ratios and 2--10keV luminosities using the time-averaged, non-contemporaneous \emph{IRAS} and BAT data, and the method from \protect\cite{2009arXiv0907.2272V} using simultaneous optical, UV and X-ray data from Swift.  The solid lines represent the desired 1:1 correlation, and empty circles show the results before correcting for host galaxy contamination of \emph{IRAS} fluxes. Host galaxy correction to \emph{IRAS}-determined IR luminosities as detailed in \S\ref{IRAScorr_L12LX}.}
\label{calibration}
\end{figure*}

\begin{figure*}
    \includegraphics[width=7cm]{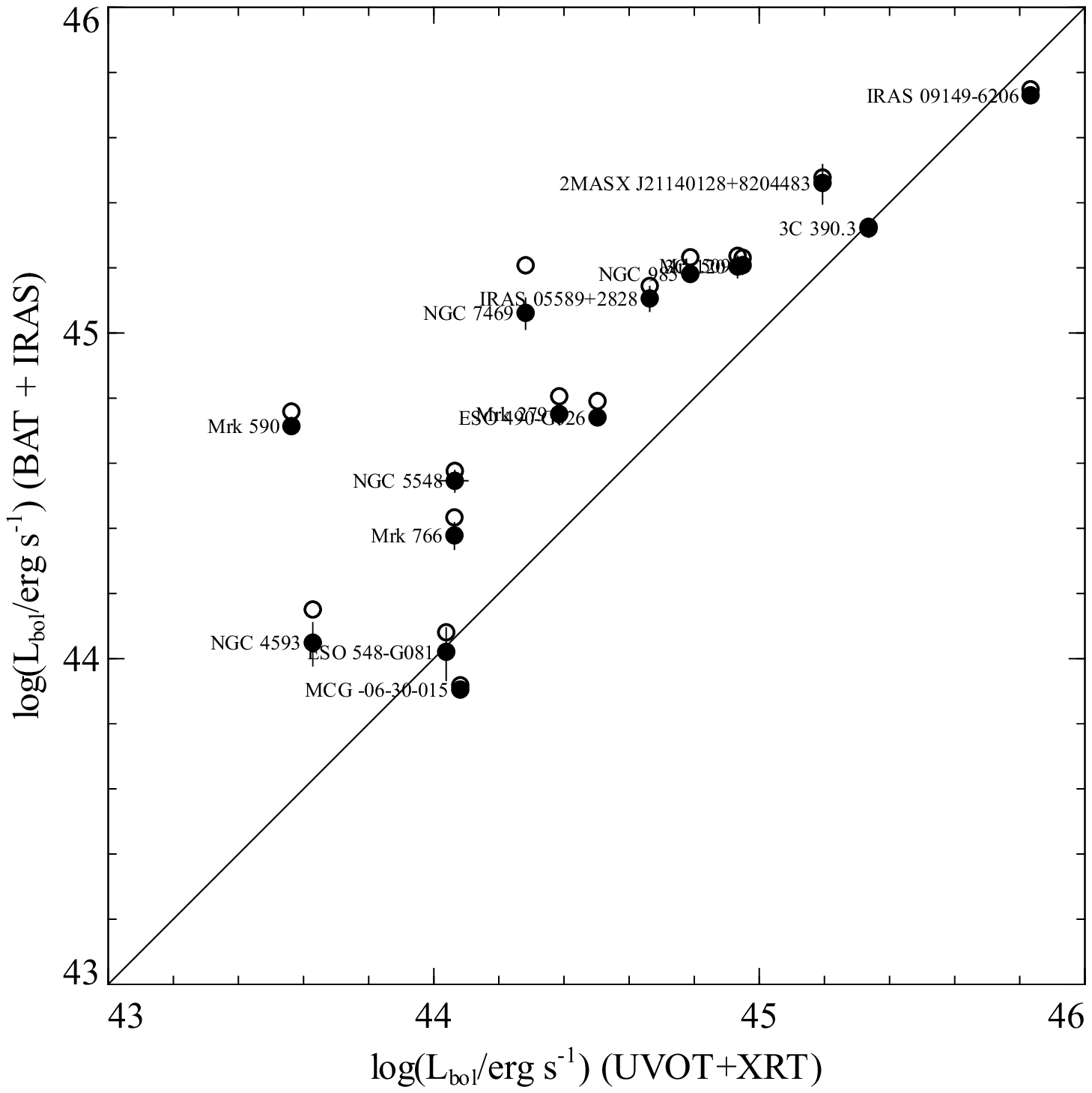}
    \includegraphics[width=7cm]{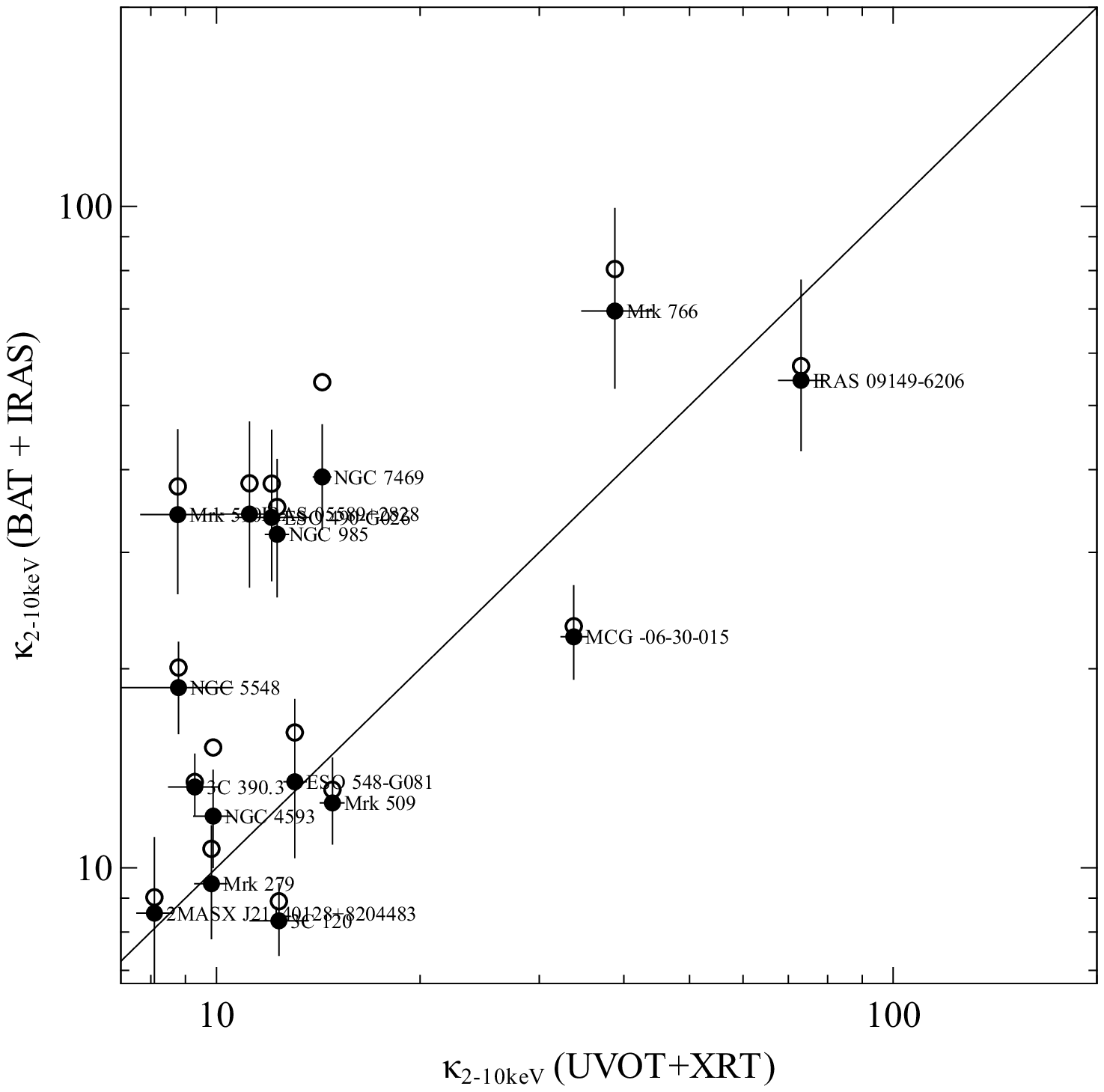}
    \includegraphics[width=7cm]{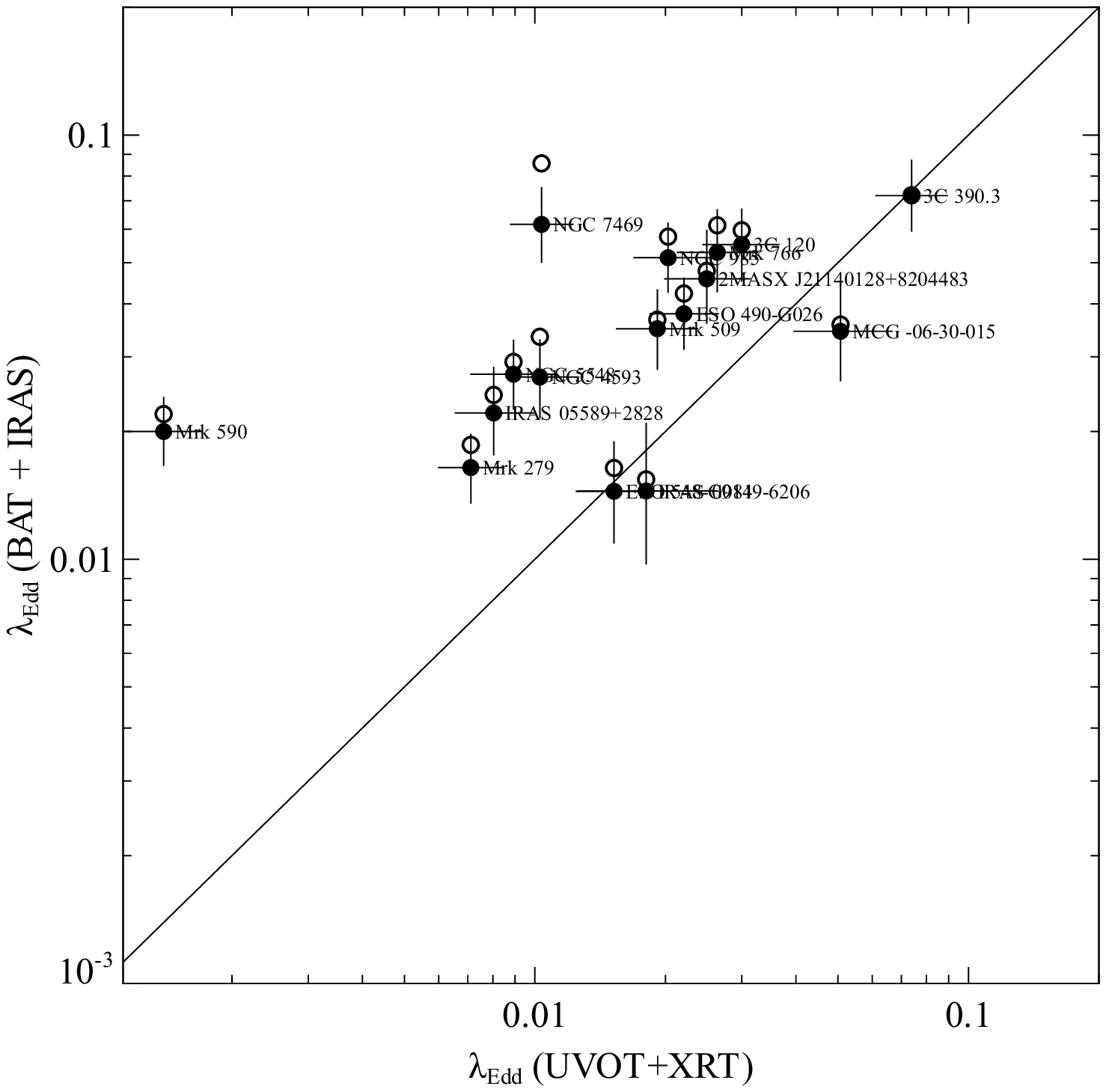}
    \caption{Comparison between values for bolometric luminosity, 2--10keV bolometric corrections, Eddington ratios and 2--10keV luminosities from \emph{IRAS} and BAT data with the results from \emph{Swift} UVOT-XRT data presented in \protect\cite{2009arXiv0907.2272V}. Host galaxy in IR accounted for as detailed in \S\ref{IRAScorr_Silva04}.  Key as in Fig.~\ref{calibration}.}
\label{calibration_silvaetalhostcorr}
\end{figure*}

We first attempt to estimate the accuracy of our bolometric luminosity calculation using IR and hard X-ray emission.  \cite{2006MNRAS.366..953B}, \cite{2009MNRAS.392.1124V} (and references therein) outline the approach of using simultaneous optical, UV and X-ray data for calculation of intrinsic SEDs, and this approach is developed further using Swift XRT and UVOT data in \cite{2009arXiv0907.2272V} for low absorption objects in the Swift/BAT 9-month catalogue.  Using IR along with long-exposure BAT data produces a long-term averaged estimate of $L_{\rm bol}$ in contrast to the `snapshot' approach from using simultaneous optical--to--X-ray data, but statistically one expects reasonable agreement between the two approaches, since the effects of variability should be averaged out when considering a large enough sample.  We present comparisons between the values of $L_{\rm bol}$, $\kappa_{\rm 2-10keV}$, $\lambda_{\rm Edd}$ and $L_{\rm 2-10keV}$ from \cite{2009arXiv0907.2272V} and this study in Figs.~\ref{calibration} and \ref{calibration_silvaetalhostcorr}, for the two different host-contamination removal methods. There are 17 objects overlapping between the two studies.

The bolometric luminosities determined from both methods agree reasonably well with those determined from UVOT and XRT fits, but both methods show a systematically larger $L_{\rm bol}$ than that determined from UVOT and XRT data. In particular, the degree of host galaxy contamination removed from fitting nuclear and host SEDs (method 2) yields estimates for $L_{\rm bol}$ about 0.3--0.4 dex larger than the optical--to--X-ray estimates.  This could be attributed to a number of factors; firstly it is possible that the simple SED fitting (in method 2) underestimates the host galaxy contribution and more correction is necessary; secondly the geometry and anisotropy corrections assumed in scaling the observed IR luminosity could be too large for many of the objects where poor agreement is reported, and thirdly, the estimates from UVOT-XRT could themselves be too small (the latter two phenomena would apply equally to method 1 and method 2).  The $L_{\rm bol}$ from the UVOT and XRT fit depends on the black hole mass and tends to increase as the black hole mass decreases; \cite{2009arXiv0907.2272V} discuss the possibility that the mass could be overestimated by about a factor of $\sim 2$ which could alleviate the discrepancy seen here.  The problem is a complex one and ultimately requires better knowledge of the torus covering fraction (which may have a very broad distribution, e.g. \citealt{2009arXiv0905.4389R}) and emission characteristics to perform this comparison more accurately.

The distributions of bolometric correction and Eddington ratio are more clustered, so a correlation between the two approaches is difficult to discern.  The finding of a low distribution of Eddington ratios in \cite{2009arXiv0907.2272V} is, however, broadly confirmed with the \emph{IRAS} and BAT method.  The notable outlier Mrk 590 is known from the literature to have a very reddened optical--UV spectrum (confirmed with UVOT), and constraining the big blue bump shape using the K-band mass estimate yields a poor fit in \cite{2009arXiv0907.2272V}.  If for this particular source, we employ the results obtained from using the reverberation mass instead, its accretion rate lies much closer to the line of one-to-one corespondence, moving to join the cluster of points around $\lambda_{\rm Edd}\sim0.02$.   Despite the scatter however, the general trend for agreement between the two methods motivates us to continue in our attempts at addressing the key questions of this study.

\begin{figure*}
\includegraphics[width=4.5cm]{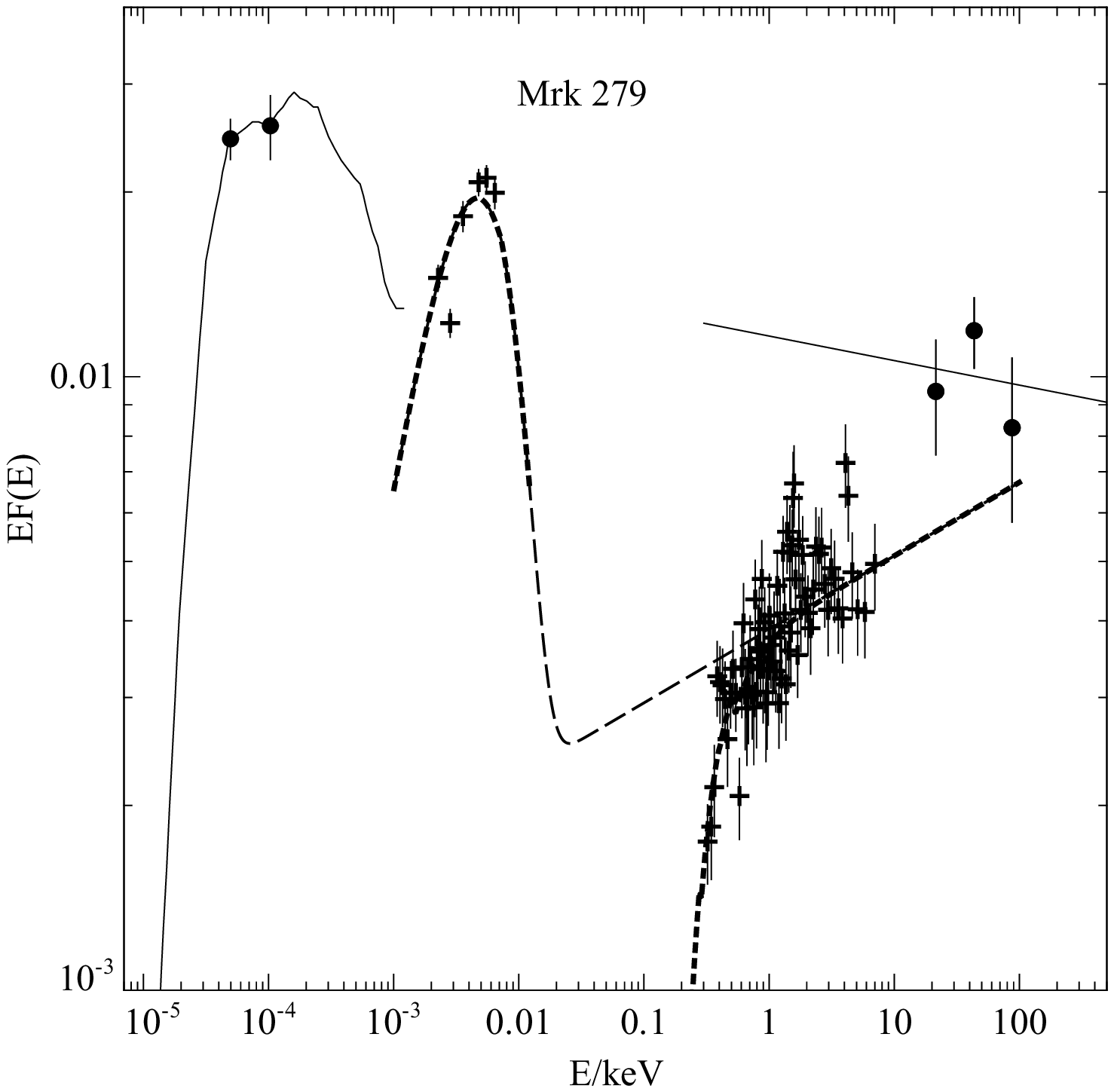}
\includegraphics[width=4.5cm]{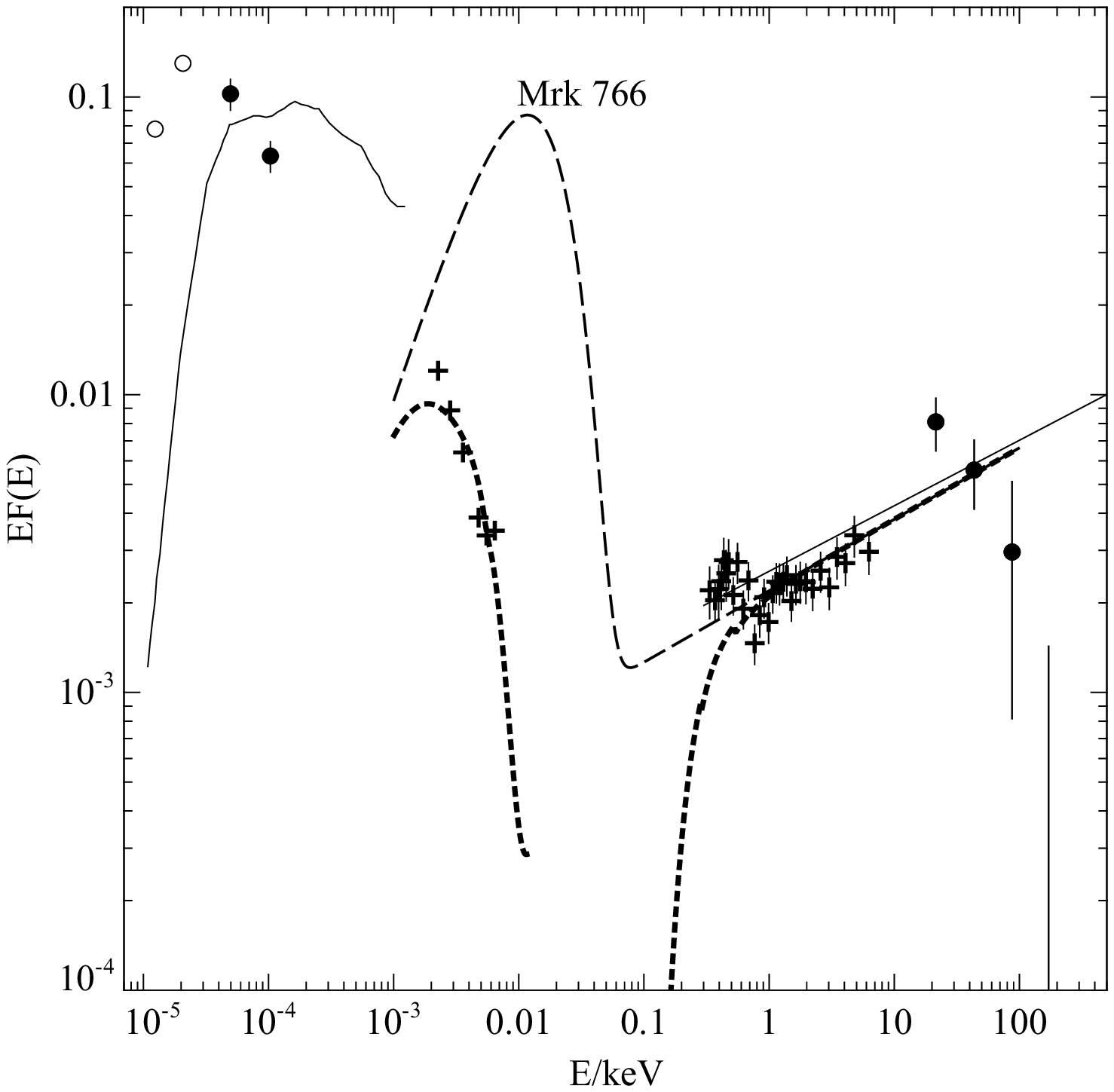}
\includegraphics[width=4.5cm]{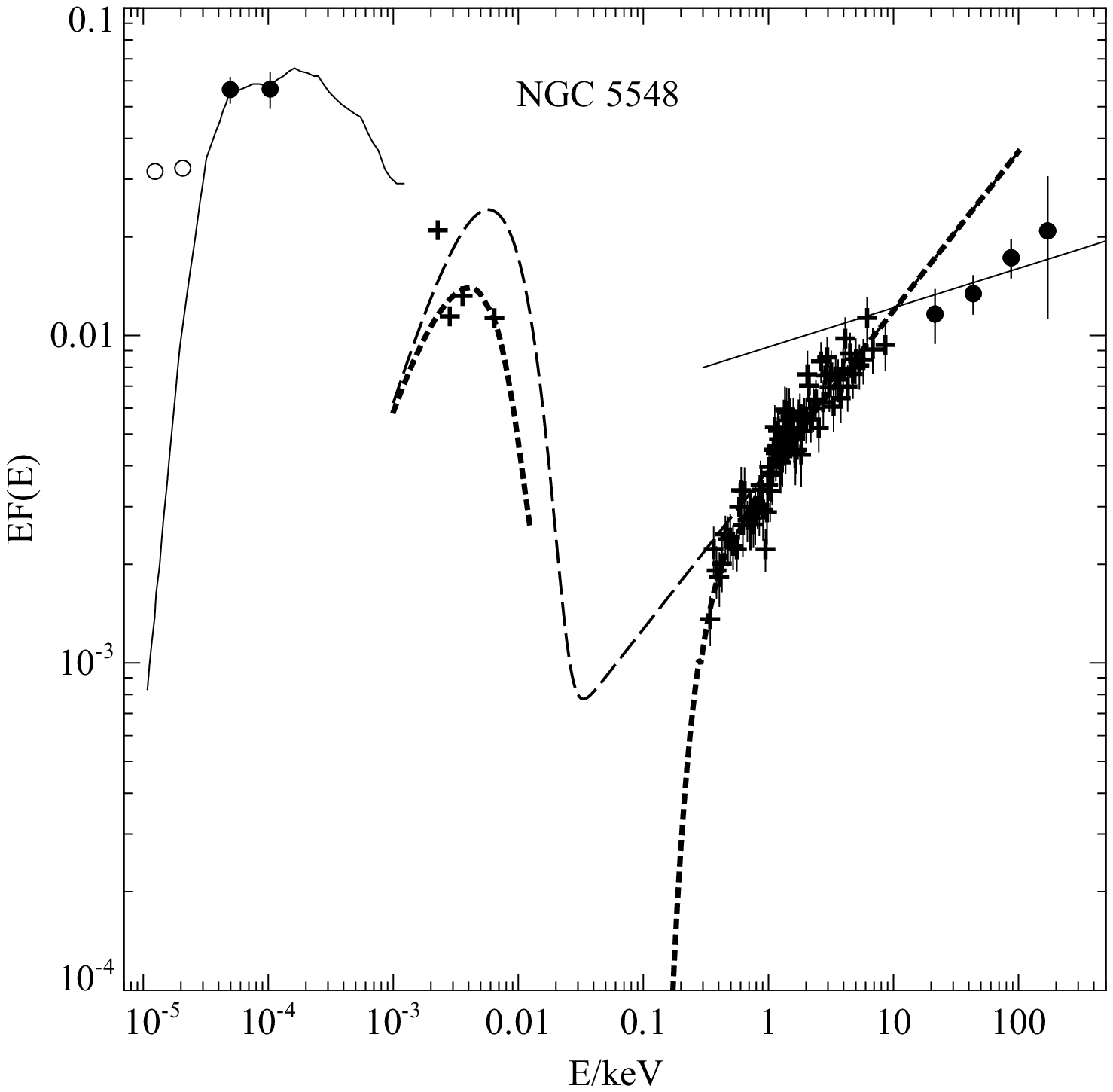}
\includegraphics[width=4.5cm]{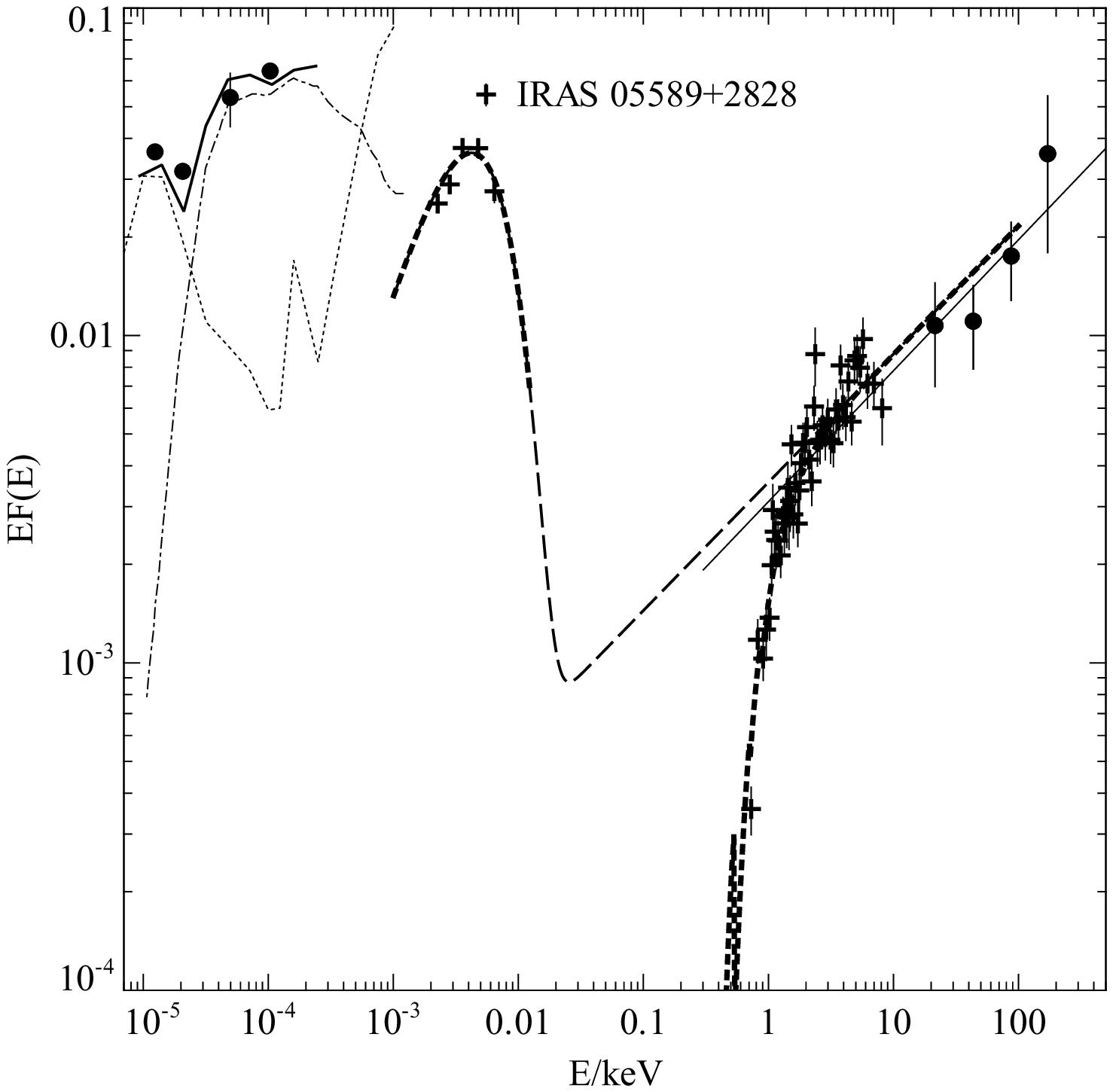}
\includegraphics[width=4.5cm]{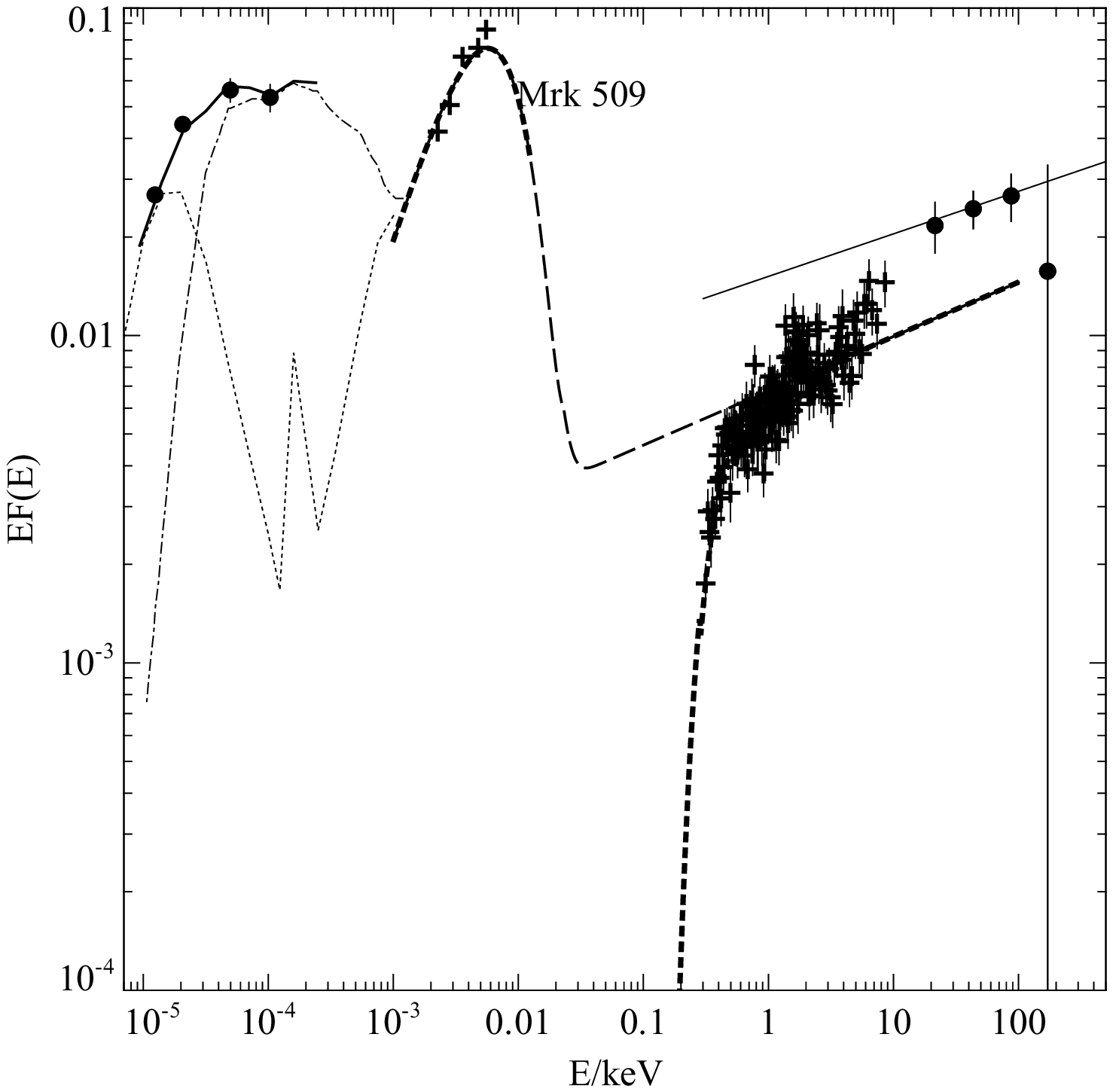}
\includegraphics[width=4.5cm]{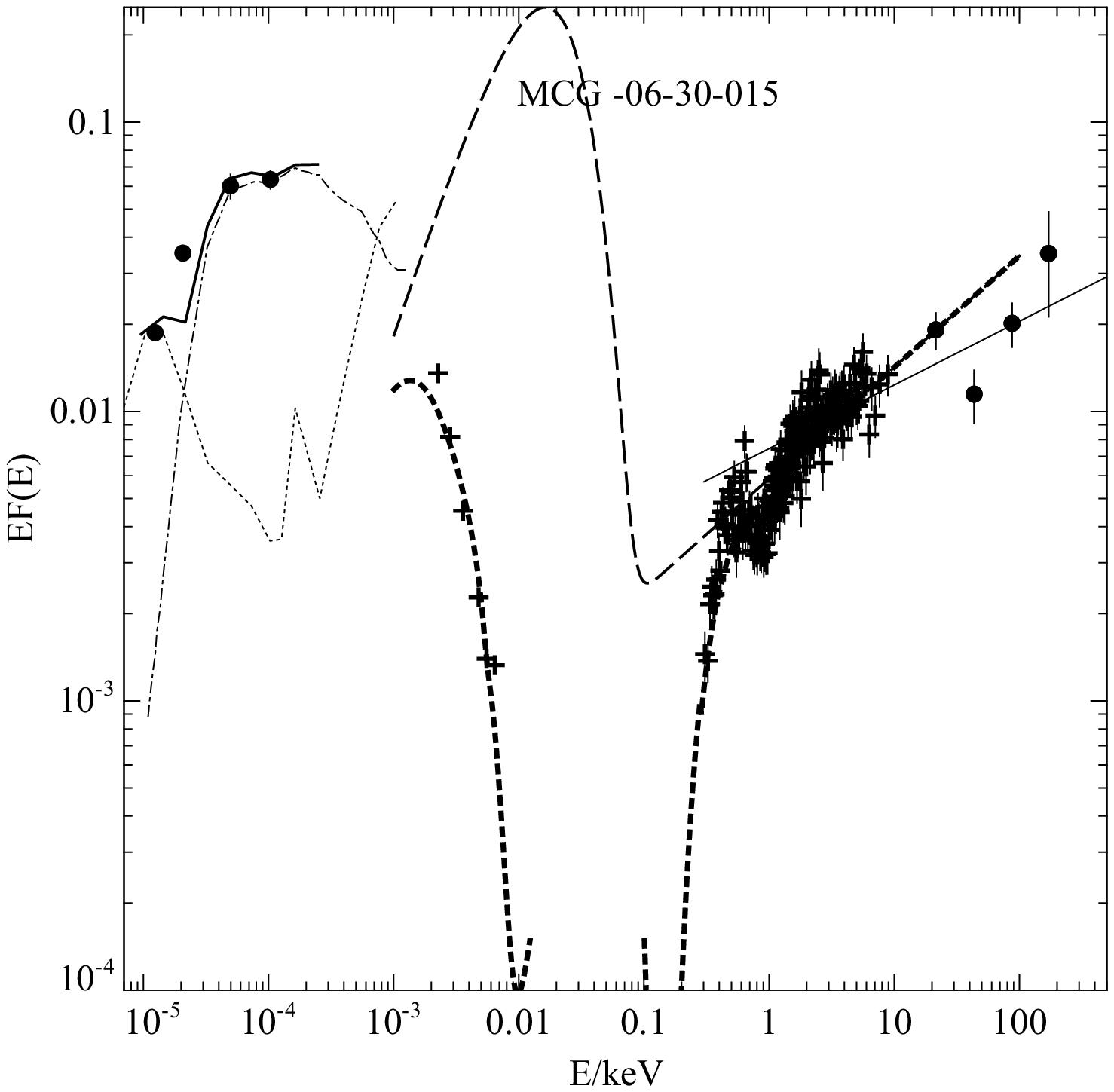}
    \caption{A selection of SEDs with IRAS, UVOT, XRT and BAT data.  Key as in Fig.~\ref{example_SEDs} with the following additions: crosses represent simultaneous UVOT and XRT data, heavy dotted lines represent the fits to those data with intrinsic absorption included (optical--UV reddening and X-ray absorption) and dashed lines represent the de-reddened, absorption corrected SED fit from which bolometric luminosities were estimated in \protect\cite{2009arXiv0907.2272V}.  The top three SEDs show the fitting of only a nuclear IR SED with host galaxy contamination corrected for afterwards as detailed in \S\ref{IRAScorr_L12LX}; the lower three show the combined fitting of nuclear and host galaxy SED templates in the IR as described in \S\ref{IRAScorr_Silva04}.}
\label{example_calibration_SEDs}
\end{figure*}

We also present some of the SEDs for the objects with XRT and UVOT data, now including the \emph{IRAS} and BAT data (Fig.~\ref{example_calibration_SEDs}).  There is some variation between the photon index used for fitting the BAT data and that found from the XRT data in \cite{2009arXiv0907.2272V}, but generally both datasets show similar values.

\section{Results and Discussion: bolometric corrections and Eddington ratios}

\begin{figure*}
    \includegraphics[width=8cm]{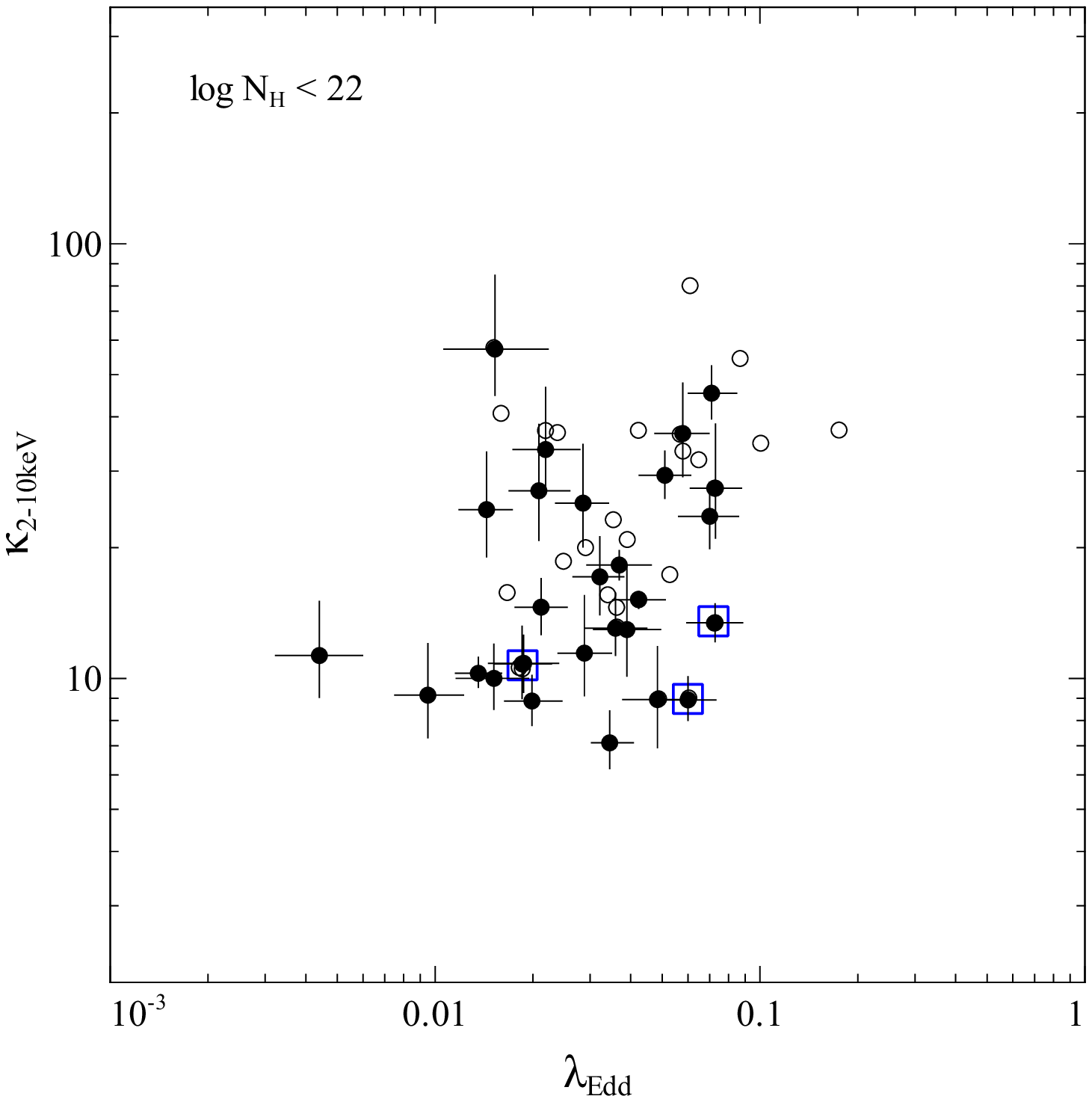}
    \includegraphics[width=8cm]{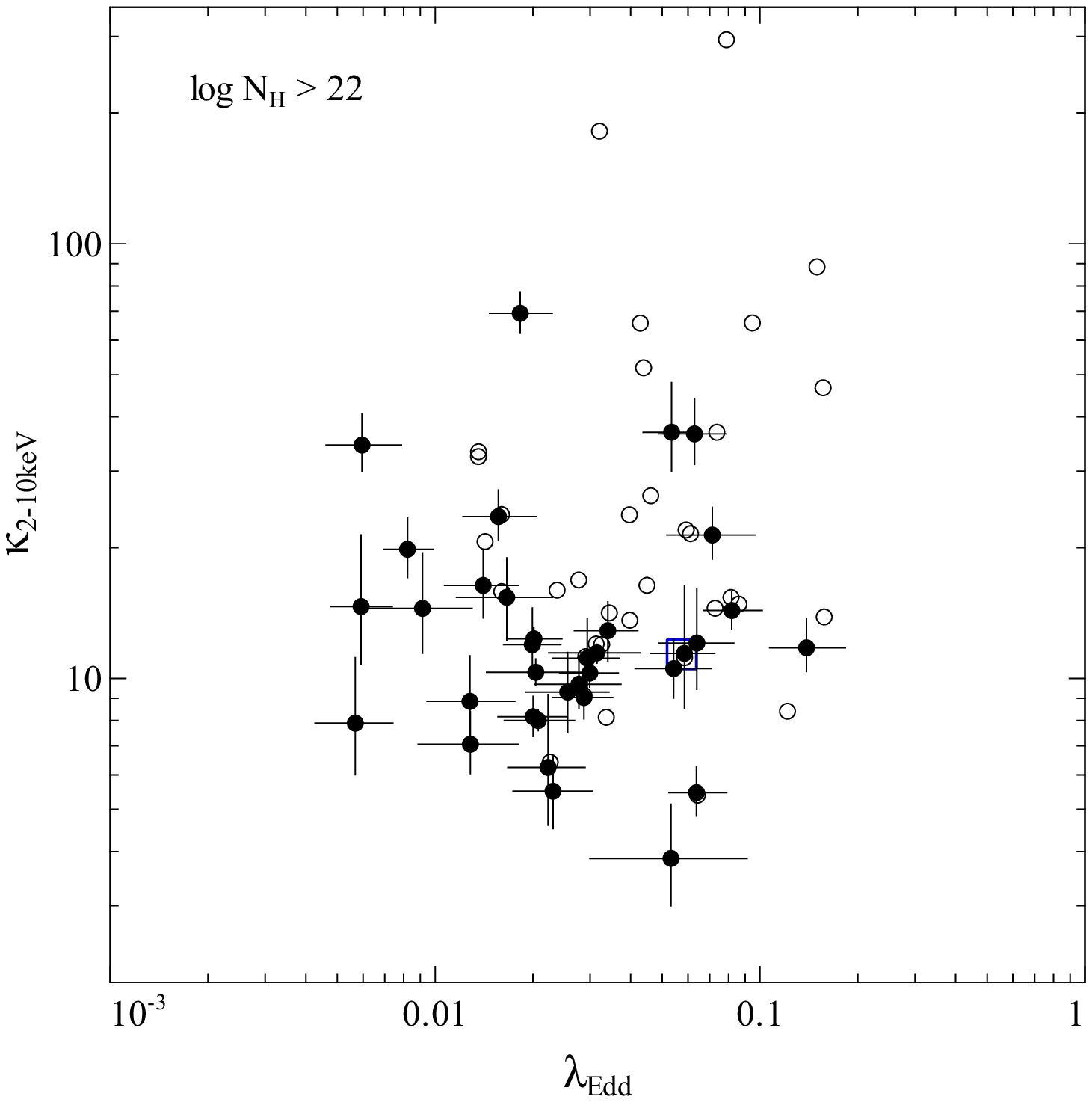}
    \caption{Bolometric corrections against Eddington ratios for low absorption (left panel) and high absorption (right panel) objects, with statistical correction for non-nuclear contamination of \emph{IRAS} fluxes (black filled points) as detailed in \S\ref{IRAScorr_L12LX}.  Results before applying the contamination correction are shown for comparison (empty circles).  Objects highlighted with blue squares are from the 3C catalogue; these radio-loud objects are demarcated to identify objects in which some jet contribution might be affecting the results.}
\label{bolcor_vs_edd_galaxycorr}
\end{figure*}

\begin{figure*}
    \includegraphics[width=8cm]{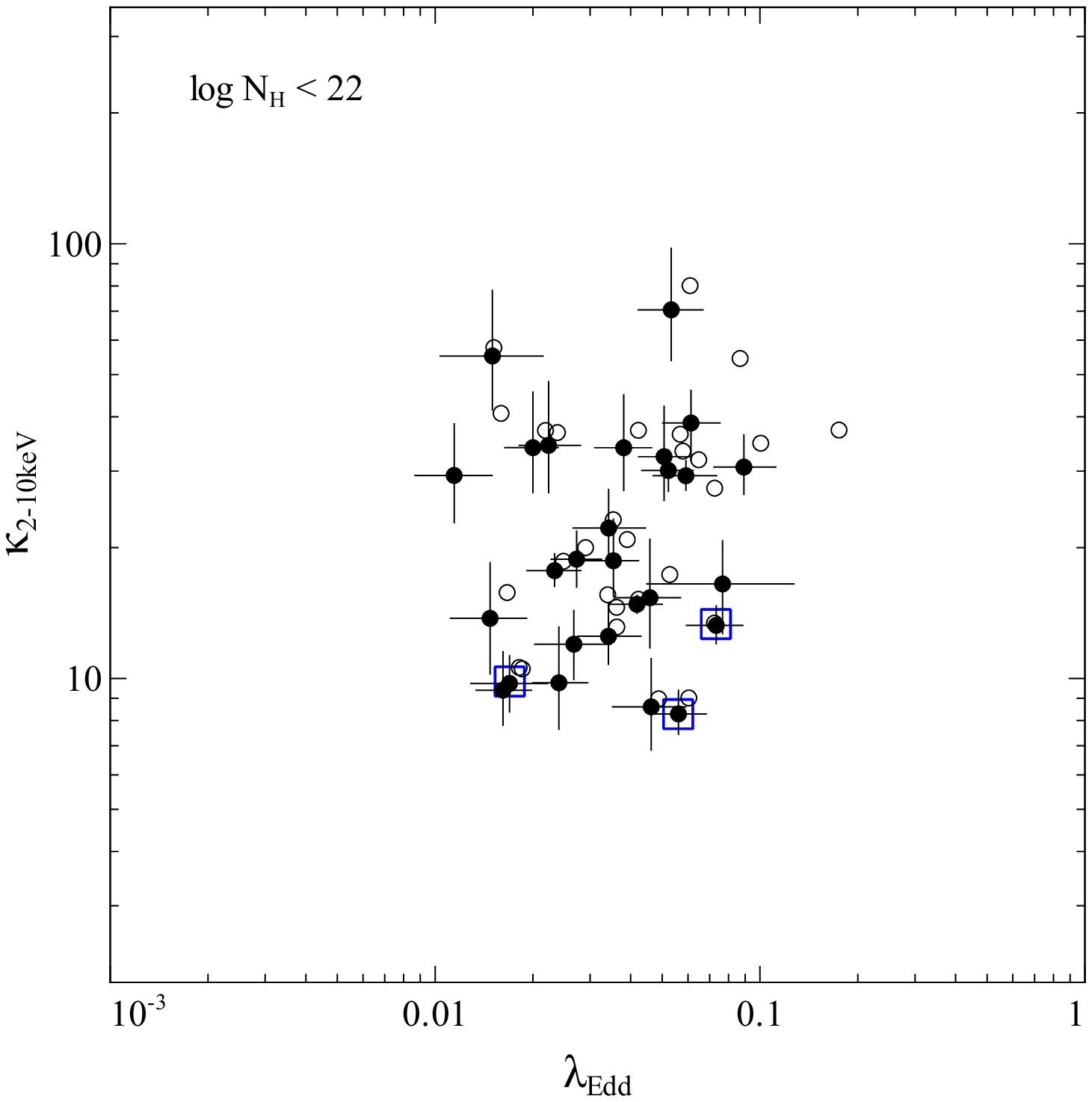}
    \includegraphics[width=8cm]{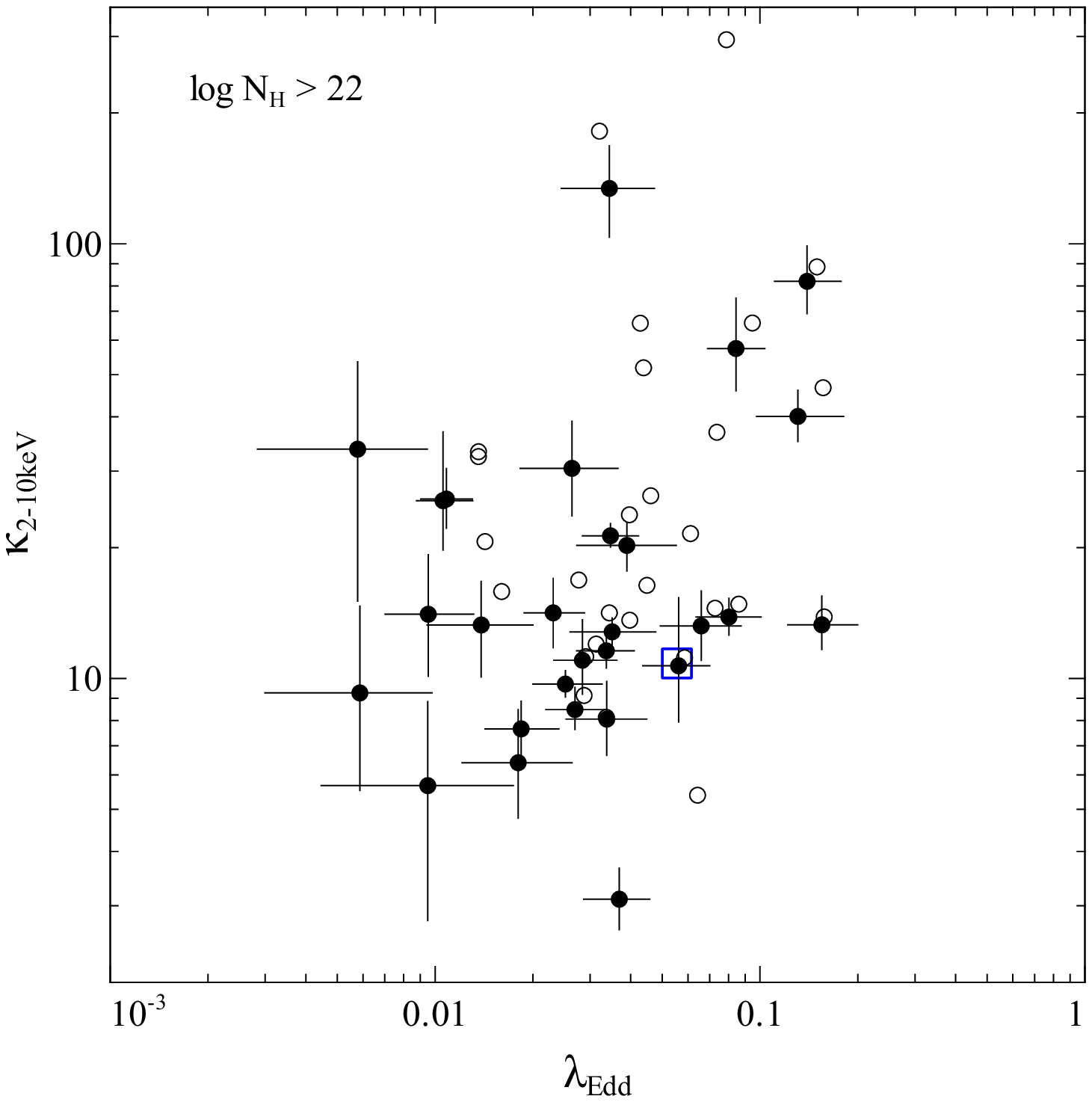}
    \caption{Bolometric corrections against Eddington ratios for low absorption (left panel) and high absorption (right panel) objects, where the host galaxy/starburst contribution in the IR has been removed by SED fitting as detailed in \S\ref{IRAScorr_Silva04}; key as in Fig.~\ref{bolcor_vs_edd_galaxycorr}}
\label{bolcor_vs_edd_silvaetal04hostcorr}
\end{figure*}

\begin{figure*}
  \includegraphics[width=8cm]{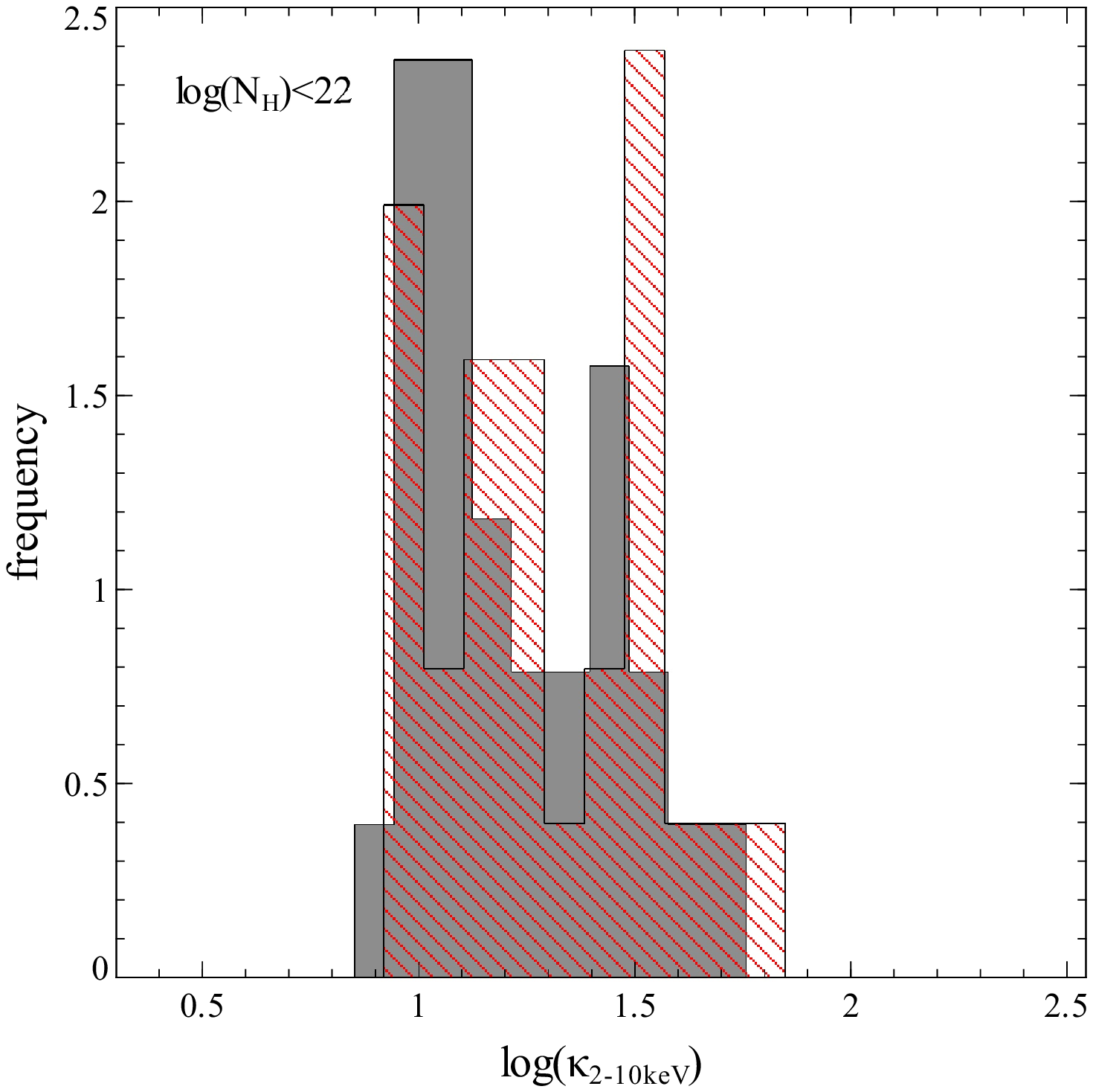}
  \includegraphics[width=8cm]{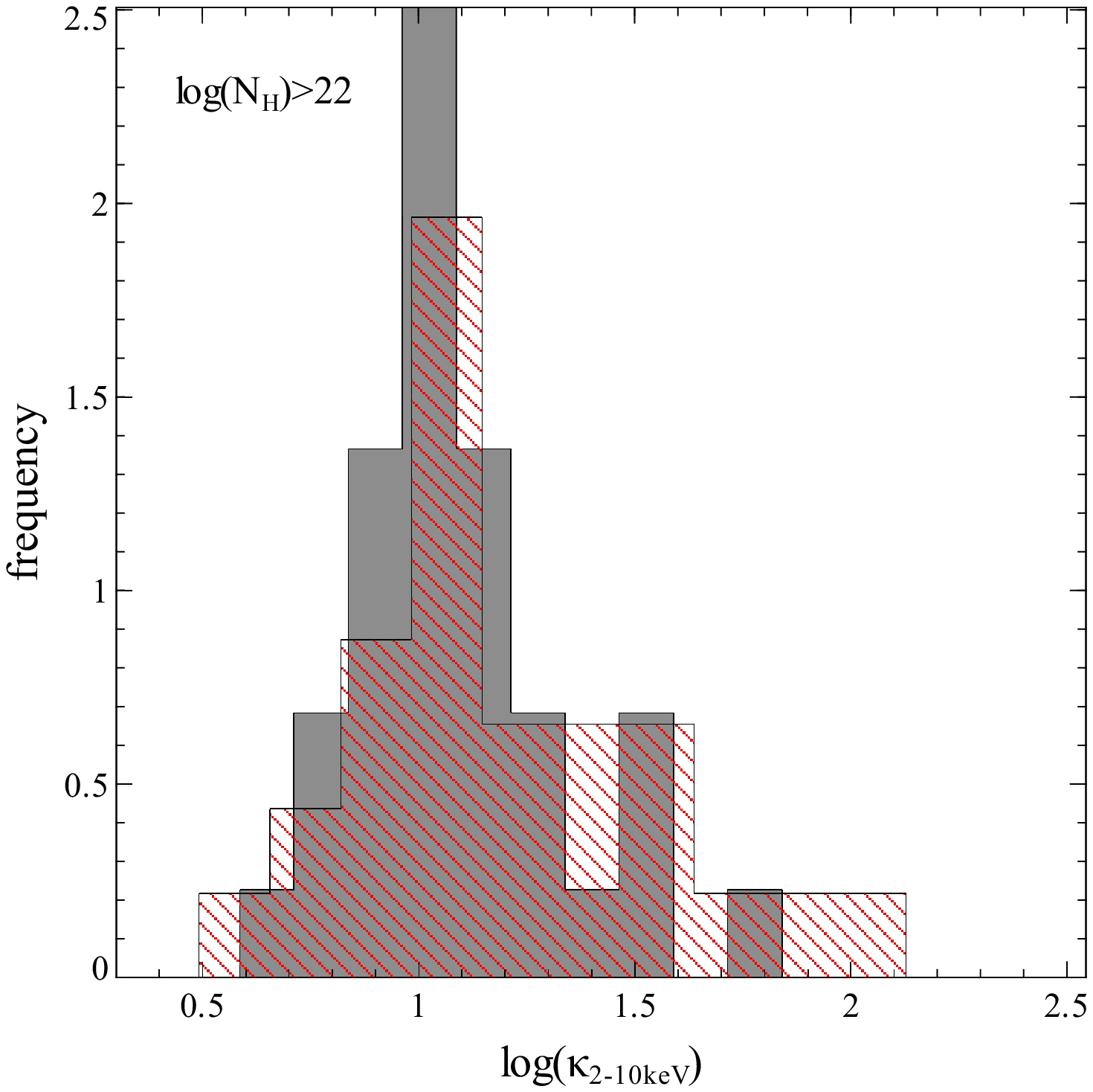}
  \includegraphics[width=8cm]{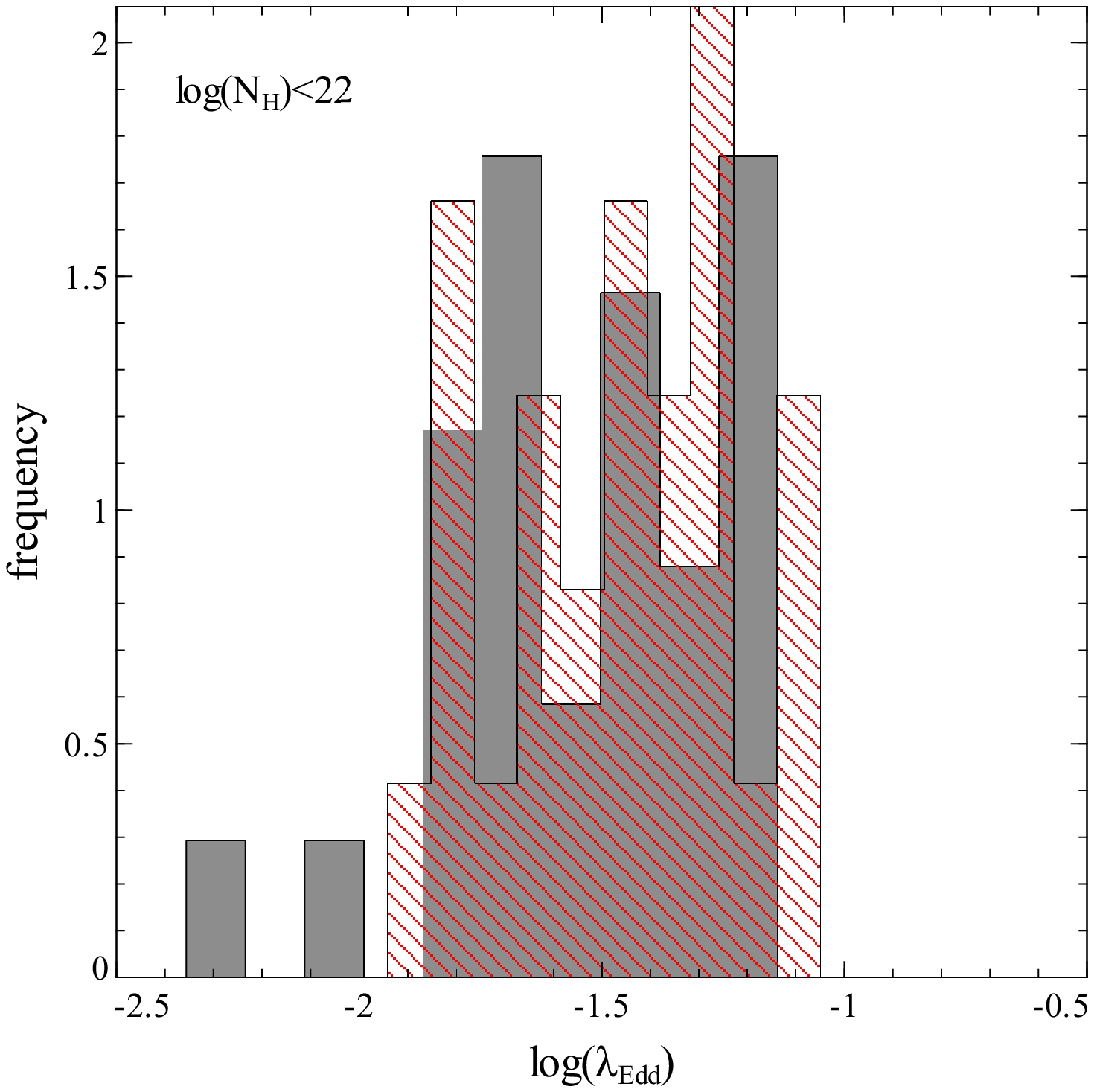}
  \includegraphics[width=8cm]{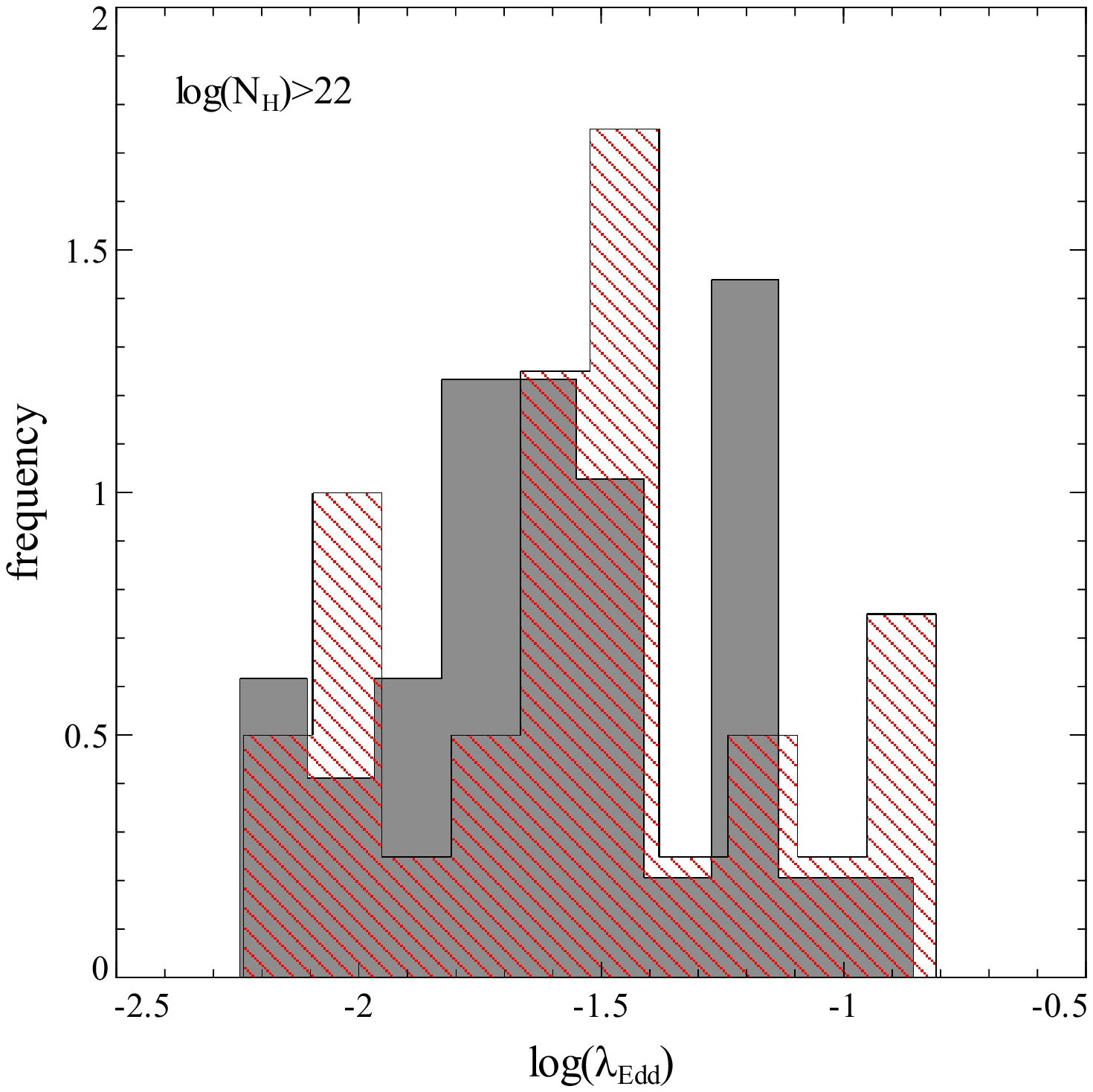}
\caption{Histograms of bolometric correction and Eddington ratio distributions.  The grey filled histograms are obtained from the statistical method of host galaxy correction in the IR, whereas the hatch-filled histograms are from host galaxy SED fitting to remove IR galaxy contamination.\label{histograms}}
\end{figure*}

We present the bolometric corrections and Eddington ratios obtained in Figs.~\ref{bolcor_vs_edd_galaxycorr} and \ref{bolcor_vs_edd_silvaetal04hostcorr}.  The distributions are plotted as histograms in Fig.~\ref{histograms} and numerical values are provided in Table~\ref{Results}. The values obtained from fitting a nuclear SED template to the IR data without any host galaxy correction are shown using empty circles for comparison, and objects from the 3C catalogues (3C 382, 3C 120, 3C 390.3 and 3C 403) are highlighted with blue squares, for easy identification in case some degree of jet contribution to the flux affects the results for these objects.

For low-absorption objects, the statistical correction method for removing the host galaxy results in bolometric corrections of $\sim19$ on average with Eddington ratios centered around $\sim0.035$ (Fig.~\ref{bolcor_vs_edd_galaxycorr}, left panel). In contrast, using the SED fitting approach yields $\langle \kappa_{\rm 2-10keV} \rangle \sim 24$ with $\langle \lambda_{\rm Edd} \rangle \sim 0.039$ (Fig.~\ref{bolcor_vs_edd_silvaetal04hostcorr}, left panel).  Both are generally consistent with the finding of low bolometric corrections for unabsorbed objects from \cite{2009arXiv0907.2272V}, but the host SED fitting approach does have a greater proportion of objects at higher bolometric corrections.

The object consistently exhibiting the largest bolometric correction in the high absorption subsample is NGC 1365, which is one of two possible borderline Compton-thick objects in that sample and known to have variable absorption as discussed in \S\ref{batdata}.  If the source was predominantly in a Compton-thick state during the BAT period of observation, the BAT flux would be substantially lower than the intrinsic AGN flux as pointed out in \S\ref{batdata}, giving an artifically high bolometric correction.  Additionally, this object is known to exhibit a nuclear starburst (\citealt{Strateva:2009xd} and references therein) so is particularly susceptible to a high level of starburst contamination in the IR fluxes, which may not have been completely removed by either method of host galaxy removal.  Both of these effects would serve to increase the bolometric correction to an abnormally high value.  A more detailed object-specific analysis is needed to check whether the bolometric correction for this object is intrinsically high.

\begin{figure*}
\includegraphics[width=4.5cm]{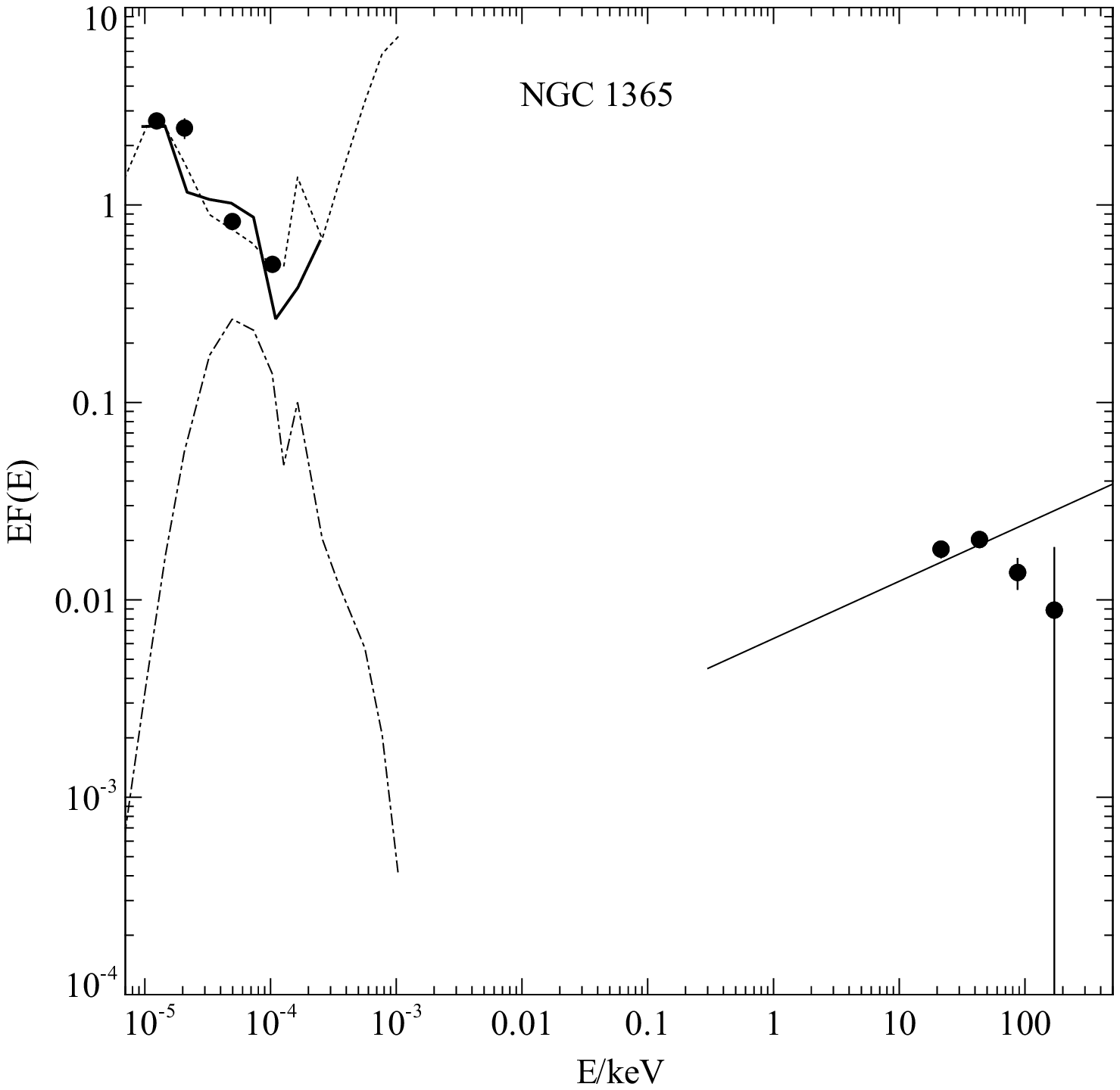}
\includegraphics[width=4.5cm]{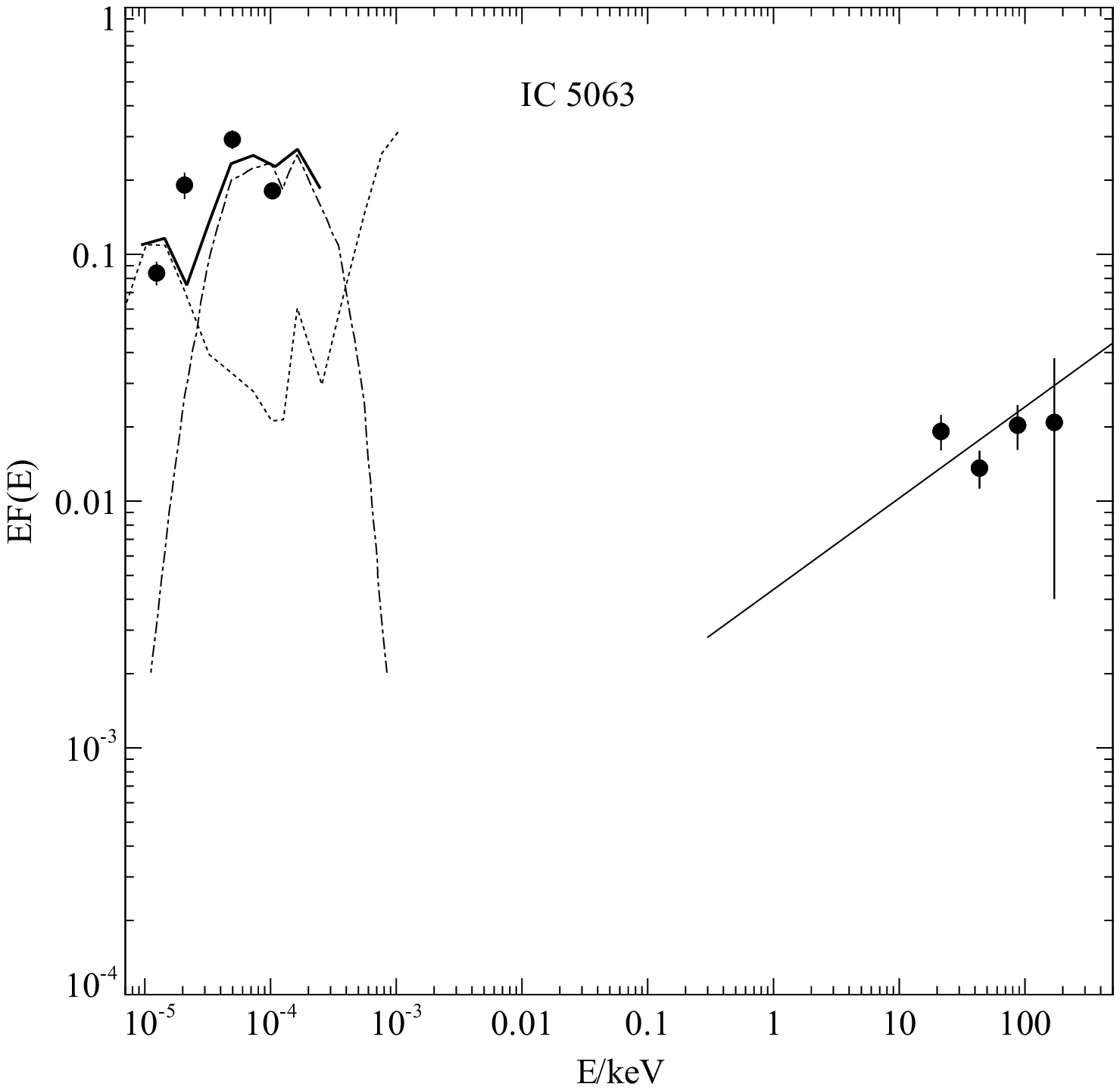}
\includegraphics[width=4.5cm]{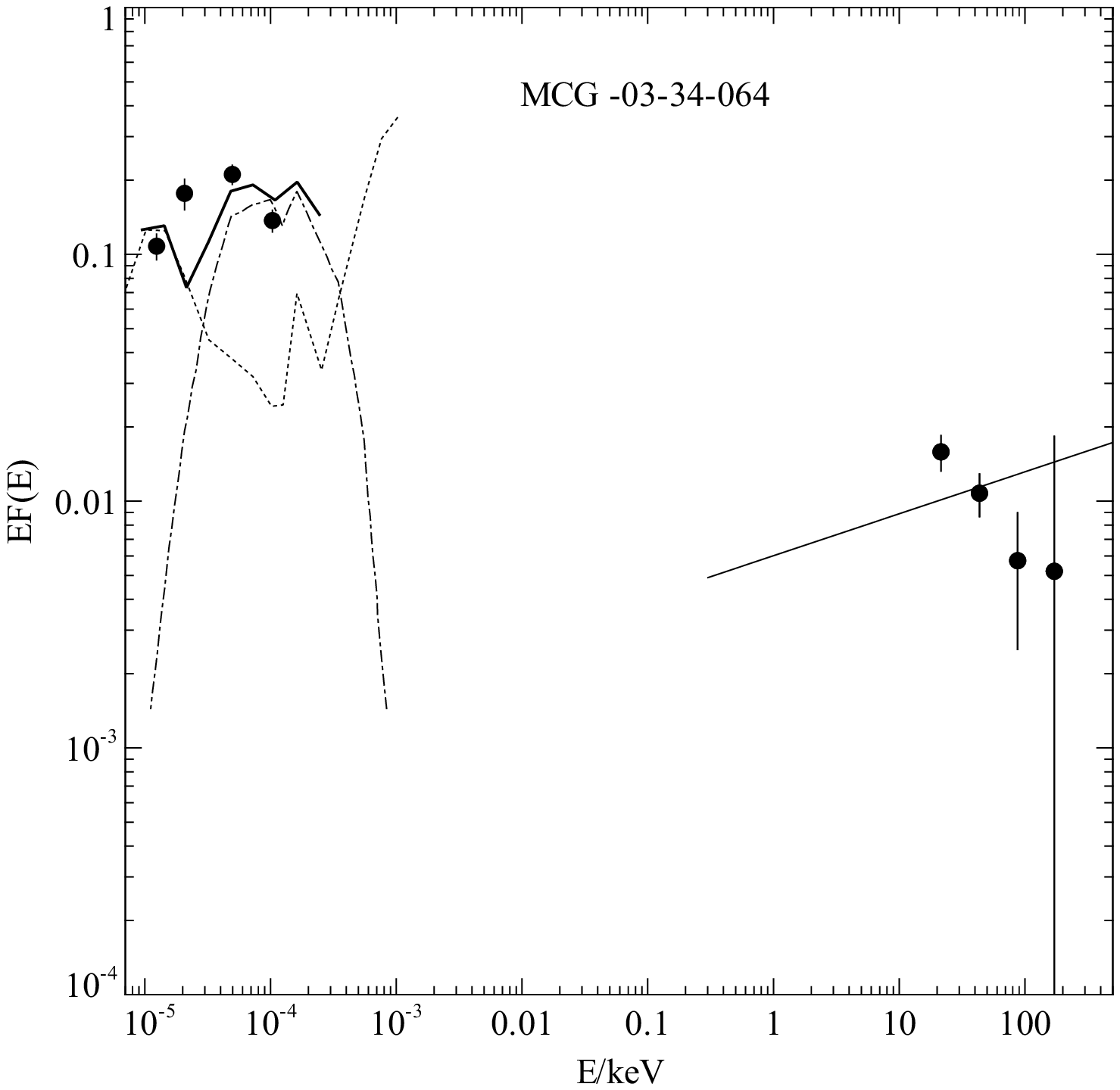}
\includegraphics[width=4.5cm]{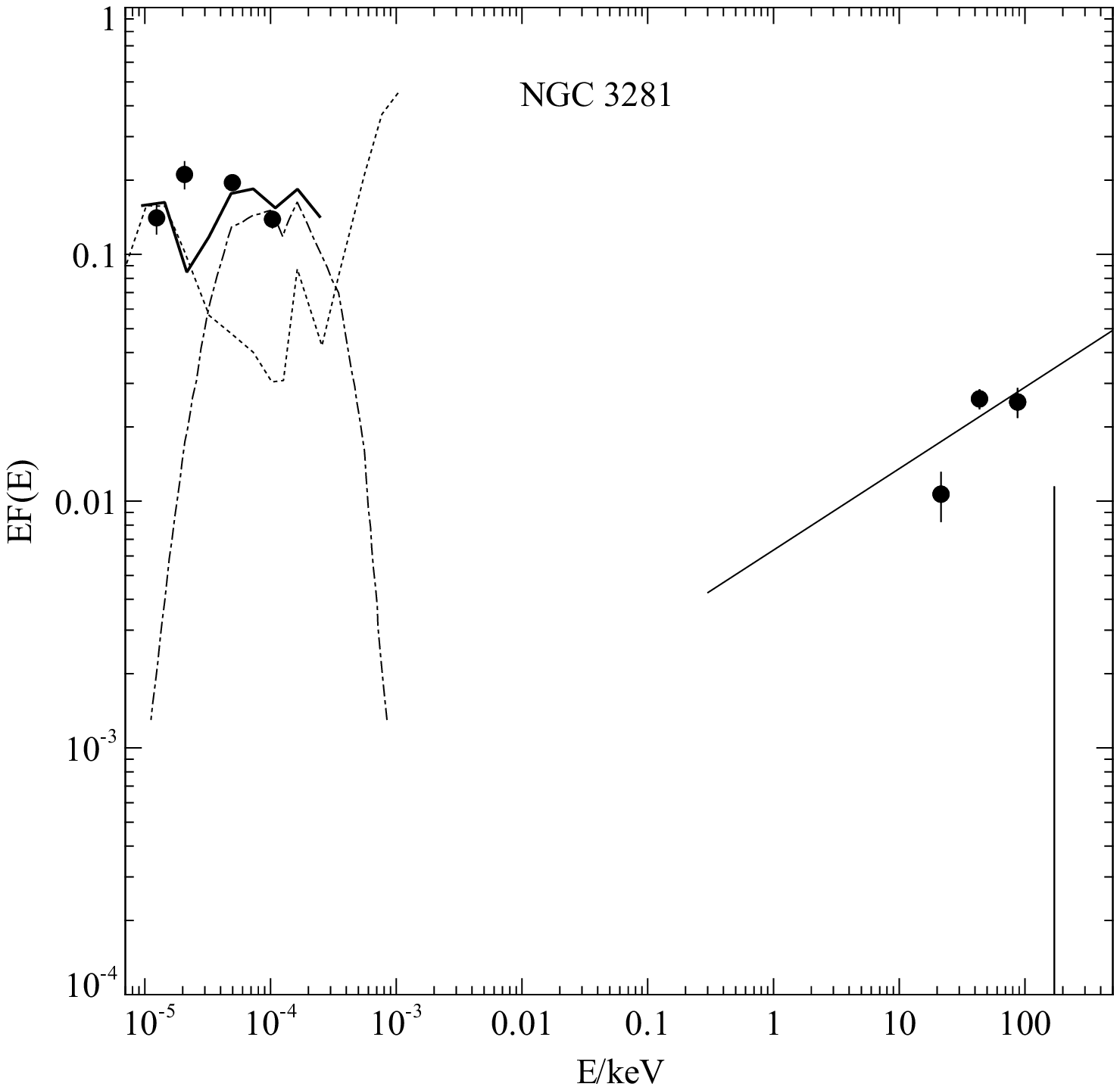}
   \caption{SEDs for the four objects with high bolometric corrections and Eddington ratios, under method 2 (\S\protect\ref{IRAScorr_Silva04}) for host galaxy removal.  Key as in Fig.~\protect\ref{example_SEDs_silvaetal04hostcorr}.}
\label{transition_SEDs}
\end{figure*}

For high-absorption objects, the statistical approach to correcting for the host galaxy (Fig.~\ref{bolcor_vs_edd_galaxycorr}, right panel) yields a wider spread of bolometric corrections than for low-absorption objects, but values are still centred on low values ($\langle \kappa_{\rm 2-10keV} \rangle \sim 15$) with a predominantly low Eddington ratio distribution ($\langle \lambda_{\rm Edd} \rangle \sim 0.034$).  However, under the host SED correction method, some notable outliers with negligible host galaxy component are located near $\lambda_{\rm Edd} \sim 0.1$ with bolometric corrections of $\sim 80$.  These outliers are (from highest bolometric correction downwards) NGC 1365, IC 5063, MCG -03-34-064 and NGC 3281.  In the case of NGC 3281, the degree of host contamination is known to be small from the comparison with VLT/VISIR fluxes in Fig.~\ref{L12micron_vs_Xray_Gandhi09compare_IRASandVISIR}, so the high accretion rate and value of $\kappa_{\rm 2-10keV}$ may be real for such objects (bearing in mind the caveats regarding NGC 1365 discussed above).  Inspection of the host and nucleus SED fits for these four objects in Fig.~\ref{transition_SEDs} implies sensible estimation of the host contamination, implying that method 1 has overestimated the nuclear contamination for these sources. The large bolometric corrections from method 2 for these objects may therefore, on average, be closer to the true values. These objects could then potentially be part of the `transition region' to high bolometric corrections at and above $\lambda_{\rm Edd} \sim 0.1$ postulated by \cite{2007MNRAS.381.1235V}, (2009).  Under method 2 for host galaxy correction, at $\lambda_{\rm Edd} < 0.03$, bolometric corrections have an average value of $\sim15$, increasing to a value of $\sim 32$ for $0.03 < \lambda_{\rm Edd} < 0.2$ (the average Eddington ratio for the whole sample is $\sim 0.043$).  This could be a preliminary indication of a significant minority of higher-accretion rate objects amongst high-absorption AGN and may provide tentative evidence that the Eddington ratio-dependent bolometric correction scheme of \cite{2007MNRAS.381.1235V} may be borne out in high-absorption AGN.  Better constraints on the nuclear MIR emission and black hole mass are needed to confirm this, however.

Under the statistical method for host-galaxy correction (method 1), fitting both 25 and 12 $\rm \mu m$ points for objects with particularly `red' 12--25 $\rm \mu m$ SEDs to a nuclear SED template results in a fit inconsistent with the 12 $\rm \mu m$ datum.  Such SED shapes probably imply that the nuclear 25$\rm \mu m$ luminosity is primarily from the host-galaxy rather than the nucleus, and fitting to only the 12$\rm \mu m$ point would be more robust.  We estimated the effect of this on the results by fitting only the 12 $\rm \mu m$ for the handful objects with such SEDs, and find that the resultant bolometric luminosities are altered on average by a few per cent (the largest change is for NGC 7582, causing a reduction of 40 per cent in the total IR luminosity).  Overall, such modifications do not produce any significant changes in the results.  More detailed model fits to better quality IR data in the future could be fit using a more considered approach.

We note that a few AGN excluded from the `host plus nucleus' SED fit subsample due to the fit yielding zero nuclear contribution (see \S\ref{totalIRlumin}) are more likely to have very small nuclear IR luminosities, therefore would be expected to lie at lower bolometric corrections and Eddington ratios; this would not be expected to alter the distributions seen significantly.

\begin{table*}
\begin{tabular}{llllllll}

\hline
AGN&${L_{\rm X}}^{a}$&${L_{\rm 1-1000 \mu m}}^{b}$&${L_{\rm bol}}^{c}$&${\kappa_{\rm 2-10keV}}^{d}$&${\kappa_{\rm 12 \mu m}}^{e}$&${M_{\rm BH}}^{f}$&${\lambda_{\rm Edd}}^{g}$\\
\hline




&&&-- (low $N_{\rm H}$) --&&&&\\
2MASX J21140128+8204483&$45.1$,$44.5$&$45.3$/$45.2$&$45.5$/$45.5$&$8.94$/$8.60$&$9.62$/$8.79$&$8.68$&$0.048$/$0.046$\\
3C 120&$44.8$,$44.3$&$45.0$/$45.0$&$45.2$/$45.2$&$8.93$/$8.29$&$8.51$/$8.67$&$8.35$&$0.060$/$0.056$\\
3C 382&$44.9$,$44.2$&$44.9$/$44.8$&$45.2$/$45.1$&$10.8$/$9.74$&$8.10$/$11.5$&$8.79$&$0.019$/$0.017$\\
3C 390.3&$44.9$,$44.2$&$45.1$/$45.1$&$45.3$/$45.3$&$13.4$/$13.2$&$9.06$/$8.90$&$8.35$&$0.073$/$0.073$\\
ESO 490-G026&$43.8$,$43.2$&$44.5$/$44.7$&$44.6$/$44.7$&$25.3$/$34.0$&$7.51$/$6.01$&$8.05$&$0.029$/$0.038$\\
ESO 511-G030&$43.9$,$43.4$&$44.2$/$44.1$&$44.4$/$44.3$&$11.4$/$9.77$&--/$8.59$&$7.84$&$0.029$/$0.024$\\
ESO 548-G081&$43.4$,$42.9$&$43.6$/$43.9$&$43.8$/$44.0$&$9.15$/$13.8$&$5.40$/$7.10$&$7.74$&$0.0095$/$0.015$\\
IC 4329A&$44.4$,$43.9$&$44.9$/$44.9$&$45.0$/$45.0$&$15.2$/$14.8$&$7.05$/$7.01$&$8.29$&$0.042$/$0.042$\\
IRAS 05589+2828&$44.2$,$43.6$&$45.0$/$45.0$&$45.1$/$45.1$&$33.6$/$34.4$&$5.56$/$6.11$&$8.64$&$0.022$/$0.022$\\
IRAS 09149-6206&$44.6$,$44.0$&$45.7$/$45.7$&$45.7$/$45.7$&$57.2$/$55.2$&$6.16$/$5.71$&$9.45$&$0.015$/$0.015$\\
MCG -06-30-015&$43.1$,$42.6$&$43.3$/$43.8$&$43.6$/$43.9$&$10.0$/$22.2$&$8.58$/$6.32$&$7.25$&$0.015$/$0.034$\\
MCG +08-11-011&$44.2$,$43.5$&$44.9$/$44.9$&$45.0$/$45.0$&$29.3$/$30.1$&$7.85$/$6.18$&$8.17$&$0.051$/$0.052$\\
Mrk 279&$44.3$,$43.8$&$44.6$/$44.6$&$44.8$/$44.8$&$10.8$/$9.40$&$7.69$/$8.05$&$8.42$&$0.019$/$0.016$\\
Mrk 290&$43.9$,$43.3$&$44.2$/$44.3$&$44.4$/$44.4$&$13.0$/$15.3$&$7.95$/$7.48$&$7.67$&$0.039$/$0.046$\\
Mrk 509&$44.7$,$44.1$&$45.1$/$45.1$&$45.2$/$45.2$&$13.1$/$12.5$&$7.46$/$7.40$&$8.56$&$0.036$/$0.034$\\
Mrk 590&$43.8$,$43.2$&$44.5$/$44.7$&$44.6$/$44.7$&$24.5$/$34.0$&$5.30$/$6.04$&$8.30$&$0.014$/$0.020$\\
Mrk 766&$43.1$,$42.5$&$43.9$/$44.4$&$44.0$/$44.4$&$27.0$/$70.5$&$8.09$/$5.56$&$7.53$&$0.021$/$0.053$\\
Mrk 79&$44.0$,$43.4$&$44.6$/$44.6$&$44.7$/$44.7$&$17.1$/$18.7$&$8.04$/$6.42$&$8.04$&$0.032$/$0.035$\\
Mrk 841&$44.3$,$43.7$&$45.1$/--&$45.1$/--&$27.4$/--&$4.81$/--&$8.17$&$0.073$/--\\
NGC 3516&$43.5$,$42.9$&$43.7$/$44.0$&$43.9$/$44.2$&$10.3$/$17.7$&$8.72$/$6.80$&$7.68$&$0.014$/$0.023$\\
NGC 3783&$43.6$,$43.0$&$44.1$/$44.4$&$44.2$/$44.4$&$18.2$/$29.3$&$9.08$/$6.22$&$7.55$&$0.037$/$0.059$\\
NGC 4051&$42.6$,$42.0$&$42.6$/$43.2$&$42.9$/$43.3$&$7.11$/$16.5$&$9.74$/$6.61$&$6.26$&$0.035$/$0.077$\\
NGC 4593&$43.5$,$43.0$&$43.7$/$43.9$&$43.9$/$44.1$&$8.87$/$12.0$&$10.4$/$7.46$&$7.51$&$0.020$/$0.027$\\
NGC 5548&$43.8$,$43.3$&$44.3$/$44.5$&$44.4$/$44.5$&$14.6$/$18.8$&$7.23$/$6.43$&$8.00$&$0.021$/$0.027$\\
NGC 7213&$42.7$,$42.1$&$42.9$/$43.5$&$43.1$/$43.5$&$11.3$/$29.3$&$6.47$/$6.02$&$7.37$&$0.0044$/$0.011$\\
NGC 7469&$44.0$,$43.5$&$45.1$/$45.0$&$45.1$/$45.1$&$45.3$/$38.7$&$8.62$/$5.70$&$8.16$&$0.071$/$0.061$\\
NGC 931&$43.8$,$43.2$&$44.5$/$44.7$&$44.6$/$44.7$&$23.6$/$30.6$&$6.31$/$6.01$&$7.65$&$0.070$/$0.089$\\
NGC 985&$44.3$,$43.7$&$45.2$/$45.1$&$45.2$/$45.2$&$36.6$/$32.4$&$6.08$/$6.17$&$8.36$&$0.058$/$0.051$\\
&&&-- (high $N_{\rm H}$) --&&&&\\
2MASX J04440903+2813003&$43.3$,$42.7$&$43.3$/--&$43.6$/--&$8.86$/--&--/--&$7.37$&$0.013$/--\\
3C 403&$44.5$,$44.1$&$45.0$/$45.0$&$45.1$/$45.1$&$11.4$/$10.7$&$5.96$/$5.49$&$8.26$&$0.059$/$0.056$\\
4U 1344-60&$43.5$,$43.0$&$43.9$/--&$44.1$/--&$12.9$/--&$7.56$/--&$7.44$&$0.034$/--\\
Cyg A&$45.0$,$44.6$&$45.7$/$45.6$&$45.7$/$45.7$&$15.4$/$13.3$&--/$5.19$&$9.41\dagger$&$0.017$/$0.014$\\
ESO 005-G004&$42.6$,$41.9$&$42.9$/$42.6$&$43.1$/$42.9$&$14.5$/$9.26$&$5.89$/$9.19$&$6.98$&$0.0091$/$0.0059$\\
ESO 103-035&$43.7$,$43.5$&$44.5$/$44.6$&$44.6$/$44.6$&$11.8$/$13.3$&$7.73$/$4.81$&$7.30$&$0.14$/$0.16$\\
ESO 297-018&$43.9$,$43.2$&$44.0$/--&$44.2$/--&$9.31$/--&--/--&$7.69$&$0.026$/--\\
ESO 506-G027&$44.2$,$43.6$&$44.4$/$44.3$&$44.6$/$44.6$&$9.04$/$8.48$&$7.64$/$7.53$&$8.02$&$0.029$/$0.027$\\
EXO 055620-3820.2&$44.3$,$43.8$&$45.2$/--&$45.2$/--&$23.6$/--&$4.74$/--&$8.90$&$0.016$/--\\
IC 5063&$43.3$,$42.8$&$44.3$/$44.7$&$44.3$/$44.7$&$36.5$/$82.0$&$6.59$/$4.45$&$7.41$&$0.063$/$0.14$\\
MCG -03-34-064&$43.4$,$43.1$&$44.6$/$44.8$&$44.7$/$44.9$&$36.8$/$57.4$&$6.23$/$4.37$&$7.81$&$0.053$/$0.084$\\
Mrk 1498&$44.6$,$44.3$&$45.2$/$45.2$&$45.3$/$45.3$&$11.1$/$11.0$&$4.65$/$5.23$&$8.72$&$0.029$/$0.028$\\
Mrk 18&$42.9$,$42.4$&$43.0$/$43.4$&$43.3$/$43.5$&$7.89$/$14.1$&$7.63$/$5.85$&$7.42$&$0.0057$/$0.0095$\\
Mrk 3&$43.9$,$43.6$&$44.7$/$44.7$&$44.8$/$44.7$&$14.3$/$13.8$&$8.50$/$5.47$&$7.72$&$0.082$/$0.080$\\
Mrk 348&$43.7$,$43.2$&$44.1$/$44.2$&$44.3$/$44.4$&$10.5$/$13.2$&$7.26$/$5.49$&$7.41$&$0.054$/$0.066$\\
Mrk 6&$43.7$,$43.0$&$44.2$/$44.4$&$44.3$/$44.5$&$19.8$/$25.9$&$6.79$/$5.67$&$8.31$&$0.0082$/$0.011$\\
NGC 1142&$44.1$,$44.2$&$44.8$/$44.5$&$44.9$/$44.7$&$5.46$/$3.11$&$6.56$/$6.24$&$7.98$&$0.064$/$0.037$\\
NGC 1365&$42.4$,$42.0$&$43.8$/$44.1$&$43.8$/$44.1$&$69.2$/$134$&$10.0$/$15.1$&$7.46*$&$0.018$/$0.034$\\
NGC 2110&$43.6$,$42.9$&$43.5$/$43.7$&$43.8$/$43.9$&$8.00$/$9.71$&$11.9$/$8.69$&$7.40$&$0.021$/$0.025$\\
NGC 2992&$43.0$,$42.3$&$43.4$/$43.7$&$43.5$/$43.8$&$16.4$/$30.4$&$7.33$/$5.54$&$7.27$&$0.014$/$0.026$\\
NGC 3227&$42.9$,$42.2$&$43.0$/$43.4$&$43.2$/$43.5$&$10.3$/$20.2$&$10.4$/$6.27$&$6.80*$&$0.020$/$0.039$\\

\hline
\emph{Continued on next page...}
\end{tabular}
\caption[Luminosities, bolometric corrections and Eddington ratios from IRAS and BAT data]{Table of results from analysis of IRAS--BAT SEDs. Luminosities $L$ are presented as log($L$) with $L$ given in $\rm erg \thinspace s^{-1}$. $a$ - Total (0.5--100 keV, first value), followed by 2--10 keV (second value) X-ray luminosity, using the photon index $\Gamma$ from \protect\cite{2009ApJ...690.1322W} (requiring $1.5<\Gamma<2.2$). Errors (in luminosity) are typically $\sim 10-20$ per cent. $b$ - total integrated nuclear IR luminosity, with geometry, anisotropy and non-nuclear contamination correction factors applied (see text).   The first value includes the statistical host galaxy correction based on $L_{\rm 2-10keV}$, and the second value is that obtained from combined nuclear and host SED template fitting.  Errors are typically $\sim5-10$ per cent. $d$ - Bolometric luminosity ($L_{\rm 0.5-100keV}+L_{\rm 1-1000\mu m}$).  Values from both methods of host galaxy removal provided, as for $c$. Errors are typically $\sim5$ per cent. $d$ - 2--10keV bolometric correction $\kappa_{\rm 2-10 keV}=L_{\rm bol}/L_{\rm 2-10 keV}$, with statistical host galaxy removal/host+nucleus SED fitting methods. $e$ - Bolometric correction for observed nuclear $12 \rm \mu m$ monochromatic luminosity $\kappa_{\rm 12 \mu m}=L_{\rm bol}/(\lambda L_{\lambda}^{\rm 12 \mu m})$, for two different methods of host galaxy removal. $f$ - logarithm of black hole mass from K-band bulge magnitude estimate (in solar masses).  Errors are typically $\sim0.1$ dex, but for a discussion of systematics see \protect\cite{2009arXiv0907.2272V}.  * Masses marked with an asterisk are likely to have the bulge luminosity, and hence black hole mass, underestimated: see text for details.  $\dagger$ We use the dynamical mass from \protect\cite{2003MNRAS.342..861T} for Cyg A; since its bulge is abnormally large by virtue of being a cD galaxy, the use of the 2MASS PSC is not valid and would underestimate the mass by a factor of $\sim30$. $g$ - Eddington ratio $\lambda_{\rm Edd}=L_{\rm bol}/L_{\rm Edd}$, using both methods of host galaxy removal.  Random errors are typically $\sim10-20$ per cent. \label{Results}}
\end{table*}
\begin{table*}
\begin{tabular}{llllllll}

\hline
AGN&${L_{\rm X}}^{a}$&${L_{\rm 1-1000 \mu m}}^{b}$&${L_{\rm bol}}^{c}$&${\kappa_{\rm 2-10keV}}^{d}$&${\kappa_{\rm 12 \mu m}}^{e}$&${M_{\rm BH}}^{f}$&${\lambda_{\rm Edd}}^{g}$\\
\hline

NGC 3281&$43.3$,$42.8$&$44.1$/$44.4$&$44.2$/$44.4$&$21.4$/$40.1$&$6.38$/$4.57$&$7.20$&$0.071$/$0.13$\\
NGC 4388&$43.0$,$42.6$&$43.6$/$43.6$&$43.7$/$43.7$&$11.4$/$12.8$&$7.83$/$5.26$&$7.07$&$0.032$/$0.035$\\
NGC 4507&$43.8$,$43.3$&$44.1$/$44.2$&$44.3$/$44.3$&$10.3$/$11.6$&$8.17$/$5.83$&$7.70$&$0.030$/$0.034$\\
NGC 526A&$43.7$,$43.3$&$44.2$/$44.3$&$44.3$/$44.4$&$12.0$/$14.2$&$6.79$/$5.87$&$7.93$&$0.020$/$0.023$\\
NGC 5506&$43.5$,$43.0$&$44.0$/$44.3$&$44.1$/$44.3$&$12.3$/$21.3$&$7.80$/$5.53$&$7.67$&$0.020$/$0.035$\\
NGC 5728&$43.3$,$42.7$&$43.5$/$43.1$&$43.7$/$43.5$&$9.70$/$6.40$&$10.3$/$10.2$&$7.15*$&$0.028$/$0.018$\\
NGC 612&$43.7$,$43.7$&$44.4$/--&$44.5$/--&$6.24$/--&$3.01$/--&$8.01$&$0.022$/--\\
NGC 6300&$42.4$,$42.1$&$42.7$/$42.6$&$42.9$/$42.8$&$7.06$/$5.67$&$8.26$/$7.17$&$6.70*$&$0.013$/$0.0095$\\
NGC 6860&$43.3$,$42.6$&$43.6$/$44.0$&$43.8$/$44.1$&$14.6$/$25.6$&$6.25$/$5.63$&$7.91$&$0.0059$/$0.011$\\
NGC 7172&$43.4$,$42.9$&$43.6$/$43.5$&$43.8$/$43.7$&$8.16$/$7.66$&$7.53$/$7.94$&$7.36$&$0.020$/$0.018$\\
NGC 7314&$42.7$,$42.4$&$42.7$/--&$43.0$/--&$3.86$/--&$11.3$/--&$6.14$&$0.053$/--\\
NGC 7582&$42.5$,$41.8$&$43.3$/$43.3$&$43.3$/$43.3$&$34.4$/$33.7$&$9.23$/$5.53$&$7.44*$&$0.0060$/$0.0058$\\
NGC 788&$43.4$,$43.1$&$43.7$/$43.9$&$43.9$/$44.0$&$5.50$/$8.06$&$6.67$/$5.56$&$7.39$&$0.023$/$0.034$\\
PGC 13946&$43.9$,$43.3$&$44.2$/--&$44.4$/--&$12.1$/--&--/--&$7.46$&$0.064$/--\\

\hline
\hline
\end{tabular}
\begin{center}
Table~\ref{Results} (continued)
\end{center}
\end{table*}


These results are illuminating for our understanding of accretion in local AGN.  Firstly, they confirm that the majority of the Swift/BAT catalogue AGN are accreting at low Eddington ratios \citep{2009ApJ...690.1322W} across the range of available absorption properties probed.  This result is not significantly modified by possible over-estimates of the black hole mass using our $M_{\rm BH}-L_{\rm bulge}$ approach, as discussed in detail in \cite{2009arXiv0907.2272V}, since if we calibrate this approach against reverberation mapping as done in their paper, we obtain an offset of a factor of $\sim2$, which would shift the Eddington ratios obtained here to correspondingly higher values.  However, the centre of the distribution would still be located at low values.  At any rate, there are geometrical uncertainties in reverberation mapping which give rise to a tolerance of a factor of $\sim 3$ in calibrating the mass estimates.  

However, we note the presence of five AGN for which the Eddington ratios could be uncertain because of their mass estimates.  \cite{2009arXiv0907.2272V} use the 2MASS Point Source Catalogue (PSC) flux as an estimate of the total unresolved nuclear and bulge light, and calculate the fraction of the 2MASS PSC flux which can be attributed to the bulge.  Inspection of the 2MASS images for all of the nearby galaxies reveal that there may be a few sources for which the assumption of an unresolved bulge is not valid, and (apart from the case of Cyg A already discussed) those overlapping with our sample are NGC 6300, NGC 3227, NGC 5728, NGC 7582 and NGC 1365.  If the bulge luminosity has been underestimated in these sources, the small resultant black hole masses will artificially increase their Eddington ratios.  If corrected, this should shift the average Eddington ratio down for this subsample.


We note that the bolometric luminosities are not dependent on the mass in the simple approach used here, in contrast to the method outlined in \cite{2007MNRAS.381.1235V} and their subsequent studies, which involves fitting a multicolour accretion disc model to optical--UV data with the model normalisation constrained by the black hole mass.  In the zeroth-order approach presented here, the main uncertainties in determining $L_{\rm bol}$ are the accuracy of the correction for non-nuclear contamination, the accuracy of the nuclear IR SED templates used and the degree of hard X-ray variability (\citealt{2007A&A...468..603P} find that using various different IR SED templates does not significantly alter the extrapolated bolometric luminosity, however). As a result, uncertainties in mass estimates only affect Eddington ratios, not bolometric luminosities.

Bolometric corrections are generally lower than those obtained for high Eddington-rate sources (\citealt{2009MNRAS.392.1124V}, 2007) and quasars \citep{1994ApJS...95....1E}.  These results provides a very interesting window onto the properties of the more complex high absorption AGN, as hinted at previously by \cite{2007A&A...468..603P}.  Our approach aims to by-pass the unwanted complexities introduced by absorption in the optical--to--X-ray regime by using the reprocessed IR and (relatively) unaffected hard X-ray BAT data. Indications of predominantly low bolometric corrections for high absorption AGN have consequences for matching the local SMBH mass density with that inferred from the X-ray background.  A significant population of highly absorbed AGN is required to fit the X-ray background spectrum (\citealt{2003MNRAS.339.1095G}, \citealt{2007A&A...463...79G}), and as discussed in \cite{2004cbhg.symp..446F}, bolometric corrections of around $\sim10-20$ are appropriate for reconciling the XRB with the local black hole mass density.

\subsection{Anisotropy and geometry corrections: smooth-vs-clumpy tori}

The similar regions of the $L_{\rm 12 \mu m}-L_{\rm 2-10keV}$ plot in Fig.~\ref{L12micron_vs_Xray_WITHgalaxycorr} occupied by both obscured and unobscured sources suggests that the $12 \rm \mu m$ luminosities are isotropic (assuming that the absorption responsible for producing the MIR emission is broadly correlated with the X-ray absorption; see \S\ref{Intro} and \S\ref{totalIRlumin}).  The SED models and anisotropy corrections used here in calculating $L_{\rm bol}$ are based ultimately on the models of \cite{1994MNRAS.268..235G} which assume a smooth dust distribution in the torus.  If the torus is clumpy however, as recent models and fits to the data suggest (\citealt{2004Natur.429...47J}, \citealt{2007A&A...474..837T}, \citealt{2008A&A...486L..17B}), the torus emission is more isotropic and would not require significant anisotropy corrections between edge-on (obscured) and face-on (unobscured) AGN; this is possibly supported by the emergence of a unified $L_{\rm 12 \mu m}-L_{\rm 2-10 keV}$ correlation for both types of AGN and its implication of isotropic MIR emission.  If we also inspect the values of $25 \rm \mu m$ luminosity plotted against X-ray luminosity we again find that obscured and unobscured objects do not occupy different regions of the plot, implying that the $25 \rm \mu m$ luminosity may also be, to a large degree, isotropic.  In any case, even if we remove the isotropy corrections for obscured objects, this would have the effect of reducing bolometric corrections and Eddington ratios further by around 20--30 per cent for that subsample (since $\kappa_{\rm 2-10 keV}=(L_{\rm X, total}+f_{\rm geo}f_{\rm anisotropy}f_{\rm starlight}L_{\rm IR}^{\rm (nuc,obs)})/L_{\rm 2-10keV}$, for geometry, anisotropy and starlight correction factors $f_{\rm geo},f_{\rm anisotropy},f_{\rm starlight}$ discussed previously).  Such a modification would not significantly alter our conclusions.

\subsection{Objects not detected by IRAS}

The non-detection of a substantial fraction of the sample warrants consideration of the expected distribution of values for $\rm \kappa_{\rm 2-10keV}$ and $\lambda_{\rm Edd}$ from the non-detected objects.  Assuming the missing objects are just below detection, the upper limits presented in Fig.~\ref{deltaL12micron_vs_Xray} imply that the majority of the missing objects would have a very similar distribution of $L_{\rm 12 \mu m}$ values to the detected objects.  If the upper limits lay significantly above the relation in comparison to the detected sample, one would expect to see higher bolometric corrections on average due to a bigger potential nuclear IR contribution, and vice versa if they lay significantly below the line.  Given that the upper limits also occupy a similar range in $L_{\rm 2-10keV}$ as the detected sample, the distribution of bolometric corrections would also be expected to be similar.  The mean black hole mass for the non-detected sample is $\rm log (M_{\rm BH}/M_{\odot})=7.82$ compared to $7.76$ for the detected sample, with very similar spreads, implying that the non-detections may lie at similar Eddington ratios as well.  However this is obviously conjecture, and future work using IR data probing to lower fluxes would be able to verify or challenge these predictions.

\begin{figure}
\includegraphics[width=7cm]{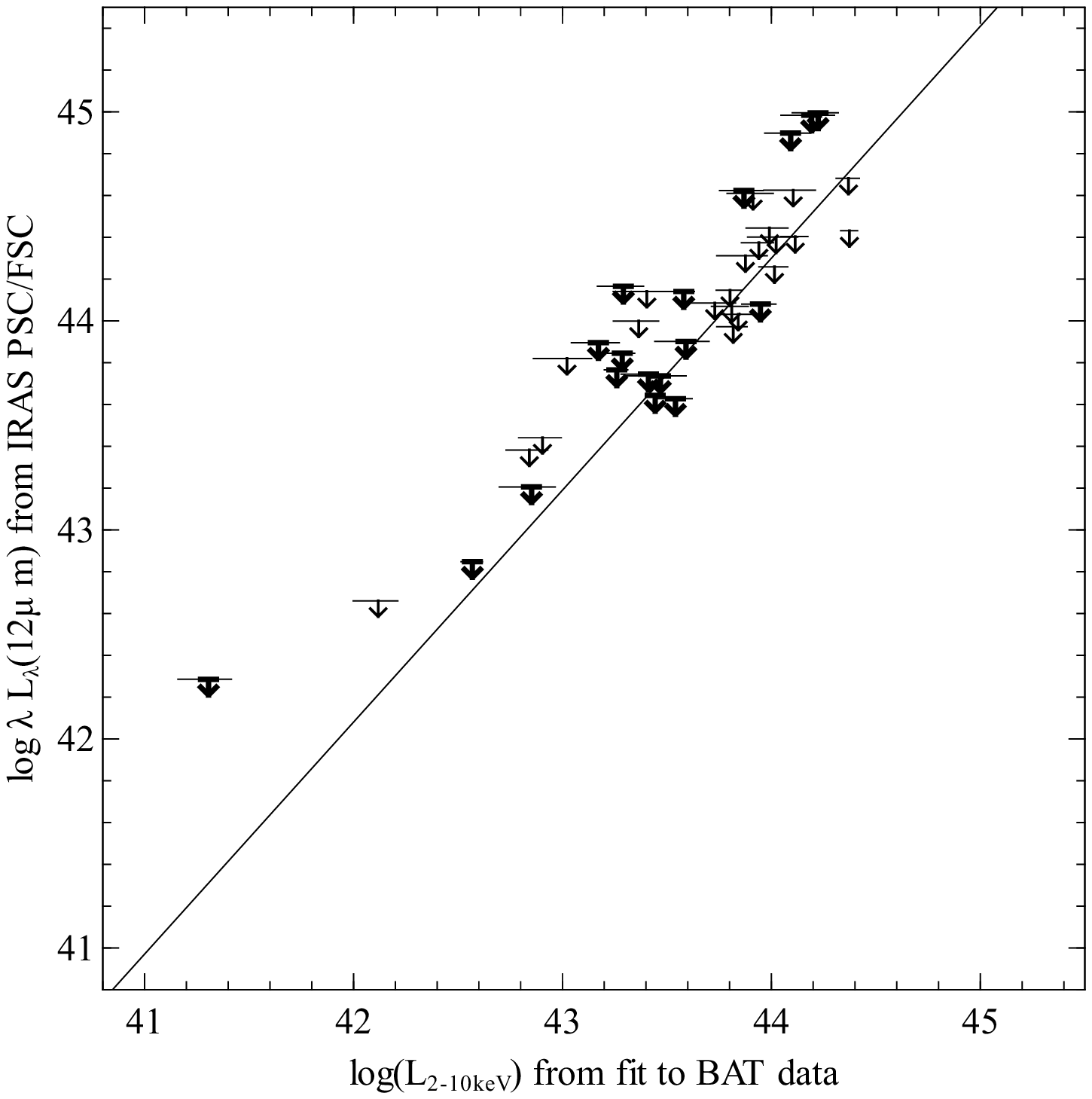}
    \caption{Upper limiting infrared 12 $\rm \mu m$ luminosity against 2--10keV luminosity from fits to BAT data, for objects without \emph{IRAS} PSC or FSC data.  The completeness limit of the \emph{IRAS} Faint Source Catalogue, 0.2 Jansky has been used in calculation of these upper limits.}
\label{L12micron_vs_Xray_NOTINIRAS}
\end{figure}

\section{Summary and Conclusions}

We have presented a simple extension of the \cite{2007A&A...468..603P} method for calculating the bolometric output of AGN using reprocessed IR emission from \emph{IRAS} along with the hard X-ray emission from \emph{Swift}-BAT.  We estimate the integrated nuclear IR luminosity by fitting of AGN nuclear SED templates from \cite{2004MNRAS.355..973S} to the IR data, and estimate non-nuclear contamination in the large aperture \emph{IRAS} fluxes via two methods: the first involves applying a correction to make the distribution of 12 $\rm \mu m$ fluxes lie on the most recent determination of the $L_{\rm X}-L_{\rm 12 \mu m}$ relation determined by \cite{2009A&A...502..457G}, and the second involves fitting a host galaxy SED template alongside the nuclear SED template.  The second method has the advantage that it does not presuppose a relation between X-ray and MIR luminosity, which will tend to bias the bolometric corrections to certain values, whereas the former method is superior in that it ought to give clean estimates of the core powers alone, on average.  We fit a simple power-law model to the hard X-ray BAT data to determine the X-ray luminosities, and then calculate the bolometric luminosity after applying corrections for torus geometry and anisotropy of emission as discussed by \cite{2007A&A...468..603P}.  We calculate Eddington ratios using black hole masses estimated from the K-band bulge luminosity as described in \cite{2009arXiv0907.2272V} (and use a well-constrained dynamical mass for the known cD galaxy Cyg A), and present the 2--10 keV bolometric corrections.  We calibrate our approach against the simultaneous optical--to--X-ray SEDs from \cite{2009arXiv0907.2272V} and find trends for agreement between the two approaches, although there are still numerous sources of scatter.

Using our zeroth-order IR and hard X-ray approach, bolometric corrections for the low-absoprtion subsample give an average of around $\sim 22$, intermediate between the averages found from the two approaches of removing the host galaxy contamination (19,24).  Eddington ratios are $\sim 0.037$ on average, from the two methods.  For the high-absorption subsample the average Eddington ratios are $0.034$ and $0.043$ from the statistical and host-SED fitting methods for host galaxy removal, respectively. The spread in bolometric corrections is greater than for low absorption objects, but for the lower part of the Eddington ratio distribution ($\lambda_{\rm Edd}<0.03$) bolometric corrections are consistently low in both methods for host galaxy contamination removal ($\sim 15$).  The errors on these average bolometric corrections are typically less than $\pm\sim3$.  Using the SED fitting method for host galaxy removal, for $0.03 < \lambda_{\rm Edd} < 0.2$ the high-absorption objects have bolometric corrections of $\sim 32 \pm 10$, possibly representing part of the transitional region from low to high bolometric corrections near Eddington ratios of $\sim 0.1$ proposed by \cite{2007MNRAS.381.1235V}, (2009), but better constraints on the host galaxy contamination and black hole masses are needed before this can be established.  These bolometric corrections for our local, high-absorption AGN represent a new angle on uncovering the accretion processes at work in this important class of object.  The objects not detected by \emph{IRAS} ($\sim 30$ per cent of the potential sample) are expected to show similar results based on considerations of their upper limiting IR luminosities, but more sensitive IR observations are needed to confirm this.

These preliminary suggestions that the majority of both high- and low-absorption \emph{IRAS}-detected objects in the \emph{Swift}-BAT catalogue are at low Eddington ratios and exhibit low bolometric corrections reinforces unified scenarios in which the processes at work in AGN of different absorption levels are fundamentally similar, but the observed properties vary chiefly based on orientation.  The possible discovery of the lower end of an Eddington-ratio dependent bolometric correction for high-absorption objects would be an interesting aspect of this finding, subject to the uncertainties already discussed.

Recently, \cite{2009arXiv0905.4439L} have studied the Eddington ratio distribution of Seyfert 2s from the optical perspective, by estimating the optical bolometric correction factor to the [\textsc{O}III] line luminosity.  For their sample of higher redshift AGN ($0.3<z<0.8$), they find that Seyfert 2s do not accrete close to the Eddington limit and Eddington ratios are $\sim 0.1$ on average.  This work reinforces previous findings of predominantly $\lambda_{\rm Edd}<0.1$ accretion in the low-redshift universe (e.g. \citealt{1989ApJ...346...68S}): here we extend this to an absorption-unbiased sample.

High absorption objects are the dominant component of the X-ray background, and the identification of predominantly low bolometric corrections for them, using a larger sample than in \cite{2007A&A...468..603P} tells us how accretion onto black holes in the past can account for the current population of dormant massive black holes (as outlined by \citealt{1982MNRAS.200..115S} and later authors).  Lower bolometric corrections for low accretion rates found here are appropriate for matching the X-ray background energy density to the local black hole mass density \citep{2004cbhg.symp..446F}. 

Additionally, the distribution of bolometric corrections in \cite{2007A&A...468..603P} ($35\pm 9$) is similar to the one we obtain for $0.03<\lambda_{\rm Edd}<0.2$ sources (using method 2 for host galaxy correction); indeed the Eddington ratios in the Pozzi et al. sample also lie predominantly in this range.  Our extension of their method down to Eddington ratios as low as $\sim 6 \times 10^{-3}$ for a representative sample of local obscured Seyferts is therefore valuable.

This study is an extension of the Pozzi et al. technique of using the IR and hard X-rays (14--195 keV) for estimating the bolometric output in a way that is less prone to the spectral complexities which affect the optical--UV and X-ray regimes, particularly for obscured objects.  However, as discussed at length, appropriate correction for the host galaxy contamination in the IR is essential.  A useful extension of this study would be to use higher resolution \emph{Spitzer} data in combination with BAT data to perform the same calculation, which would provide a more accurate measure of the nuclear reprocessed IR without having to apply the large contamination corrections needed here (e.g. \citealt{2008ApJ...689...95M}).  The uncertainties in the covering fraction also represent a limitation of the study, and future work determining the appropriate distribution of covering fractions to use will be very valuable.

Some of the uncertainties in the black hole mass estimates should be addressed in the analyses of Winter et al. and Koss et al. (in prep) which will provide $H\beta$ linewidth-based estimates of the black hole masses in the catalogue, which can be used to refine the position of the sources on the Eddington ratio axis.

One class of object for which our approach may not be appropriate is the Ultraluminous Infrared Galaxy (ULIRG) class.  In these, the mid-infrared continuum is not directly attributable to the AGN (e.g. \citealt{2006ApJ...640..167A}) and so they should be corrected for or excluded when considering the IR continuum of large samples of AGN, as discussed by \cite{2009ApJ...697..182T}.  However the redshift distribution of ULIRGs significantly differs from the BAT catalogue, so we expect contamination to be minimal, with none of the objects presented in our study being classified as ULIRGs in NED.

Questions remain as to exactly what the IR continuum due to reprocessing in AGN is.  The continuum produced depends on the dust distribution, grain size and torus geometry; in particular the clumpiness or smoothness of the dust distribution is important.  Another useful extension to this work would be to explore how different continuum models, for example those for clumpy tori (\citealt{2006A&A...452..459H}, \citealt{2008ApJ...685..147N}, \citealt{2008ApJ...685..160N}), would alter the values of $L_{\rm bol}$ found here, especially as recent observational evidence strongly favours the clumpy torus scenario. The recent work of \cite{2009arXiv0906.5368R}, who fit clumpy torus models to mid-IR SEDs for a small sample of AGN, paves the way for further studies in this area.

The 9-month BAT catalogue \citep{2008ApJ...681..113T} has provided numerous opportunities to understand the accretion processes at work in local AGN (e.g. \citealt{2008arXiv0807.4695M}, \citealt{2009ApJ...690.1322W}, \citealt{2009arXiv0907.2272V}).  Similar follow-up studies on the recently-released 22-month BAT catalogue will afford the chance to widen these analyses to a larger, more representative sample of local AGN.

\section{Acknowledgements}

RVV acknowledges support fom the Science and Technology Facilities Council (STFC), ACF thanks the Royal Society for Support and PG acknowledges a RIKEN Foreign Postdoctoral Research fellowship.  We thank the \emph{Swift}/BAT team for the 9-month AGN catalogue BAT data.  We thank Richard McMahon for help with processing and understanding the \emph{IRAS} data.  We thank Laura Silva for providing us with the SED templates and clarifying related issues.  We also thank the anonymous referee for helpful comments and suggestions that improved this paper.  This research has made use of the NASA Extragalactic Database (NED) and the NASA/IPAC Infrared Science Archive, which are operated by the Jet Propulsion Laboratory, California Institute of Technology, under contract with the National Aeronautics and Space Administration.

\bibliographystyle{mnras} 
\bibliography{swiftIRASBATseds}

\end{document}